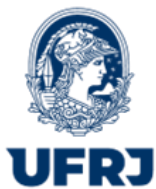

UNIVERSIDADE FEDERAL DO RIO DE JANEIRO
CENTRO DE CIÊNCIAS DA SAÚDE
INSTITUTO DE PESQUISAS DE PRODUTOS
NATURAIS WALTER MORS

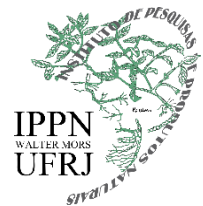

CONSTITUINTES QUÍMICOS E ATIVIDADES BIOLÓGICAS DAS FOLHAS E
CASCAS DO CAULE DE *Ficus maxima* Mill. (MORACEAE)

Felipe Costa Cardoso

Rio de Janeiro

2023

Felipe Costa Cardoso

CONSTITUINTES QUÍMICOS E ATIVIDADES BIOLÓGICAS DAS FOLHAS E CASCAS DO CAULE DE *Ficus maxima* Mill. (MORACEAE)

Dissertação de mestrado apresentada ao Programa de Pós-Graduação de Química de Produtos Naturais, Instituto de Pesquisas de Produtos Naturais Walter Mors, Universidade Federal do Rio de Janeiro, como requisito parcial à obtenção do título de Mestre em Ciências.

Orientador: Roberto Carlos Campos Martins

Rio de Janeiro

2023

## CIP - Catalogação na Publicação

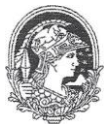

**UFRJ**
UNIVERSIDADE FEDERAL
DO RIO DE JANEIRO

FELIPE COSTA CARDOSO
(AUTOR)

PROF. ROBERTO CARLOS CAMPOS MARTINS (ORIENTADOR)

Dissertação de mestrado submetida ao Programa de Pós - Graduação em Química de Produtos Naturais, Instituto de Pesquisas de Produtos Naturais da Universidade Federal do Rio de Janeiro - UFRJ, como parte dos requisitos necessários à obtenção do título de Mestre em Ciências.

Aprovada por:

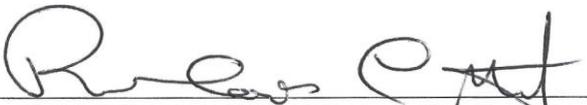

Prof. Roberto Carlos Campos Martins

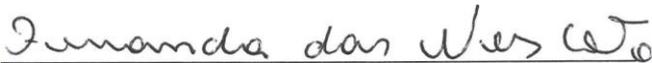

Profª. Fernanda das Neves Costa

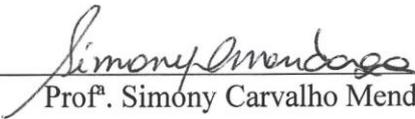

Profª. Simony Carvalho Mendonça

Rio de Janeiro
Setembro/2023

# AGRADECIMENTOS

"A ciência é um processo contínuo. Nunca termina. Não existe uma única e definitiva verdade a ser alcançada, após a qual todos os cientistas poderão se aposentar."

*Carl Sagan – Cosmos, 1980.*

# RESUMO

*Ficus maxima*, pertencente à família Moraceae, subgênero *Pharmacosycea*, é uma planta medicinal popularmente conhecida no Brasil por "caxinguba", termo do tupi-guarani que significa "árvore que dá seiva contra verme". No Brasil, essa espécie ocorre naturalmente nos estados da região Norte e Mato Grosso. Folhas e frutos de *F. maxima* são importantes fontes alimentares para aves e mamíferos e são utilizadas por povos indígenas das Américas Central e do Sul para tratamento de parasitas intestinais, gengivite, inflamações internas e picadas de cobras. Apesar de sua importância farmacológica, a literatura carece de informações e estudos acerca dos constituintes químicos e atividades biológicas dessa espécie. Diante disso, o presente trabalho objetivou realizar a caracterização química de folhas e cascas do caule de *F. maxima* através da técnica de Cromatografia Líquida de Ultra Eficiência associada à Espectrometria de Massas sequencial (CLUE-EM-EM) e testar a atividade antinociceptiva e de inibição de enzimas CYP1A a partir dos extratos etanólicos e frações oriundas destes. O material vegetal foi coletado nas ilhas de Abaetetuba, Pará, Brasil em outubro de 2013, e submetidos à extração e posterior fracionamento em cromatografia em coluna de gel de sílica. As análises por CLUE-EM-EM foram realizadas com ionização por *eletrospray* (IES) nos modos positivo e negativo, e a identificação putativa dos metabólitos foi realizada com auxílio das bibliotecas de EM-EM e redes moleculares da plataforma *Global Natural Products Social Networking* (GNPS). No total, 45 metabólitos das classes dos flavonoides, triterpenos, cumarinas, ácidos graxos poliinsaturados, aminoácidos e alcaloides puderam ser identificados putativamente no extrato etanólico e frações das folhas de *F. maxima*. Foi possível identificar **100**, o triterpeno **51**, os flavonoides **75** e **70** e o alcaloide **29** como constituintes majoritários no extrato das folhas. O extrato etanólico das cascas do caule apresentou potente atividade antinociceptiva (AA) na fase inflamatória do teste da formalina, com mecanismo de ação nas vias muscarínicas; apresentou também 62,6 ± 9,2% de AA, com mecanismo de ação atuante na via dos opioides, no teste da placa quente. A fração em diclorometano (DCM) das folhas de *F. maxima* apresentou potente atividade inibitória de CYP1A1 *in vitro*. Os resultados do trabalho permitiram confirmar o uso da *F. maxima* como agente anti-inflamatório e correlacionar os constituintes majoritários analisados por CLUE-EM-EM com as atividades observadas. Contudo, é importante ressaltar a necessidade de mais estudos para isolamento e caracterização de substâncias ainda não identificadas em *F. maxima* para incluir a espécie como potencial fonte terapêutica alternativa na medicina popular.

Palavras-chave: *Ficus maxima;* inflamação; CYP1A; CLUE-EM-EM.

# ABSTRACT

*Ficus maxima*, Moraceae family, subgenus *Pharmacosycea*, is a medicinal plant popularly known in Brazil as "caxinguba", a Tupi-Guarani term that means "tree that gives sap against worms". In Brazil, this species occurs naturally in the states of the North region and Mato Grosso state. *F. maxima* leaves and fruits are important food sources for birds and mammals and are used by indigenous peoples of Central and South America to treat intestinal parasites, gingivitis, internal inflammation, and snakebites. Despite its pharmacological importance, there's a lack of information and studies in the literature, about the chemical constituents and biological activities, on this species. In view of this, the present work aimed to carry out the chemical characterization of leaves and stem bark of *F. maxima* through the technique of Ultra-Performance Liquid Chromatography-tandem Mass Spectrometry (UPLC-MS-MS) and to test the antinociceptive activity and CYP1A inhibition from ethanolic extracts and species fractions. The plant material was collected on the islands of Abaetetuba, Pará, Brazil in October 2013, and subjected to extraction and subsequent fractionation in silica gel column chromatography. UPLC-MS-MS analysis were performed with electrospray ionization (ESI) in positive and negative modes, and the putative identification of metabolites was performed supported by MS-MS libraries and molecular networks within the Global Natural Products Social Networking platform (GNPS). In total, 45 metabolites from the classes of flavonoids, triterpenes, coumarins, polyunsaturated fatty acids, amino acids and alkaloids could be putatively identified in the ethanolic extract and fractions of the leaves of *F. maxima*. It was possible to identify the **100**, the triterpene **51** and the flavonoids **75** and **70**, and the alkaloid **29** as the main constituents in the leaves extract. The stem bark ethanolic extract showed potent antinociceptive activity (AA) in the inflammatory phase in formalin test, with a mechanism of action in the muscarinic pathways; it also showed $62.6 \pm 9.2\%$ of AA, with a mechanism of action active in the opioid pathway, in the hot plate test. The dichloromethane fraction (DCM) of *F. maxima* leaves showed potent CYP1A1 inhibitory activity *in vitro*. The results of the work led to the confirmation of the use of *F. maxima* as an anti-inflammatory agent and to the correlation of the major constituents analyzed by UPLC-MS-MS with the observed activities. However, it is important to emphasize that further studies are needed to isolate and characterize substances not yet identified in *F. maxima* to include the species as a potential alternative therapeutic source in folk medicine.

Keywords: *Ficus maxima;* inflammatory activity; CYP1A; UPLC-MS-MS

# LISTA DE ABREVIATURAS E UNIDADES

| | |
|---|---|
| ACN | Acetonitrila |
| AcOEt | Acetato de etila |
| CaxCsc-AcOEt | Fração em acetato de etila das cascas do caule de *F. maxima* |
| CaxCsc-DCM | Fração em diclorometano das cascas do caule de *F. maxima* |
| CaxCsc-EB | Extrato etanólico das cascas do caule de *F. maxima* |
| CaxCsc-Hex | Fração em hexano das cascas do caule de *F. maxima* |
| CaxFls-AcOEt | Fração em acetato de etila das folhas de *F. maxima* |
| CaxFls-DCM | Fração em diclorometano das folhas de *F. maxima* |
| CaxFls-EB | Extrato etanólico das folhas de *F. maxima* |
| CaxFls-Hex | Fração em Hexano das folhas de *F. maxima* |
| CCD | Cromatografia em Camada Delgada |
| $CD_3OD$ | Metanol deuterado |
| $CDCl_3$ | Clorofórmio deuterado |
| CLUE-EM-EMAR | Cromatografia Liquida de Ultra Eficiência associada à Espectrometria de Massas sequencial de Alta Resolução |
| d | Dupleto |
| Da | Dalton |
| DCM | Diclorometano |
| DMSO | Dimetilsulfóxido |
| EM | Espectrometria de massas |
| EM-EM | Espectrometria de massas sequencial |
| EtOH | Etanol |
| FMBN | *Feature Based Molecular Networking* |
| GNPS | *Global Natural Products Social Molecular Networking* |
| $H_2SO_4$ | Ácido sulfúrico |
| Hex | Hexano |
| HMBC | *Heteronuclear Multiple Bond Correlation* |
| HSQC | *Heteronuclear Multiple Quantum Coherence* |
| Hz | Hertz |
| i.p. | intraperitoneal |
| $J$ | Constante de acoplamento em Hertz |
| $m/z$ | Razão massa/carga |
| MeOH | Metanol |
| N/D | Não-disponível |
| nm | Nanômetro |
| ppm | Partes por milhão |
| RMN $^{13}$C | Ressonância Magnética Nuclear de Carbono 13 |
| RMN $^{1}$H | Ressonância Magnética Nuclear de Hidrogênio 1 |
| TR | Tempo de Retenção em minutos |
| UV | Ultravioleta |
| δ | Deslocamento químico em ppm |

# LISTA DE FIGURAS

# LISTA DE TABELAS

# LISTA DE ESQUEMAS

# LISTA DE GRÁFICOS

# LISTA DE FLUXOGRAMAS

# SUMÁRIO

# 1 INTRODUÇÃO

## 1.1 Os produtos naturais

A natureza tem suprido as necessidades básicas dos seres humanos ao longo dos tempos, sobretudo de medicamentos para tratamento de inúmeras doenças. A maioria dos fármacos tem sido descoberta em vez de planejada, sendo essa a principal razão pela qual grande parte das substâncias medicinais são produtos naturais ou derivados. De acordo com a Organização Mundial da Saúde (OMS), cerca de 80% da população mundial faz uso de extratos vegetais ou de seus princípios ativos como terapia na medicina tradicional, principalmente em países em desenvolvimento (WERMUTH *et al.*, 2015).

Considerada uma prática milenar, a experimentação popular de diversas plantas, principalmente, contribuiu para a diferenciar as espécies com efeitos benéficos daquelas que, ou têm efeitos tóxicos para o homem ou são ineficazes (PHILLIPSON; ANDERSON, 1989; CLARK, 1996; MACIEL *et al.*, 2002).

Neste contexto, é o reino vegetal quem tem contribuído majoritariamente para o fornecimento de substâncias utilizáveis no tratamento de doenças. Os produtos do metabolismo secundário das plantas têm apresentado grande potencial farmacológico (ANDRADE; CASALI, 1999; MONTANARI; BOLZANI, 2001; VEIGA JUNIOR; PINTO; MACIEL, 2005).

Como exemplo, pode-se citar o isolamento do agente analgésico e indutor de sono morfina (um alcaloide), em 1804, a partir do ópio extraído da planta *Papaver somniferum*, da família Papaveraceae. Este marco impulsionou intensas investigações em plantas medicinais no século XIX que culminaram no isolamento e purificação de uma diversidade de substâncias, em sua maioria alcalóides, como quinina (**1**), cafeína (**2**), nicotina (**3**), colchicina (**4**) etc. (Figura 1) (BERNARDINI *et al.*, 2017).

Figura 1 - Estruturas de alcaloides extraídos de plantas.

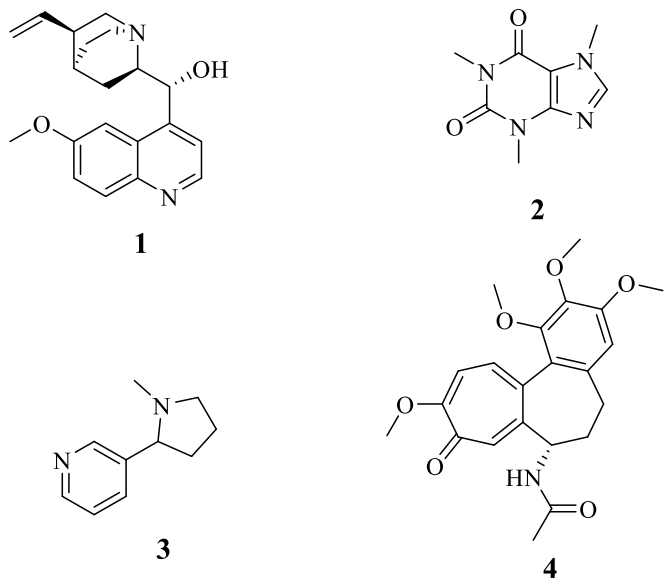

Fonte: Elaborado pelo autor.

No Brasil, um país de grandes dimensões territoriais, a diversidade de ambientes ecogeográficos torna-se propícia ao cultivo de uma vasta quantidade de plantas medicinais que impulsionam o desenvolvimento de terapias com produtos naturais. Contudo, muitas plantas da flora nativa acabam sendo consumidas com pouco ou nenhum conhecimento acerca de suas propriedades farmacológicas, uma cultura que é extensamente propagada por gerações (MACIEL *et al.*, 2002; VEIGA JUNIOR; PINTO; MACIEL, 2005; DONATO *et al.*, 2020).

Um grande exemplo de espécies vegetais muito importantes e que têm feito parte da história da civilização humana são as figueiras, pertencentes ao gênero *Ficus* da família Moraceae. A figueira comum (*Ficus carica*), produtora do fruto comestível, era cultivada em todas as civilizações do Mediterrâneo na Antiguidade. Essa espécie é a primeira planta descrita na Bíblia, cujas folhas foram utilizadas por Adão e Eva para se cobrirem. Presentes em praticamente todos os continentes, exceto a Antártica, as figueiras foram incorporadas em diversas culturas dos povos desses locais, sobretudo em tradições *religiosa*s (GIBBONS, 2006). Outras espécies importantes do gênero, como *Ficus religiosa, Ficus benghalensis* e *Ficus racemosa* têm despertado o interesse de pesquisadores devido aos usos históricos dessas plantas na medicina popular para o tratamento de várias enfermidades, doenças infecciosas e até mesmo câncer (NAWAZ; WAHEED; NAWAZ, 2019).

Diante do exposto, fica evidente a crescente necessidade do estudo químico e farmacológico dessas plantas e outras espécies do gênero a fim de se compreender suas propriedades terapêuticas e conhecer seus compostos bioativos.

## 1.2    A família Moraceae

A família Moraceae (Morácea), comumente apelidada de "família da figueira", é uma família botânica que inclui árvores, arbustos, lianas e ervas. Agrupando cerca de 38 gêneros e mais de 1.100 espécies, as Moráceas distribuem-se naturalmente nas regiões tropicais e subtropicais do globo. A família inclui importantes espécies produtoras de frutos comestíveis com considerável relevância econômica, como, por exemplo, a jaqueira (*Artocarpus heterophyllus*), a fruta-pão (*Artocarpus altilis*), amoreira (*Morus spp.*) e, como já mencionado anteriormente, a figueira (*Ficus carica*) (NAWAZ; WAHEED; NAWAZ, 2019; SALEHI *et al.*, 2020).

Os principais gêneros incluídos nesta família são *Ficus* (mais de 800 espécies) e *Dorstenia* (cerca de 110 espécies). No Brasil, estão presentes 21 gêneros de Moraceae e, pelo menos, 71 espécies endêmicas das 230 reconhecidas (PORTO, 2011; PEDERNEIRAS; MACHADO; SANTOS, 2022).

A maioria dos trabalhos publicados com espécies da família Moraceae descreve o isolamento e caracterização de metabólitos secundários das classes dos terpenos, descrevendo suas atividades anticancerígenas e de compostos fenólicos, como os flavonoides, com atividades antidiabéticas e anti-inflamatórias (SIENIAWSKA *et al.*, 2022; PORTO, 2011).

### 1.2.1    O Gênero *Ficus*

O gênero *Ficus* apresenta aproximadamente 850 espécies com plantas extremamente diferentes. Desde árvores com mais de trinta metros de altura até pequenos arbustos e espécies rasteiras (SALEHI *et al.*, 2020). A figueira comum (*F. carica*) é uma das espécies do gênero mais abundantes em florestas tropicais e tem sido amplamente cultivada no mundo todo. Os frutos de *Ficus,* bem característicos para o gênero, são um recurso chave por servirem de alimento para diversos frugívoros, incluindo morcegos.

O gênero ainda pode ser dividido em pelo menos seis subgêneros, a saber: *Ficus, Pharmacosycea, Sycidium, Sycomorus, Synoecia e Urostigma* (BERG, 1989).

Dentro do gênero *Ficus spp.*, as principais classes de metabólitos já identificados e reportados na literatura incluem cumarinas (RAMADAN *et al.*, 2009; RAHEEL *et al.*, 2017), aminoácidos (SWAMI; BISHT, 1996; SINGH *et al.*, 2020), flavonoides (GASPAR

DIAZ *et al.*, 1997; KHANAL; PATIL, 2020; DESTA; SHUMBAHRI; GEBREHIWOT, 2020) e alcaloides (SWAMI; BISHT, 1996; DURESHAWAR *et al.*, 2019; SINGH *et al.*, 2020), extraídos de todas as partes da planta. A Tabela 1 apresenta os principais metabólitos e espécies de *Ficus* estudadas mais recentemente.

Tabela 1 - Classes de metabólitos extraídos de espécies de *Ficus*.

| Classe química | Espécie botânica | Partes estudadas | Referência |
|---|---|---|---|
| Monoterpenos | *F. exasperata* | Cascas da raíz | Oladosu *et al.*, 2009 |
| Diterpenos e Triterpenos | *F. ulmifolia; F. fistulosa; F. microcarpa; F. benghalensis F. pandurata; F. carica* | Folhas; cascas do caule; raízes aéreas; látex | Ragasa e Tsai. 2009 (a); Tuyen *et al.*, 1999; Kuo e Chaiang, 1999; Khanal e Patil, 2020; Ramadan *et al.*, 2009; Desta *et al.*, 2020 |
| Esteroides | *F. benghalensis; F. racemosa; F. pandurata; F. benjamina* | Folhas; cascas do caule | Bhar e Thakur, 1981; Ragasa e Tsai, 2009; Ramadan *et al.*, 2009; Swami e Bisht, 1996 |
| Cumarinas | *F. benghalensis;* | Cascas do caule | Raheel *et al.*, 2017. |
| Flavonoides | *F. benghalensis; F. maxima; F. carica* | Cascas; Folhas | Khanal e Patil, 2020; Gaspar *et al*, 1997; Desta *et al.*, 2020 |
| Aminoácidos | *F. benghalensis; F. religiosa* | Partes aéreas | Daniel *et al.*, 1998; Singh *et al.*, 2020 |
| Alcaloides | *F. carica; F. religiosa* | Folhas; partes aéreas | Dureshahwar *et al.*, 2019; Singh *et al.*, 2020 |

### 1.2.2 A espécie *Ficus maxima* Mill. *e* sua importância

Quase todas as espécies de *Ficus* têm sido usadas tradicionalmente na medicina popular para tratamento de doenças respiratórias e doenças da pele com condições inflamatórias principalmente (NAWAZ; WAHEED; NAWAZ, 2019). Uma dessas espécies de relevante interesse, porém pouco estudada, é a *F. maxima* Mill., do subgênero *Pharmacosycea*, conhecida popularmente no Brasil por "caxinguba" ou "caxinguba da folha grande". O nome "caxinguba" vem do tupi-guarani "kuaxingýua" e significa "árvore que dá xarope ou seiva medicinal contra verme".

Com árvores que variam de cinco a trinta metros de altura, a espécie *F. maxima* se distribui geograficamente do norte do Caribe ao sudoeste da América do Sul (Figura 2). No Brasil, são confirmadas ocorrências da espécie nos estados da região Norte e no estado do Mato Grosso, na região Centro-Oeste (PEDERNEIRAS; MACHADO; SANTOS, 2022).

Figura 2 - (a) Árvore da espécie *F. maxima*. (b) Detalhe das folhas e caule de *F. maxima*.

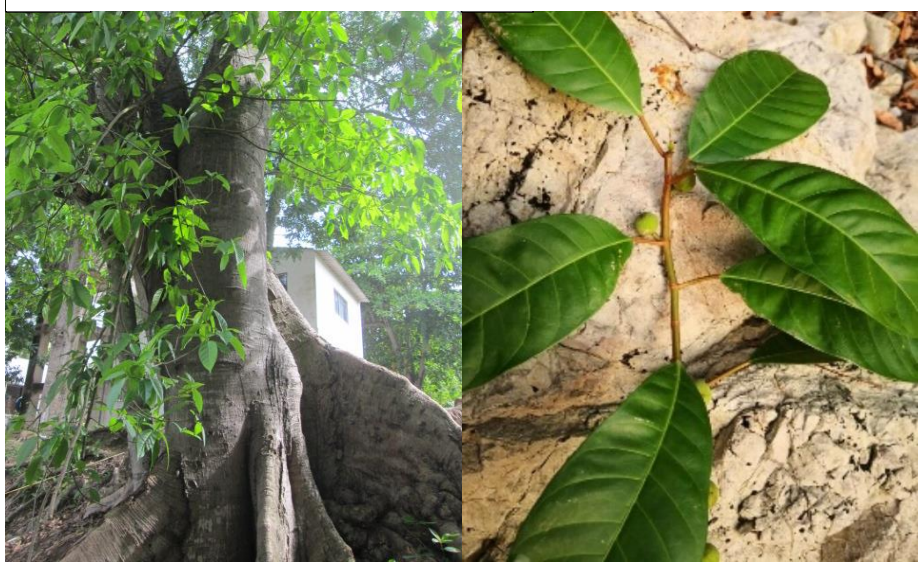

Fonte: https://identify.plantnet.org/es/k-world-flora/species/*Ficus*%20maxima%20Mill./data. Acesso: 1 de set. 2023.

As folhas e frutos de *F. maxima* (Figura 3) são importantes fontes alimentares para aves e mamíferos, sendo também utilizadas por alguns povos indígenas para tratar inflamações internas, parasitas intestinais e gengivite (LENTZ, 1993; TENE *et al.*, 2007). Uma das ações medicinais da caxinguba vem da casca e do látex utilizado como chá para tratar verminoses (MARTINS *et al.*, 2005). Informações etnofarmacológicas afirmam que esse látex, porém, deve ser utilizado com cautela, pois é cáustico, devido à presença do princípio ativo da classe dos alcaloides denominado de "caxinguvina". Outros usuários da espécie relatam que o chá/decocção das folhas também é utilizado para tratar vermes, diarreia, febre e dores no estômago (BOURDY *et al.*, 2000).

Figura 3 - (a) Figos em corte de *F. maxima*; (b) Folhas e parte do caule de *F. maxima*; (c) Tronco e corte da casca do tronco de *F. maxima*.

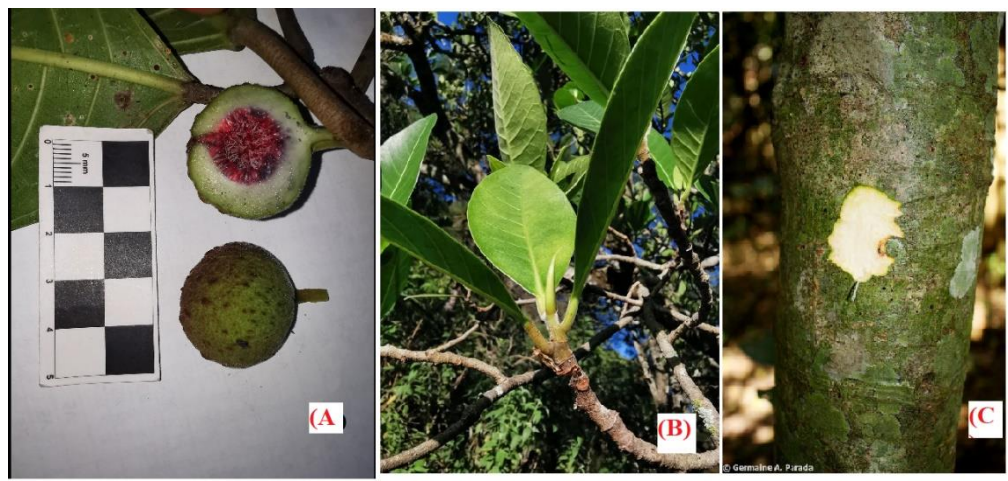

Adaptado da galeria de fotos do *Global Biodiversity Information Facility* (GBIF), autores: Pablo Carrillo-Reyes, Germaine A. Parada. disponível em: https://www.gbif.org/pt/occurrence/gallery?taxon_key=5571386. Acesso: 25 de julho de 2022.

As folhas de *F. maxima* também são utilizadas pelo povo Maya Lacandon para o tratamento de picadas de cobra. As folhas são umedecidas pela mastigação e depois aplicadas no local da picada (KASHANIPOUR; MCGEE, 2004). A caxinguba também é muito utilizada na medicina popular como agente anti-helmíntico e antirreumático (GASPAR DIAZ *et al.*, 1997).

Realizando uma busca por investigações fitoquímicas de *F. maxima* na literatura é retornado um único trabalho datado de 1997 citando o isolamento e a caracterização de quatro flavonoides *O*-metilados (**5 – 8**) a partir das folhas, sendo um deles (**8**) não descrito em espécies de Moraceae à época (Figura 4) (GASPAR DIAZ *et al.*, 1997)

Figura 4 - Estruturas de flavonoides O-metilados extraídos das folhas de *F. maxima*.

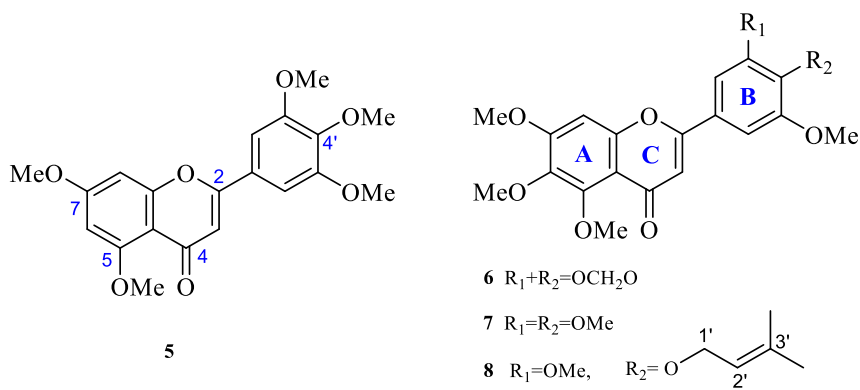

Adaptado de Gaspar Diaz *et al*, 1997.

Até o momento, não foram encontrados outros estudos abrangendo dados acerca da composição química ou atividades biológicas da espécie em questão.

Embora haja uma grande diversidade de espécies de *Ficus* farmacologicamente importantes, em grande parte oriundas dos continentes Africano e Asiático (por exemplo, *F. carica*, *F. racemosa, F. religiosa, F. benghalensis*), poucos são os registros de estudos fitoquímicos e de atividades biológicas de espécies brasileiras. Uma busca pelas bases dados mais populares como *Scopus, Web of Science, PubMed Central* e SciElo utilizando os termos "*Ficus maxima*" retorna apenas algumas dezenas de trabalhos nos últimos vinte anos. Destes, os destaques são para a identificação e catalogação de novas espécies.

Levando em consideração os diversos grupos de metabólitos secundários biossintetizados pelas espécies de *Ficus*, como flavonoides, cumarinas, terpenos e alcaloides com atividades biológicas promissoras e a escassez de trabalhos acerca de estudos químicos de espécies deste gênero no território brasileiro, o presente trabalho visou realizar o estudo fitoquímico e de atividades antinociceptiva e de inibição de CYP1A1 dos extratos etanólicos das folhas e cascas do caule de *Ficus maxima* Mill. coletadas na região Amazônica do estado do Pará.

## 2   OBJETIVOS

### 2.1  Objetivo geral

- Realizar o estudo químico do extrato alcoólico das folhas e cascas do caule da espécie *Ficus maxima* e avaliar a sua possível atividade anti-inflamatória, antinociceptiva e inibitória de enzimas relacionadas à metabolização de fármacos no organismo humano (CYPs).

### 2.2  Objetivos específicos

- Isolamento e purificação dos metabólitos a partir dos extratos por técnicas de partição líquido-líquido em hexano (Hex), diclorometano (DCM), acetato de etila (AcOEt) e fracionamento dos extratos por técnica de cromatografia líquida em coluna (CC) aberta de gel de sílica;

- Identificação putativa das substâncias através de análise de cromatografia líquida de ultra eficiência associada à espectrometria de massas de alta resolução sequencial (CLUE-EM-EMAR) com auxílio da biblioteca espectral da plataforma GNPS;

- Caracterização das substâncias isoladas por técnica de Ressonância Magnética Nuclear de hidrogênio e carbono 13 uni e bidimensionais (RMN $^1$H e $^{13}$C 1D e 2D);

- Avaliação de atividade antinociceptiva *in vivo* dos extratos e frações;

- Avaliação da inibição de enzimas do sistema Citocromo P450 pelos extratos e frações.

## 3  METODOLOGIA

### 3.1  Coleta e identificação botânica

Folhas frescas e cascas do caule de *F. maxima* (massa total não-disponível) foram coletadas na região das ilhas de Abaetetuba, Pará, Brasil (lat.: -1.62527; long.: -48.954167) em outubro de 2013. A identificação da espécie foi realizada pelo parataxonomista Carlos Alberto Santos da Silva, do Museu Paraense Emílio Goeldi. Uma exsicata foi produzida e registrada sob o número *voucher* MFS002121 no Herbário Profª. Drª. Marlene Freitas da Silva (MFS) da Universidade do Estado do Pará (UEPA). A coleta da espécie possui autorização do Conselho de Gestão do Patrimônio Genético (CGEN) sob o número 148/2013.

### 3.2  Preparo dos extratos e obtenção das frações

O material vegetal foi secado em estufa, a 40 °C por 48h, e em seguida triturado em moinho de facas até ser reduzido a pó. Após isso, foi submetido à extração a frio por maceração com etanol, obtendo-se, assim, uma massa de 5,3 g de extrato etanólico bruto das folhas (código CaxFls-EB) e 1,8 g das cascas (código CaxCsc-EB). Foram separados também aproximadamente de 50 a 80 mg de cada extrato para ensaios biológicos.

A massa de 5,3 g de CaxFls-EB foi redissolvida em solução de metanol:água (MeOH:$H_2O$) na proporção de 3:7 e logo após submetida ao particionamento líquido-líquido utilizando como solventes de extração hexano (Hex), diclorometano (DCM) e acetato de etila (AcOEt) das marcas AcmaLabs® e Neon®, gerando três frações orgânicas: CaxFls-Hex, CaxFls-DCM, CaxFls-AcOEt e uma hidroalcoólica. De modo semelhante, a massa de 1,8 g de CaxCsc-EB foi particionada gerando as frações CaxCsc-Hex, CaxCsc-DCM e CaxCsc-AcOEt. As partições foram realizadas três vezes com um volume aproximado de 200 mL de solvente cada. As frações orgânicas obtidas foram secadas com sulfato de sódio anidro ($Na_2SO_4$), filtradas por gravidade e concentradas em evaporador rotativo até completa remoção do solvente.

### 3.3  Análise das frações por cromatografia em camada delgada (CCD)

As frações oriundas do particionamento foram analisadas por cromatografia em camada delgada (CCD), em um sistema de solventes Hex:AcOEt 7:3 para avaliação de seu perfil químico. Foram utilizadas cromatoplacas de gel de sílica em suporte de alumínio da marca Silicycle de 200 μm e marcador de fluorescência a 254 nm (F254). Como método de revelação física não-destrutiva das amostras de CCD foi utilizada irradiação com a lâmpada ultravioleta em câmara escura nos comprimentos de onda de 254 nm e 365 nm. Como métodos de revelação química das amostras analisadas por CCD foram utilizados como reveladores: solução de ácido sulfúrico 10% (v/v) em etanol; e solução de vanilina 1% (p/v) em metanol (Neon®) e posterior etapa de aquecimento a 110 °C em manta aquecedora IKA® C-MAG HS7 de 2 a 5 minutos.

### 3.4  Análise fitoquímica por ensaio colorimétrico

Os extratos brutos da casca e folhas de *F. maxima*, bem como suas partições foram submetidos a ensaio colorimétrico adaptado com soluções reveladoras empregadas em CCD para detecção de esteroides e terpenos, e de alcaloides (MATOS, 2009). Para tanto, aproximadamente 2 mg das amostras foram solubilizadas em 2 mL de solvente apropriado (DCM ou MeOH) e transferidas para tubos de ensaio ou frascos de vidro.

Para análise de triterpenos, foi preparada uma solução, sob banho de gelo, de 50 mL de anidrido acético (Tedia Brazil®) contendo 5 mL de ácido sulfúrico 97% (Vetec ®). A solução foi deixada refrigerada a 4 °C por 30 minutos antes do uso. Em seguida, foi adicionado 1 mL desta solução aos tubos de ensaio ou frascos contendo as amostras e observadas as mudanças ou não de coloração. A formação de coloração verde forte ou verde-azulada indica a presença de esteroides ou terpenos.

Para análise de alcaloides, foi utilizada uma solução do reagente de Dragendorff comercial da marca ACS Científica. Do mesmo modo, 1 mL do reagente foi adicionado aos tubos de ensaio ou frascos contendo as amostras. A formação de precipitado de coloração laranja ou laranja-avermelhado indica a presença de alcaloides.

### 3.5  Análise fitoquímica por cromatografia líquida de ultra eficiência associada à espectrometria de massas sequencial (EM-EM) de alta resolução (CLUE-EM-EMAR)

As análises por CLUE-EM-EMAR foram feitas no Laboratório de Metabolômica (LabMeta) do LADETEC UFRJ utilizando o cromatógrafo líquido Dionex UltiMate 3000 UHPLC da Thermo Scientific® empregando uma coluna Zorbax® Eclipse plus C18 (2,1 mm; 50 mm; 1,8 μm) mantida a 40 °C. O sistema foi associado a um espectrômetro de massas híbrido Quadrupolo-Orbitrap, modelo Q-Exactive Plus (Thermo Scientific®), com uma fonte de ionização por eletrospray (IES) operando no modo de ionização positivo e negativo na faixa de 100-1000 Da.

Cada amostra foi preparada na concentração de 1 mg/mL em ACN ou MeOH com solução de 1 μL do padrão interno 3-fluorofenilalanina (1 mg/mL) sob agitação em misturador de vórtice. As amostras foram centrifugadas a 12.500 rotações por minutos (RPM) durante 1:30 min para remoção de partículas e os sobrenadantes transferidos para *vials*. As separações cromatográficas foram realizadas em coluna analítica de fase reversa com uma taxa de fluxo fixada em 1 mL/min. A coluna foi mantida a 40 °C, o volume de amostra injetado foi de 5 μL e a temperatura do amostrador automático foi mantida a 4 °C. Um gradiente de eluição envolvendo uma programação binária de água acidificada (0,1% de ácido fórmico, v/v) e ACN foi aplicado. O gradiente de eluição começou com 35% de ACN e foi mantido por 5 min, depois atingiu 100% de ACN (rampa linear) em 8 min e foi mantido até os 19 min; em seguida, voltou a 35% ACN e foi mantido por 2,9 min, totalizando um tempo de execução de 22 min. A aquisição dos espectros de massa (EM) foram realizadas nos modos de ionização positivo e negativo e faixa de varredura completa (*m/z* 100 a 1000). Os dados brutos foram analisados com *software* Thermo Xcalibur Qual Browser 4.2.47 (Thermo Scientific®) e MZmine 2.53 extraindo-se os tempos de retenção (TR) e razão *m/z*.

## 3.6    Análises de perfil químico das frações e extrato no GNPS

As amostras CaxFls-EB, CaxFls-Hex, CaxFls-DCM e CaxFls-AcOEt também foram analisadas com auxílio da base de dados de espectrometria de massas através da plataforma de acesso aberto *Global Natural Products Social Networking* (GNPS) para identificação putativa dos metabólitos presentes. CaxCsc-EB e suas frações encontram-se em fase de análises. Os arquivos em formato "raw" das análises de CLUE-EMAR foram convertidos para o formato "mzML" com o *software* MSConvert® 3.0.22257-7de06d7 e carregados para a plataforma, onde a pesquisa das substâncias foi realizada com o fluxo de trabalho *Molecular-Librarysearch-V2*. Foram também criadas redes moleculares com o fluxo de trabalho *Feature Based Molecular Networking* (FBMN). Para tanto, os dados de EM-EM foram primeiro processados com o *software* MZmine 2.53 e em seguida carregados para o GNPS para análise. Os dados foram filtrados removendo todos os íons do fragmento EM-EM dentro de +/- 17 Da do íon precursor. Os espectros de EM-EM foram filtrados, escolhendo apenas os seis principais íons de fragmento na janela de +/- 50 Da em todo o Espectro. A tolerância da massa do íon precursor foi ajustada para 0,02 Da e a tolerância do íon do fragmento EM-EM para 0,02 Da. Uma rede molecular foi então criada onde as arestas foram filtradas para ter uma pontuação de cosseno acima de 0,7 e mais de seis picos correspondentes. As arestas entre dois nodos foram mantidas na rede se, e somente se, cada um dos nodos apareceu nos respectivos dez principais nodos mais semelhantes. Finalmente, o tamanho máximo de uma família molecular foi definido como 100 e as arestas de menor pontuação foram removidas das famílias moleculares até que o tamanho da família molecular estivesse abaixo desse limite. Os espectros na rede foram então pesquisados em bibliotecas espectrais do GNPS. Os espectros da biblioteca foram filtrados da mesma maneira que os dados de entrada. Todas as correspondências mantidas entre espectros da rede e espectros da biblioteca deveriam ter uma pontuação mínima de 0,7 e pelo menos seis picos correspondentes. O fluxo de trabalho *Dereplicator* foi usado para anotar espectros EM-EM. As redes moleculares foram, então, exportadas e visualizadas usando o *software* Cytoscape 3.10.

### 3.7    Análises por Ressonância Magnética Nuclear (RMN)

Os espectros de Ressonância Magnética Nuclear (RMN) foram obtidos no Laboratório Multiusuário de Análises por RMN (LAMAR) do IPPN-UFRJ em espectrômetro Varian® VMR400 ($^{1}H$ 399,739 MHz, $^{13}C$ 100,523 MHz). Utilizou-se como solvente para as amostras o clorofórmio deuterado ($CDCl_3$) (Acros Organics®) ou metanol deuterado ($CD_3OD$) (Sigma-Aldrich®). Como referência interna foi utilizado o tetrametilsilano (TMS). Os deslocamentos químicos ($\delta$) foram expressos em partes por milhão (ppm) e as constantes de acoplamento ($J$) em Hertz (Hz). As áreas relativas dos sinais foram obtidas por integração eletrônica, a calibração dos espectros foi obtida a partir do sinal do TMS. O processamento dos espectros foi realizado utilizando o *software* MestReNova versão 6.0.2.

### 3.8    Ensaio de atividade antinociceptiva via teste de formalina

A atividade antinociceptiva do extrato bruto das folhas, cascas do caule e frações de *F. maxima* foi avaliada no laboratório coordenado pelo Prof. Dr. Guilherme Carneiro Montes, do Departamento de Farmacologia da UERJ. Inicialmente, foi utilizado o teste da formalina, baseado no método de Dubuisson e Dennis (1977 *apud* AZEVEDO *et al.*, 2007) e adaptado para camundongos por HUNSKAAR *et al.,* 1985. Neste experimento, foram utilizados grupos de dez camundongos tratados com o veículo (DMSO) ou CaxFls-EB, CaxCsc-EB nas doses de 3, 10 e 30 mg/kg, intraperitoneal (i.p.). A morfina (1 mg/kg, i.p) e ácido acetilsalicílico (AAS) (150 mg/kg i.p) foram utilizados como substâncias de referência. Trinta minutos após o tratamento com veículo ou as amostras, os camundongos receberam 20 µL de uma solução de formalina (solução de formaldeído 2,5% em salina), através de injeção i.p. na pata traseira direita. O tempo gasto pelos animais em lamber e/ou morder a pata injetada, indicativo de resposta nociceptiva, (SOUZA *et al.*, 2000) foi cronometrado por um período de 30 min.

O tempo gasto (em segundos) pelo animal lambendo ou mordendo a pata durante os primeiros cinco minutos (1ª Fase) após a injeção da formalina é resultante da estimulação direta dos nociceptores, levando a uma resposta neurogênica. A inibição dessa fase é indicativa de fármacos analgésicos que atuam a nível central. Em seguida, há uma interface de aproximadamente 10 minutos caracterizada por mecanismos inibitórios da dor e posteriormente vem a segunda fase. O tempo de lambida da pata durante o intervalo entre 15 e 30 min é um indicador de fase inflamatória gerada tanto pela estimulação de nociceptores como pela liberação de mediadores inflamatórios (HUNSKAAR; HOLE, 1987).

As duas fases foram registradas separadamente para investigar o tipo de efeito do derivado sobre a resposta nociceptiva induzida pela formalina (AZEVEDO *et al.*, 2007; HEIDARI *et al.*, 2009). O comportamento nociceptivo foi quantificado pela reatividade, em segundos, de lamber ou morder a pata que recebeu a injeção. Uma redução significativa na duração de lambida da pata, pelos animais tratados com determinada substância em comparação a um grupo controle, na primeira fase é considerada como uma resposta antinociceptiva neurogênica (BHANDARE *et al.*, 2010). Aos animais que apresentam a resposta comportamental reduzida na segunda fase pode-se aferir uma resposta antinociceptiva inflamatória periférica (KAYSER *et al.*, 2007).

**3.9     Ensaio de atividade antinociceptiva em placa quente**

A placa quente (modelo 7406 – LE) permite avaliar a atividade de substâncias antinociceptivas através de um aparelho cuja temperatura de sua placa, localizada na superfície superior, pode ser controlada. Neste aparelho, um cronômetro acoplado é ativado por um pedal externo, permitindo a medida precisa do tempo do início da reação do animal ao estímulo térmico.

Este teste, descrito por Eddy e Leimback (1953), representa uma modificação do modelo original de Woolfe e MacDonald (1944). Consiste em quantificar a latência para o primeiro sinal de lamber a pata ou saltar para evitar o calor. As reações do animal ao estímulo térmico quando é colocado sobre a placa quente ($52 \pm 0,5$ °C) são respostas indicativas de nocicepção (ALMEIDA; OLIVEIRA, 2006a). O estímulo térmico da placa quente é utilizado para avaliar a atividade analgésica mediada por mecanismos centrais (AL-GHAMDI, 2001). A triagem inicial teve como finalidade avaliar a sensibilidade dos animais ao estímulo térmico. Somente os animais com latência entre 8 e 15 s foram usados. Para esta abordagem experimental, cada grupo experimental (n=10) recebeu os seguintes tratamentos via i.p.: CaxCsc-EB 3, 10 e 30 mg/kg; veículo (DMSO) e morfina (10 mg/kg). As medidas no aparelho foram feitas nos tempos 5, 15, 30, 40, 50, 60, 70, 80, 90 e 100 minutos após administração das substâncias.

Os resultados foram expressos em porcentagem de analgesia e a resposta antinociceptiva, ou seja, a atividade analgésica (% AA), foi calculada através da seguinte fórmula:

$$\%AA = \frac{(Latência\ observada - Latência\ controle)}{(Cutt - off\ -\ Latência\ controle)}\ x\ 100\%$$

O tempo máximo de permanência na placa (*cut-off*) foi de 35 s para se evitar qualquer dano à pata do animal.

## 3.10  Ensaio de atividade inibitória de CYP1A

A avaliação da atividade inibitória de CYP1A foi conduzida pelos colaboradores prof. Dr. Carlos Henrique Ramos da FIOCRUZ e prof. Dr. Davyson de Lima Moreira do Instituto de Pesquisas Jardim Botânico do Rio de Janeiro.

### 3.10.1  Tratamento de camundongos tipo DBA/2

Foram utilizadas durante o tratamento camundongos tipo DBA/2 em idade adulta, pesando entre 23 e 25 g, fornecidas pelo Centro de Criação de Animais de Laboratório (CECAL) da Fiocruz. Os animais receberam suprimento de água e ração ad libitum (ração para camundongos e ratos Nuvital CR1, Nuvilab® Curitiba, PR, Brasil). No biotério do LTA, os animais foram acomodados em gaiolas de plástico com tampa de aço inoxidável e cama de maravalha de pinho branco. A temperatura e a umidade relativa do ar, $23 \pm 2°C$ e aproximadamente 70%, respectivamente, foram mantidas estáveis no ambiente e os animais permaneceram em ciclo claro/escuro de 12h (ciclo claro, das 7:00 às 19:00h). Este trabalho foi aprovado pelo comitê de ética em uso no uso de animais (CEUA) da FIOCRUZ (P-41/19-1).

### 3.10.2  Tratamento

O tratamento dos animais seguiu de acordo Rodrigues e Prough (1991), da seguinte forma: Grupo 1 (n=3) com β-naftoflavona em suspensão de óleo de milho (MKBF2096V, Sigma-Aldrich) por 4 dias consecutivos e por via intraperitoneal (i.p.). Foram administradas diariamente dose única de 80 mg/kg de peso corporal, no período da manhã. O Grupo 2 (n=3) recebeu doses de óleo de milho i.p. pelo mesmo período, caracterizado como grupo controle para comparar o potencial de indução do CYP1A nos roedores. Grupo 3 (n=3) foi tratado com Eup-5 em suspensão de óleo de milho, diariamente, por 4 dias, dose única de 80 mg/kg de peso corporal i.p. para verificar seu comportamento na atividade in vivo do CYP1A.

### 3.10.3   Eutanásia e remoção de órgãos

Todos os animais de ambos os tratamentos foram mortos por deslocamento cervical no quinto dia após jejum de 24 h a contar da última dose. Imediatamente após a eutanásia, foi feita uma ampla incisão longitudinal e outra transversal no abdômen do animal, sendo o fígado rapidamente retirado, resfriado em banho de gelo e pesado. Os fígados foram embalados individualmente em papel alumínio, congelados e armazenados em nitrogênio líquido até o momento do preparo da fração microssomal.

### 3.10.4 Preparo das frações microssomais hepáticas

As frações microssomais hepáticas das ratas tratadas e controles foram obtidas segundo De-Oliveira e colaboradores (1997). Os fígados foram retirados do nitrogênio líquido e imediatamente transferidos para o banho de gelo para descongelar. A partir de então, todo o procedimento e suas etapas subsequentes seguiram em temperatura igual ou inferior a 4 °C. Os fígados foram pesados, lavados com solução de sacarose 250 mM e secados com papel de filtro. Em seguida, os órgãos foram homogeneizados em solução tampão Tris 100 mM com KCl 150 mM (pH 7,4), usando homogeneizador de vidro (capacidade de 50 mL) e pistilo de teflon, a uma velocidade aproximada de 1.200 x g. O volume da solução tampão correspondeu a quatro vezes o peso do órgão. Em seguida, o homogeneizado hepático foi levado à centrifugação em ultracentrífuga Beckman® XL-90 (rotor 70.1 Ti, gentilmente autorizado pelo Instituto Nacional de Controle de Qualidade em Saúde - INCQS) a 4 °C e a 9.000 x g por 30 min. Após centrifugação, o sedimento contendo núcleo, mitocôndrias e restos celulares foi desprezado. O sobrenadante obtido foi filtrado em gaze e centrifugado a 100.000 x g a 4 °C por 1 h na ultracentrífuga Beckman®. Após a segunda ultracentrifugação, o sobrenadante foi desprezado e o sedimento contendo as enzimas ligadas ao reticulo endoplasmático liso (microssomos) foi homogeneizado em tampão fosfato de potássio dibásico 100 mM com 20% de glicerol e EDTA 1 mM (pH 7,4), a uma velocidade de aproximadamente 250 x g. Os microssomos, obtidos foram distribuídos em tubos para criogenia e congelados em nitrogênio líquido até o momento de uso. O motor do homogeneizador usado para a preparação microssomal foi da Novatécnica® (Agitador Mecânico 110v modelo NT136 50W – 0,5A 60Hz).

### 3.10.5 Quantificação de proteínas na fração microssomal

A concentração total de proteínas nas amostras de fração microssomal das ratas controles e tratadas foi determinada, separadamente, usando o método de Bradford (1976), que se baseia na ligação do corante azul de Coomassie (Reagente de Bradford, SLBP3810V, Sigma-Aldrich) à proteína, sendo a intensidade da cor do corante proporcional à concentração de proteínas da amostra.

Para a obtenção da curva padrão para quantificação, foram adicionados 250 µL do Reagente de Bradford aos poços contendo 5 µL das diferentes concentrações de albumina sérica bovina (SLBK9549V, Sigma-Aldrich): 0,28, 0,56, 0,84 e 1,4 mg/mL. A formação do complexo proteína-corante é rápida e estável no período de 5 a 60 minutos e, por tal, foi adotado um intervalo constante de 30 min entre a adição do reagente às proteínas diluídas e a leitura de absorbância, realizada por espectrofotometria a 595 nm, em leitora de microplacas EZ Read 2000 Biochrom®. Todas as determinações foram realizadas em triplicata.

A concentração total de proteínas nas amostras foi expressa em mg/ mL. Os resultados representam a média de três leituras das concentrações do padrão e das amostras.

A curva padrão para quantificação de proteínas na fração microssomal do grupo de animais tratados, do grupo de animais controle e do superssoma rhCYP2C9*1, foi feita com diferentes concentrações de albumina sérica bovina. O coeficiente de correlação foi superior a 0,99 ($r^2 = 0,995$) e, portanto, essa curva padrão foi usada para a quantificação das proteínas. A equação da curva-padrão obtida foi y=0,428x+0,019.

### 3.10.6 *Determinação da atividade da etoxiresorufina-O-desetilase (EROD)*

A etoxiresorufina (EROD) é metabolizada preferencialmente por isoenzimas da subfamília CYP1A. Esta reação leva à formação da resorufina, cujo acúmulo pode ser medido pela intensidade da fluorescência avaliada por um espectrofluorímetro (PARKINSON, 2001; NERURKAR *et al.*, 1993).

A determinação da atividade EROD na fração microssomal hepática foi realizada como descrito por Burke e colaboradores (1985), exceto pela substituição do NADPH (fosfato de nicotinamida adenina dinucleotídeo reduzido) por um sistema regenerador de NADPH que funciona como co-fator para várias reações de biotransformação (DE OLIVEIRA *et al.*, 1999). A reações ocorreram em tempo real em cubeta de quartzo e foi medida em espectrofluorímetro Shimadzu RF-5301PC, adaptado a banho-maria com termostatização para manter a temperatura estável em 37 °C, cuja leitura foi realizada com excitação de 550 nm e emissão de 582 nm. Os dados foram manipulados no programa RF-530 XPC.

Inicialmente foram adicionados à cubeta de quartzo solução-tampão de $K_2HPO_4$ 100 mM (pH 7,8) em quantidade necessária para completar o volume final de 2000 µL e pré-incubação por 3 min a 37 °C. Após o tempo de pré-incubação, foram adicionados ao preparo microssomal (1 mg/mL) em solução tampão $K_2HPO_4$ 100 mM pH (7,8) para fornecer 0,5 mg de proteína na cubeta; substrato específico (EROD = etoxiresorufina 1mM em DMSO 1% p/v) para uma concentração final de 5 μM na cubeta. Após 2 min, a reação teve início com a adição de sistema regenerador de NADPH (glicose-6-fosfato 5 mM; β-NADP 0,25 mM; solução de $MgCl_2$ 2,5 mM; glicose- 6-fosfato-desidrogenase 0,5 U/mL). A reação ocorreu durante 60 s e o produto resorufina, substância fluorescente, foi quantificado em tempo real.

Os valores de fluorescência obtidos foram convertidos para quantidade de resorufina produzida pela curva padrão e expresso em picomoles (pmol) de resorufina/ mg de proteína/ min da reação.

### 3.10.7   *Curva padrão de resorufina*

Para a determinação da atividade do CYP1A foi construida uma curva padrão de resorufina a partir de uma solução-mãe de resorufina 1 μM em tampão $K_2HPO_4$ 100 mM (pH 7,8) para obtenção da equação de cálculo da fluorescência em diferentes concentrações. As concentrações de resorufina variaram de 0,0 a 100,0 picomoles. Equação da curva-padrão obtida y=0,209x-0,221.

## 4    RESULTADOS E DISCUSSÕES

### 4.1    Perfil químico e análise fitoquímica do extrato orgânico de folhas de *F. maxima*

Inicialmente, o extrato bruto das folhas CaxFls-EB (5,35 g) de *F. maxima* foi analisado por CCD (cromatogramas não disponíveis) em sistema de eluição Hex:AcOEt 7:3 para avaliação do perfil químico e polaridade dos seus constituintes químicos. A amostra também foi analisada por RMN [1]H (Figura 5), HSQC e HMBC (espectros não mostrados).

Deste modo, foi realizado um particionamento líquido-líquido de CaxFls-EB para obtenção de três frações orgânicas de polaridade crescente com Hex, DCM e AcOEt, gerando as frações CaxFls-Hex (1,16 g; 22,15% rendimento), CaxFls-DCM (194,7 mg; 3,70% rendimento), CaxFls-AcOEt (671,11 mg; 12,75% rendimento) e uma fração aquosa (não pesada), a fim de se estudar sua composição química.

Figura 5 - Espectro de RMN [1]H (400 MHz, CDCl₃) do extrato bruto etanólico das folhas de *F. maxima*.

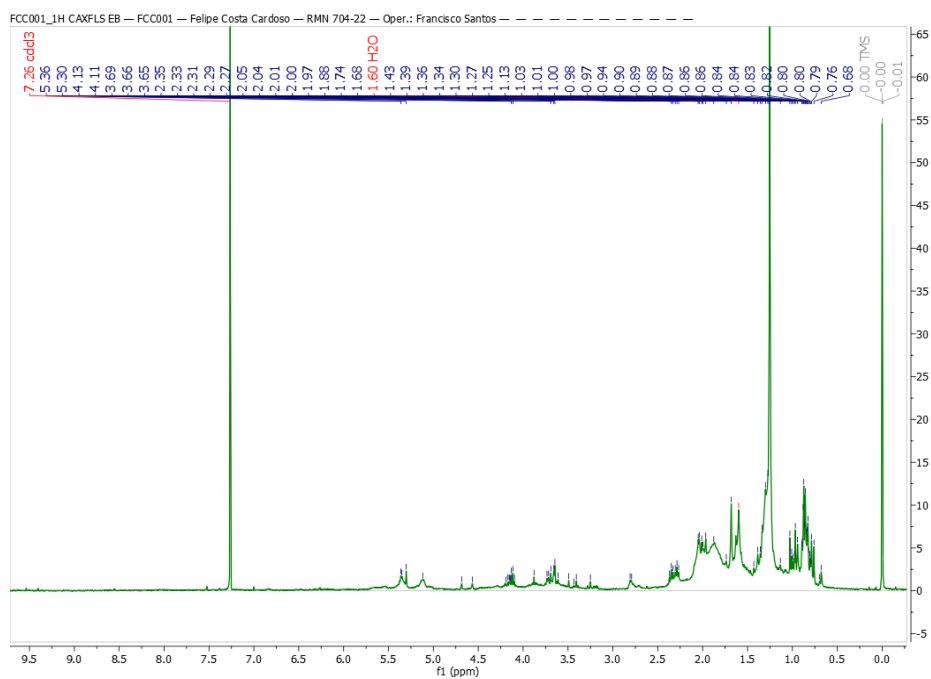

Fonte: Elaborado pelo autor.

Após o particionamento, as frações obtidas foram novamente avaliadas por CCD e por RMN [1]H. Na Figura 7 é possível observar os cromatogramas do extrato bruto e das partições das folhas de *F. maxima* eluídas no sistema Hex:AcOEt 7:3.

Figura 7 - Cromatogramas de CCD do extrato bruto e frações em Hex, DCM, AcOEt das folhas de *F. maxima*

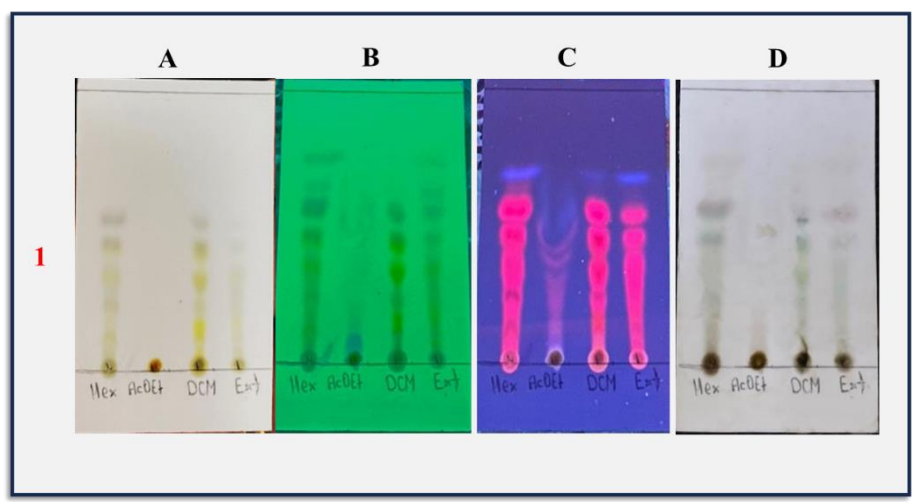

Em A1, é mostrado o cromatograma sob luz visível; em B1, cromatograma sob luz ultravioleta no comprimento de 254 nm em câmara escura; em C1, cromatograma sob luz ultravioleta de 365 nm em câmara escura e em D1, cromatograma após revelação com solução de ácido sulfúrico 10% (v/v) em etanol e posterior aquecimento da placa cromatográfica a 110 °C de 2 a 5 min.

Fonte: Elaborado pelo autor.

Na Figura 6 é possível observar os Cromatogramas do extrato alcoólico das cascas do caule de *F. maxima*, também eluídas em sistema Hex:AcOEt 7:3.

Figura 6 - Cromatogramas das frações em Hex, DCM e AcOEt das cascas do caule de *F. maxima*.

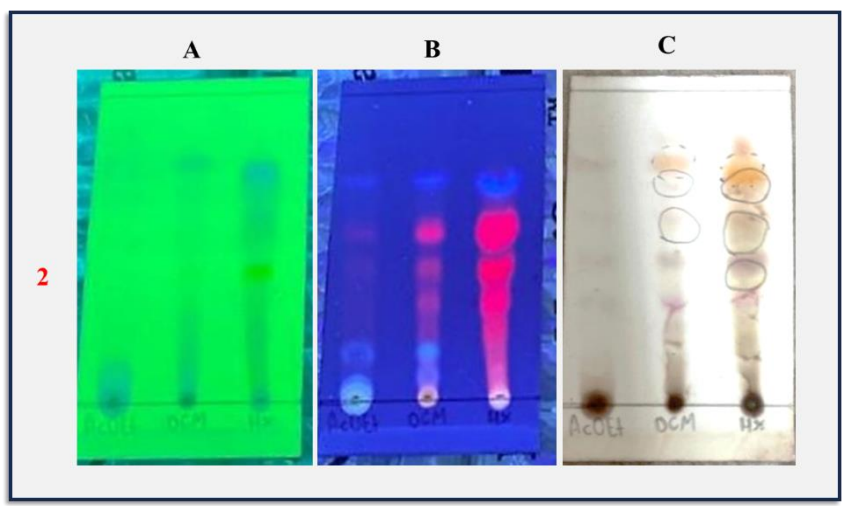

Em A2 é mostrado o cromatograma sob luz UV de 254 nm; em B2 é mostrado o cromatograma sob luz UV de 365 nm e em C2 é mostrado o cromatograma após revelação com solução de ácido sulfúrico 10% (v/v) em etanol e posterior aquecimento da placa a 110 °C.

Fonte: Elaborado pelo autor.

A revelação não-destrutiva sob luz UV, mais específica para revelação de compostos fenólicos e policonjugados, sobretudo no comprimento de 365 nm, demonstra que

CaxFls-EB e suas frações de baixa e média polaridade (Hex e DCM) possui grande quantidade de substâncias policonjugadas como possíveis constituintes majoritários. O mesmo padrão também é observado para CaxCsc-EB. A revelação com $H_2SO_4$ 10% (v/v) é utilizada para detecção geral, contudo mais específica para esteróis e esteroides. Nesse caso, substâncias da classe dos terpenos, principalmente, e seus ésteres, após borrifo com solução desse revelador apresentam *spots* de coloração amarronzada e outros de coloração preta, como podem ser observados no cromatograma D1 e no cromatograma C2 (STAHL, 2013), indicando presença desses metabólitos nas amostras. As cascas do caule de *F. maxima*, porém, indicam presença de compostos mais polares tanto no extrato quanto nas suas partições, em comparação com as folhas.

Na análise por RMN [1]H do extrato bruto e das partições das folhas (Figura 8, Figura 9, Figura 10) é possível observar a presença majoritária de sinais de hidrogênios alquílicos (na faixa de 0,8 a 1,5 ppm) característicos de hidrocarbonetos, como também de hidrogênios aromáticos (na faixa de 6,5 a 8,0 ppm), corroborando a possível presença de substâncias das classes dos terpenos e flavonoides ou ácidos fenólicos, respectivamente, principalmente na partição em Hex e DCM. Infelizmente, a fração CaxFls-AcOEt apresentou muitos problemas de solubilidade na maioria dos solventes testados, sendo parcialmente solúvel em MeOH e parte somente solúvel em $H_2O$, deixando as análises por RMN [1]H um pouco mais difíceis; visto que a intensidade dos sinais dos possíveis constituintes foi diminuída devido aos sinais do solvente utilizado ($CD_3OD$) para análise e da pouca solubilidade da amostra nesse último. Contudo, no espectro adquirido para CaxFls-AcOEt (Figura 10), foi possível observar a presença de alguns poucos sinais de hidrogênios aromáticos (6,5 – 8,0 ppm) e hidrogênios de álcoois (entre 3,0 e 4,0 ppm), o que pode indicar a presença de fenóis glicosilados como constituintes majoritários nessa fração.

Figura 8 - Espectro de RMN $^1$H (400 MHz, CDCl$_3$) de CaxFls-Hex.

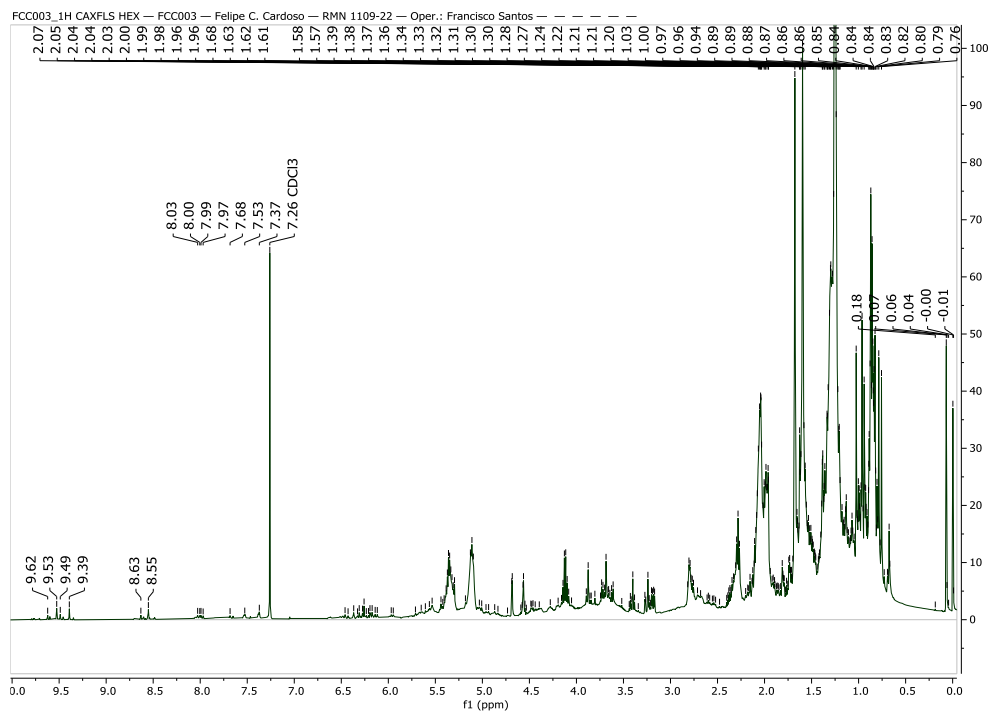

Fonte: Elaborado pelo autor.

Figura 9 - Espectro de RMN $^1$H (400 MHz, CDCl$_3$) de CaxFls-DCM.

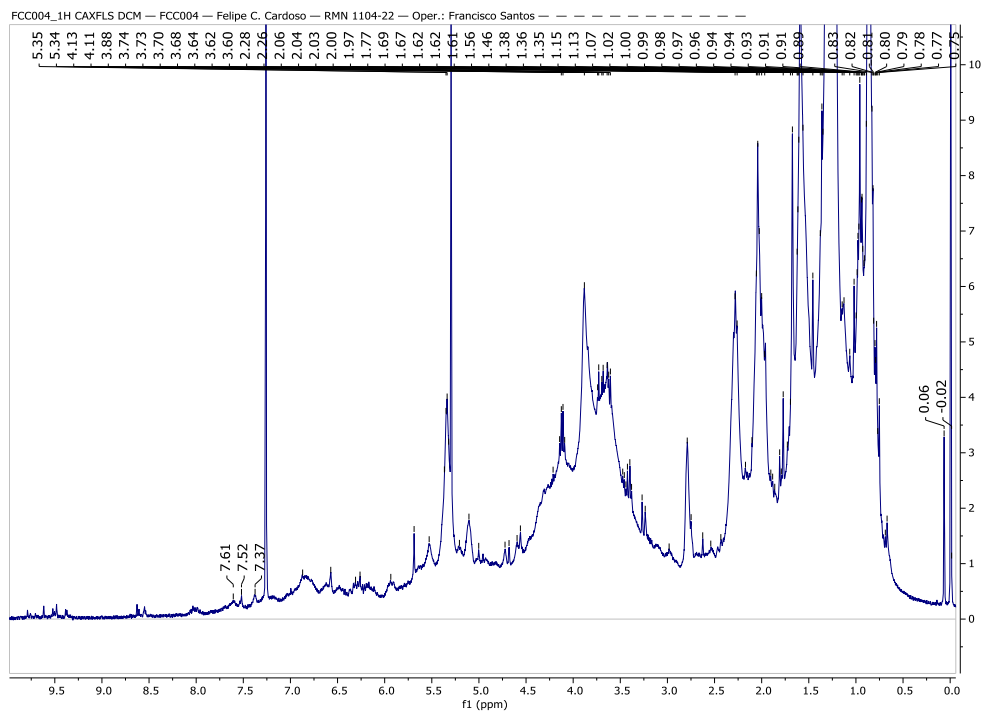

Fonte: Elaborado pelo autor.

Figura 10 - Espectro de RMN [1]H (400 MHz, CDCl₃) de CaxFls-AcOEt.

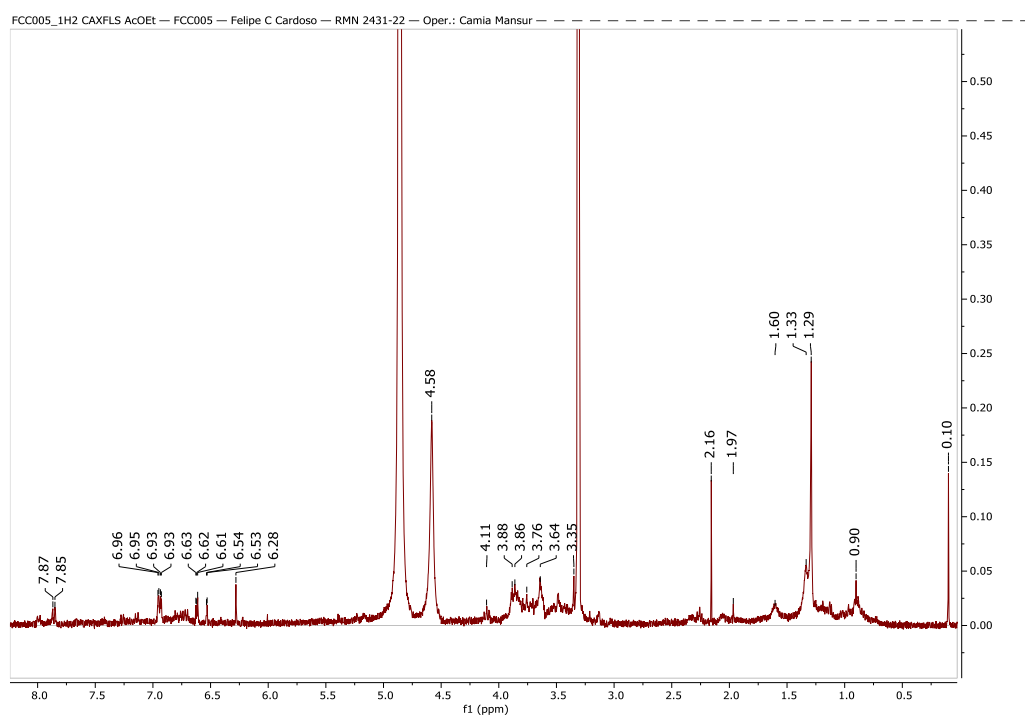

Fonte: Elaborado pelo autor.

Com base na literatura e reportagens das classes majoritárias constituintes do gênero *Ficus*, uma análise fitoquímica colorimétrica foi conduzida a fim de se avaliar a presença de terpenos e alcaloides nas frações com o uso de reveladores disponíveis no laboratório.

No teste colorimétrico para detecção de terpenos, a solução conhecida como revelador de Liebermann-Burchard (LB) utilizada é específica para detecção de Δ5-3-esterol, estereoides e terpenos glicosilados. A mudança de coloração é devida ao grupo hidroxila dessas classes, que reagem com a solução, ocasionando uma ampliação da conjugação no anel adjacente ao anel hidroxilado (XIONG; WILSON; PANG, 2007).

A solução converte rapidamente esteroides e análogos em seus derivados acetato e sulfato, com a presença de insaturações em posições variadas do esqueleto carbônico (XIONG; WILSON; PANG, 2007). Após repetidas etapas de dessaturação, são formados polienos, que, gradualmente, rearranjam para núcleos aromáticos produzindo a coloração verde ou verde-azulada característica.

Assim, aproximadamente 2 mg amostras foram transferidas para tubos de ensaio e dissolvidas em 2 mL de DCM. Em seguida, 1 mL da solução de LB foi adicionado a cada tubo e observada a mudança ou não de coloração (Figura 11).

Figura 11 - Ensaio colorimétrico para detecção de esteroides e terpenos em extratos e frações de *F. maxima* com solução de LB.

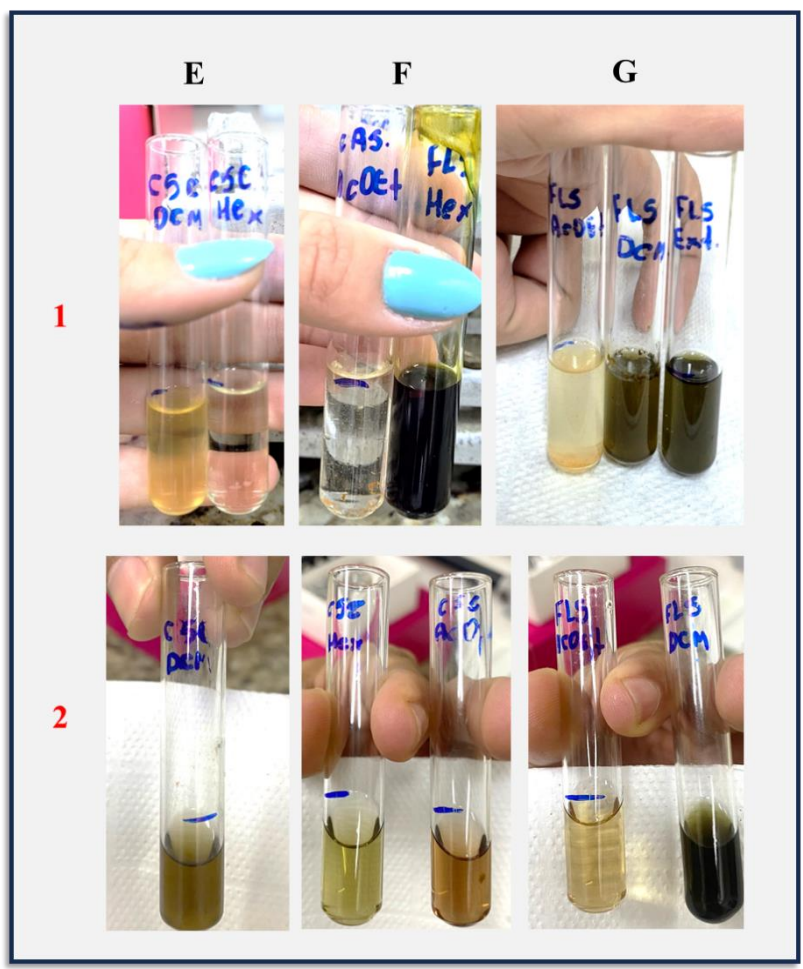

A coloração verde foi adquirida mais intensamente em CaxFls-Hex e CaxFls-DCM (G2), seguida por uma coloração verde de média intensidade em CaxCsc-DCM (E2) sugerindo que essas frações possuem concentração mais elevada das classes de substâncias testadas.

Fonte: Elaborado pelo autor.

Foi detectada a presença de esteroides e terpenos em ambos os extratos CaxFls-EB e CaxCsc-EB pelo teste de LB e em todas as partições dos dois extratos de *F. maxima*.

Para a detecção de alcaloides nas amostras, foi utilizado o reagente de Dragendorff comercial. Nesse caso, o reagente é uma mistura de solução concentrada de iodeto de potássio com subnitrato de bismuto, em ácido diluído, produzindo um complexo solúvel de tetraiodobismutato de potássio que tem coloração laranja. Este complexo em solução reage com o grupo amina dos alcaloides presentes, formando, por sua vez, um complexo insolúvel de coloração variando de laranja-avermelhado a marrom (RAAL *et al.*, 2020).

A presença de alcaloides foi detectada em praticamente todas as amostras, sendo que os precipitados com coloração marrom mais intensa foram observados em CaxFls-EB e CaxFls-DCM, frascos identificados por FLSEXT e FLSDCM, respectivamente na Figura 12.

Figura 12 - Ensaio colorimétrico para detecção de alcaloides em extratos e frações de *F. maxima* com reagente de Dragendorff.

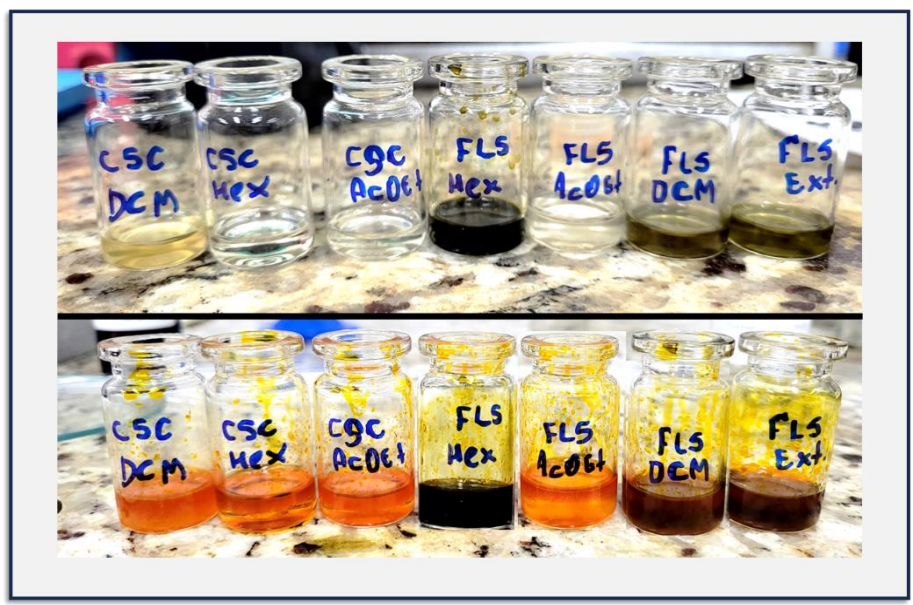

(Superior) amostras não tratadas. (Inferior) amostras tratadas com o reagente.
Fonte: Elaborado pelo autor.

O reagente de Dragendorff pode, algumas vezes, detectar compostos nitrogenados que não são necessariamente da classe dos alcaloides, como sais de amônio quaternário e lactamas, por exemplo. Contudo, foi um bom indicativo da presença destas classes no extrato de *F. maxima*.

## 4.2    Fracionamento da partição em hexano e diclorometano

Devido às características observadas nas análises qualitativas supracitadas, e pela quantidade de massa disponível, optou-se, inicialmente pelo subfracionamento das partições CaxFls-Hex, seguida do subfracionamento de CaxFls-DCM (Fluxograma 1). Assim, foi realizado o fracionamento de CaxFls-Hex por técnica de cromatografia líquida de adsorção em coluna de gel de sílica, utilizando um sistema de eluição Hex:AcOEt em gradiente de 0 - 100%, com grau crescente de polaridade, resultando em 106 subfrações, que, após análises por CCD, foram reunidas de acordo com a similaridade e padrão cromatográfico, resultando em 15 subfrações (CaxFls-HEx R1-R15) (Figura 13). Destas, duas subfrações (CaxFls-Hex R11) e (CaxFls-Hex R14) destacadas nos Cromatogramas

K1 e K2, apresentaram perfil cromatográfico interessante, indicando prováveis substâncias majoritárias isoladas do extrato bruto.

Fluxograma 1 - Fluxo de trabalho de fracionamento das folhas de *F. maxima*.

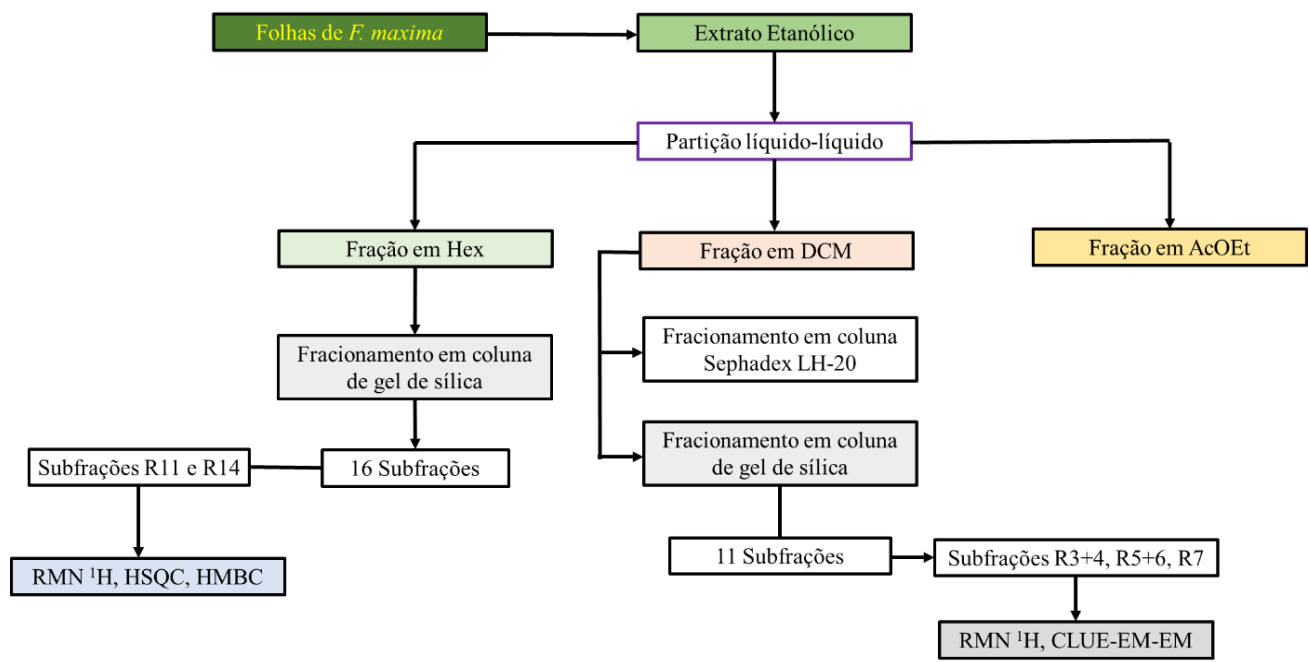

Figura 13 - Cromatogramas das reuniões das frações de CaxFls-Hex obtidas por cromatografia em coluna de gel de sílica.

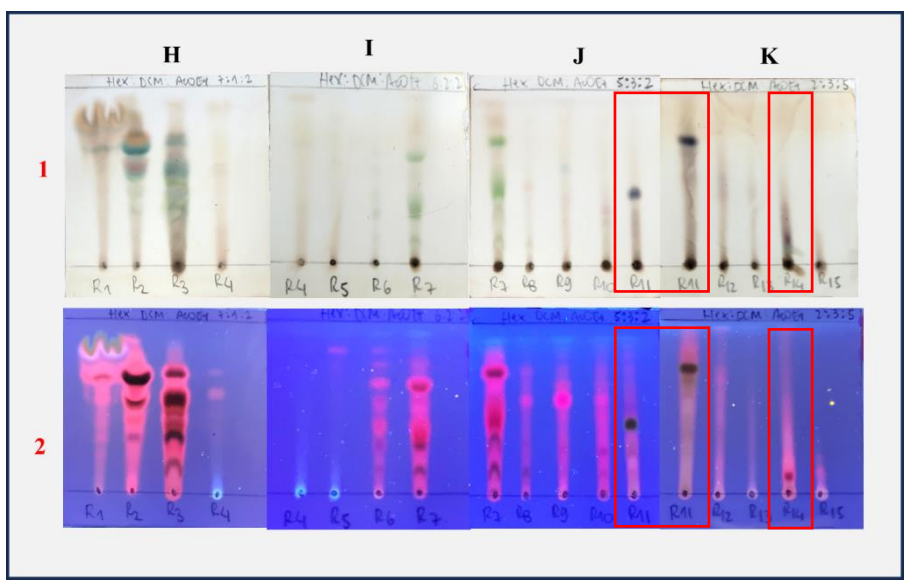

Cromatogramas H1 e H2: sistema de eluição Hex:DCM:AcOEt 7:1:2; Cromatogramas I1 e I2: sistema de eluição Hex:DCM:AcOEt 5:2:2; Cromatogramas J1 e J2: sistema de eluição Hex:DCM:AcOEt 5:3:2; Cromatogramas K1 e K2: sistema de eluição Hex:DCM:AcOEt 2:5:5. Cromatogramas H1-K1 revelados em H2SO4 10% (v/v); Cromatogramas H2-K2 revelados em luz UV 365 nm. Fonte: Elaborado pelo autor.

A subfração CaxFls-Hex R11, embora não aparentasse estar totalmente pura, foi separada e analisada por RMN $^1$H, HSQC, HMBC. No momento desta dissertação, ainda se aguardam os resultados relativos à análise da fração CaxFls-Hex R14.

Após as análises realizadas por CCD das frações obtidas, algumas foram reunidas, gerando novas reuniões. A Tabela 2 sumariza as massas obtidas das novas reuniões e das frações.

Tabela 2 - Códigos, sistema de eluição e massas obtidas das reuniões de CaxFls-Hex.

| Código fração | Proporção sistema de recolhimento | Massa reunião (mg) |
|---|---|---|
| CaxFls-Hex R1 | Hex 100% | 111,8 |
| CaxFls-Hex R2,3 | Hex:AcOEt 9:1 | 157,6 |
| CaxFls-Hex R4 | Hex:AcOEt 85:15 | 34,7 |
| CaxFls-Hex R5 | Hex:AcOEt 8:2 | 15,7 |
| CaxFls-Hex R6 | Hex:AcOEt 8:2 | 10,2 |
| CaxFls-Hex R7 | Hex:AcOEt 7:3 | 17,8 |
| CaxFls-Hex R8,9 | Hex:AcOEt 1:1 | 19,5 |
| CaxFls-Hex R10 | Hex:AcOEt 1:1 | 11,9 |
| CaxFls-Hex R11 | Hex:AcOEt 1:1 | 15,9 |
| CaxFls-Hex R12 | Hex:AcOEt 1:1 | 8,2 |
| CaxFls-Hex R13 | AcOEt 100% | 5,4 |
| CaxFls-Hex R14 | AcOEt 100% | 15,0 |
| CaxFls-Hex R15 | AcOEt:MeOH 1:1 | 5,2 |
| CaxFls-Hex R16 | MeOH 100% | 145,5 |

A fração CaxFls-DCM foi a que mostrou melhor atividade nos ensaios biológicos realizados para inibição de CYP1A (resultados discutidos na seção 4.8). Diante dos resultados promissores, uma massa de 97 mg da amostra CaxFls-DCM foi submetida a fracionamento em coluna cromatográfica de Sephadex LH-20, utilizando um sistema de eluição Acetona:MeOH 7:3, resultando em 10 subfrações. Contudo a separação não se mostrou exitosa, e as subfrações foram reunidas para serem submetidas a um novo fracionamento. Dessa nova reunião, uma massa de 50 mg foi submetida a fracionamento por coluna cromatográfica de gel de sílica *flash*, utilizando um sistema de eluição Hex:DCM:AcOEt em gradiente de 0 - 100%, com grau crescente de polaridade. Foram recolhidas 125 subfrações, que após análises por CCD foram reunidas de acordo com grau de similaridade e padrão cromatográfico. Após reunião, foram obtidas 11 subfrações (CaxFls-DCM R1 – R11) (Figura 14), (Tabela 3).

Figura 14 - Cromatogramas das reuniões das frações de CaxFls-DCM obtidas por cromatografia em coluna de gel de sílica.

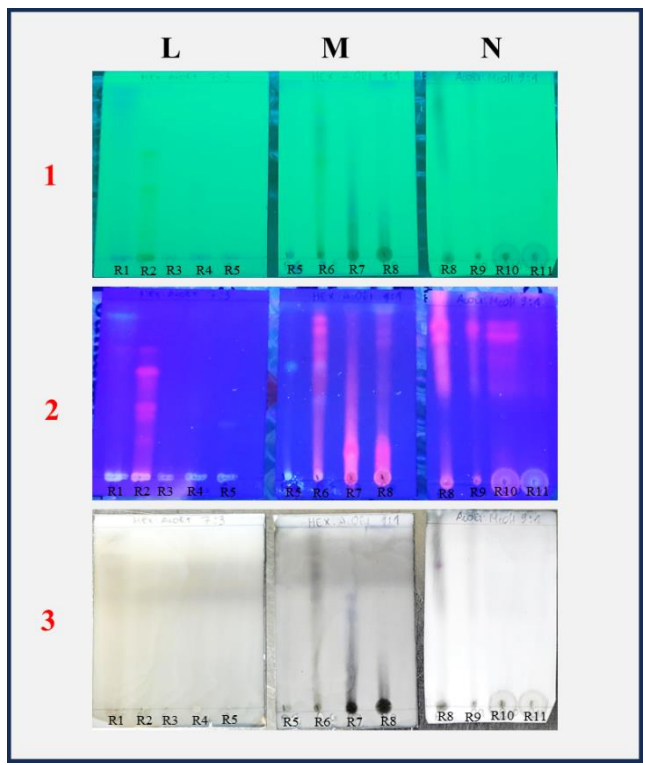

Cromatogramas L1-L3: sistema de eluição Hex:AcOEt 7:3; Cromatogramas M1-M3: sistema de eluição Hex:AcOEt 1:1; Cromatogramas N1-N3: sistema de eluição AcOEt:MeOH 9:1; Cromatogramas L1-N1 revelados luz UV 254 nm; Cromatogramas L2-N2 revelados em luz UV 365 nm; Cromatogramas L3-N3 revelados em vanilina sulfúrica 1% em EtOH. Fonte: Elaborado pelo autor.

Tabela 3 - Códigos, sistemas de eluição e massas obtidas das reuniões de CaxFls-DCM.

| Código fração | Proporção sistema de recolhimento | Massa reunião (mg) |
|---|---|---|
| CaxFls-DCM R1 | Hex 100% | 11 |
| CaxFls-DCM R2 | Hex:DCM:AcOEt 9:0,5:0,5 | 0,8 |
| CaxFls-DCM R3 | Hex:DCM:AcOEt 9:0,5:0,5 | 0,3 |
| CaxFls-DCM R4 | Hex:DCM:AcOEt 7:2:1 | 1,6 |
| CaxFls-DCM R5 | Hex:DCM:AcOEt 5:4:1 | 0,5 |
| CaxFls-DCM R6 | Hex:DCM:AcOEt 3:4:3 | 3,1 |
| CaxFls-DCM R7 | AcOEt:MeOH 9:1 | 4,0 |
| CaxFls-DCM R8 | AcOEt:MeOH 7:3 | 14 |
| CaxFls-DCM R9 | AcOEt:MeOH 1:1 | 0,1 |
| CaxFls-DCM R10 | MeOH 100% | 0,9 |

| CaxFls-DCM R11 | MeOH:H$_2$O 1:1 | 0,4 |

As subfrações CaxFls-DCM R3 a R7 mostraram perfis cromatográficos apresentando substâncias majoritárias que aparentavam não estar presentes nas demais frações. Pela análise em CCD, CaxFls-DCM R3 foi reunida com a CaxFls-DCM R4, gerando uma nova reunião CaxFls-DCM R3+4, e as CaxFls-DCM R5 e R6 também foram reunidas, gerando a nova reunião CaxFls-DCM R5+6. A subfração CaxFls-DCM R7 foi mantida intacta. Essas subfrações foram separadas para análises de CLUE-EM-EM, para triagem das substâncias e construção de rede molecular no GNPS; e análises de RMN [1]H, HSQC e HMBC. Nesse momento, encontram-se em fase de aquisição dos espectros.

O Fluxograma 2 mostra o fluxo de trabalho das frações de *F. maxima* que foram submetidas aos ensaios biológicos. A metodologia aplicada e os resultados são discutidos nas seções 4.6 – 4.8.

Fluxograma 2 - Fluxo de trabalho das frações de *F. maxima* selecionadas para ensaios biológicos.

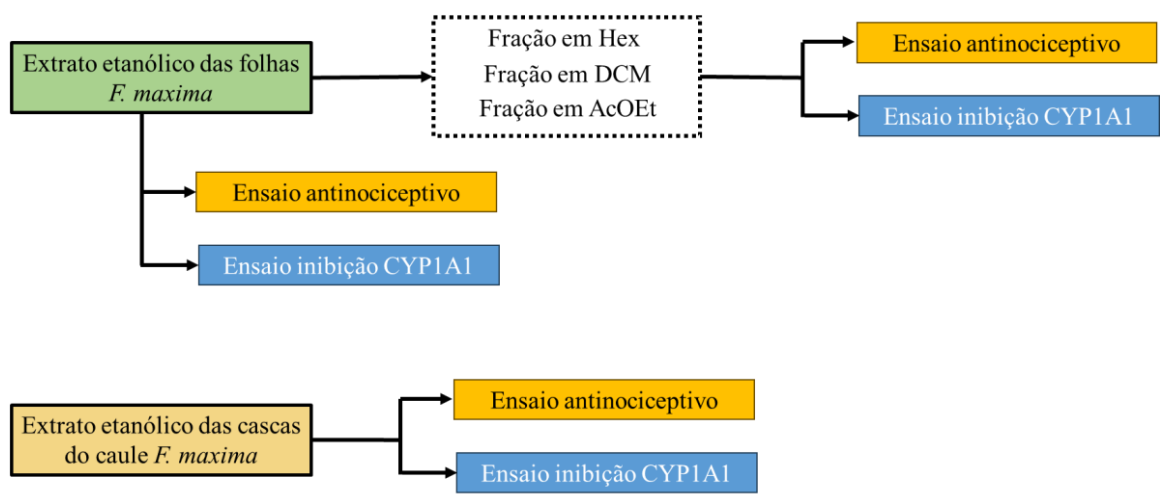

## 4.3 Análise por RMN da subfração CaxFls-Hex R11

A subfração CaxFls-Hex R11 foi recolhida em sistema Hex:AcOEt 1:1 e foi submetida a análises de RMN [1]H para avaliação da pureza e isolamento de substância nela presente.

O espectro de RMN [1]H de CaxFls-Hex R11 (Figura 15) mostrou sinais sobrepostos com deslocamentos químicos na região de baixa frequência, típicos de esteroides e terpenos, assim como os simpletos entre $\delta_H$ 0,79 e 1,29 relativos a prótons metílicos característicos desta última classe química. Os dados de deslocamentos químicos de $^{13}$C foram diretamente avaliados dos espectros de correlação bidimensional HSQC e HMBC.

Figura 15 - Espectro de RMN $^1$H (400 MHz, CDCl$_3$) editado da subfração CaxFls-Hex R11.

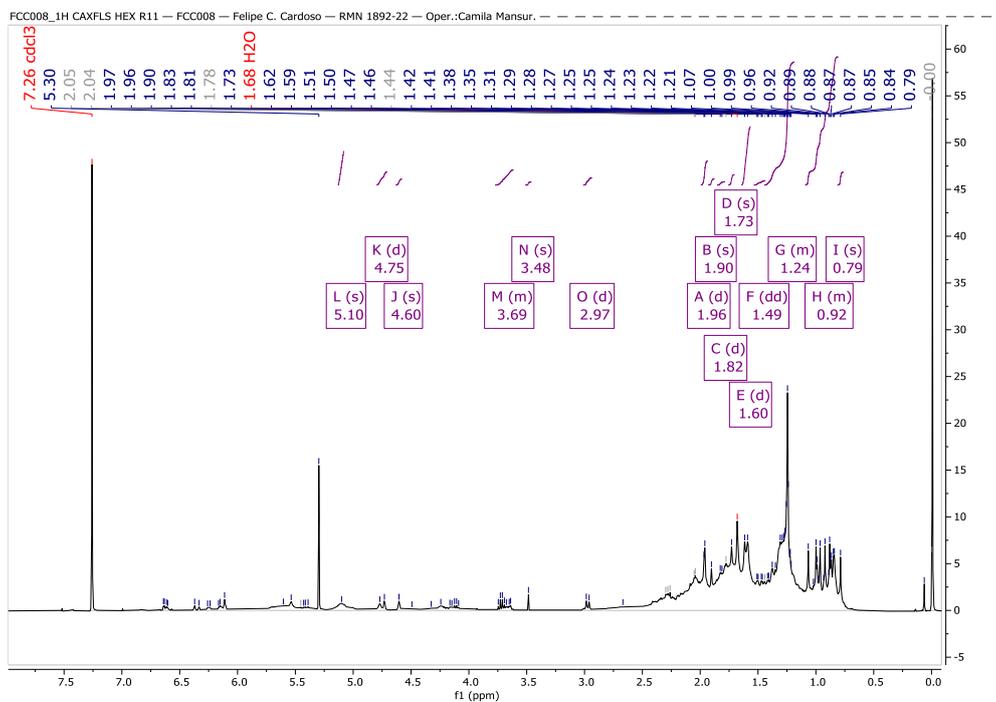

A maioria dos sinais de $^{13}$C da amostra se acumulam abaixo de 60 ppm, característico de terpenos, com C, CH, CH$_2$ e CH$_3$ alicíclicos. Um sinal em $\delta_c$ 208,7 observado no espectro de correlação HMBC sugere a presença de carbono carbonílico (C=O) de cetona ou aldeído. A ausência de sinais de hidrogênio próximo a 10 ppm, no entanto, descarta a possibilidade da função aldeído, neste caso. Ademais, um sinal $\delta_c$ 109,72 com correlações com prótons em $\delta_H$ 4,63 e 4,75 vistas no espectro de HSQC (não-mostrados) sugerem a presença de hidrogênios vinílicos, comumente observado em terpenos pentacíclicos. Outra informação interessante obtida pela análise dos espectros de RMN foi a presença de outro sinal em $\delta_H$ 6,13 (d, $J$ = 15,7 Hz, 1H) atribuído a hidrogênios vinílicos próximo à carbonilas (cetona $\alpha,\beta$-insaturada) indicando uma provável porção ciclohexenona na substância. Comparando-se esses dados com aqueles provenientes de substâncias naturais isoladas na literatura, foi sugerida a presença de um triterpeno pentacíclico com esqueleto semelhante ao do lupano (**13**) (Figura 16), em que as prováveis substituições estariam nas posições C-3 e C-28, com insaturação em algum dos anéis A, C ou D.

Figura 16 - Estrutura de triterpeno pentacíclico tipo lupano (**13**).

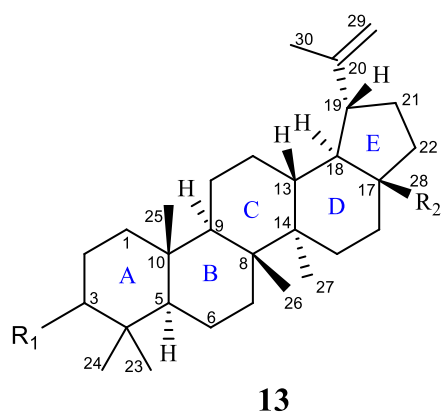

**13**

Uma provável substância com essas características é a glochidona (**14**), um triterpeno tipo lupano (Figura 17), ou derivado deste, que pode estar presente na subfração CaxFls-Hex R11. Triterpenos do tipo **13**, como lupeol (**51**) e acetato de lupanila, têm sido isolados de espécies vegetais de Moraceae, como *F. benghalensis* (RAJAN *et al.*, 2023), *M. alba* e *D. arifolia*, por exemplo (FINGOLO *et al.*, 2013).

Figura 17 - Estrutura do triterpeno glochidona (**14**).

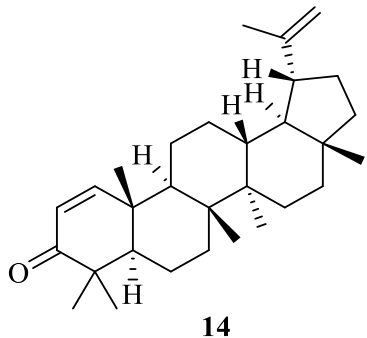

**14**

Uma predição do espectro de RMN [1]H de **14** (em azul) foi realizada no *software* MestReNova para comparação com o espectro experimental (em vermelho) da subfração em análise (Figura 18). A comparação entre os dois mostra que a maioria dos sinais discutidos anteriormente e preditos para **14** estão presentes no espectro experimental, com exceção do sinal de H-2 que está ausente. No entanto, as constantes de acoplamento *J* esperadas para os hidrogênios vinílicos de C-29 e da ligação C-1/C-2 (*J*=10 Hz, acoplamento *cis*) não estão em conformidade com a estrutura proposta e os dados reportados na literatura (PUAPAIROJ *et al.*, 2005). Tendo em vista que a subfração ainda contém muitos sinais sobrepostos, de prováveis impurezas, alguns sinais importantes podem ter tido suas intensidades diminuídas, o que dificulta uma análise precisa para elucidação da estrutura, embora a substância **14** tenha sido anotada nas análises por CLUE, discutidas nas seções a seguir. Adicionalmente, a presença de outros triterpenos com o mesmo núcleo, mas substituições em posições diferentes podem acarretar as diferenças observadas para os deslocamentos químicos. Sendo assim, novas etapas de purificação e isolamento dessa subfração, bem como análises por EM e RMN são necessárias para a caracterização inequívoca da(s) substância(s) presente(s).

Figura 18 - Espectros de RMN [1]H (400 MHz, CDCl$_3$) predito (em verde) e experimental (em vermelho) de **14**.

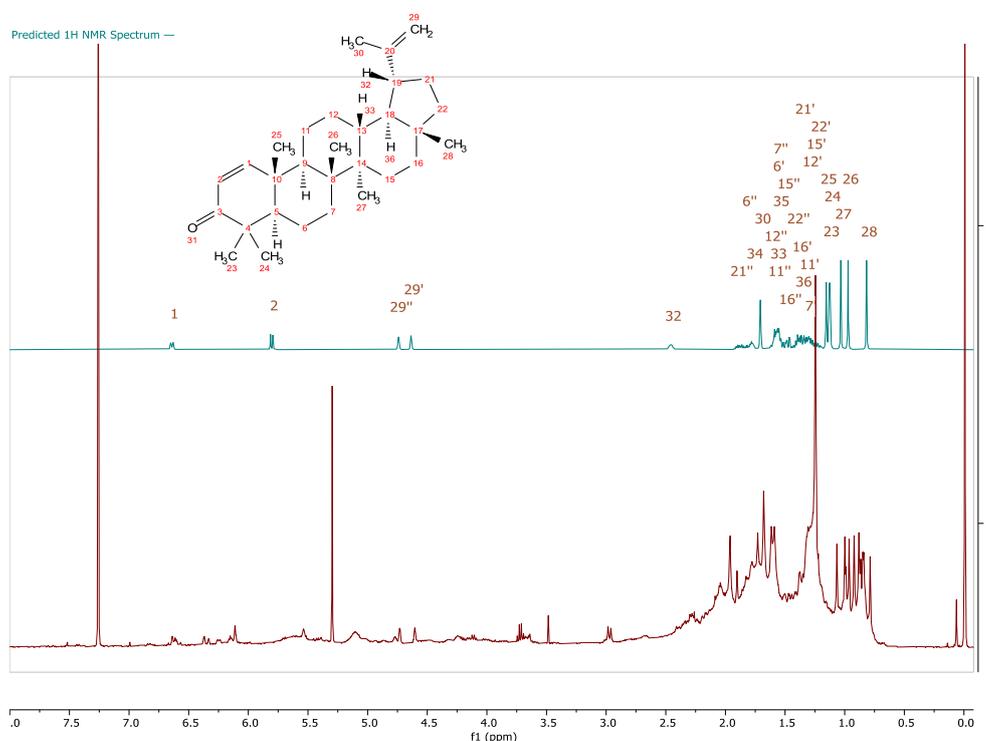

## 4.4    Análise fitoquímica por CLUE-EM-EM e criação de redes moleculares

As técnicas cromatográficas, como a cromatografia líquida (CL), associadas à espectrometria de massas sequencial de alta resolução (EM-EMAR) representam uma das opções mais populares para estudos, identificação e caracterização de substâncias bioativas em espécies vegetais (ALVAREZ-RIVERA *et al.*, 2019). A associação dessas técnicas tem o objetivo de aumentar a eficiência e rapidez das análises quando comparadas às técnicas isoladas, fornecendo informações fidedignas sobre as estruturas químicas das substâncias em estudo, principalmente em matrizes complexas (STORION; GONÇALVES; MARCUCCI, 2020).

A CLUE-EM-EM foi aplicada mais recentemente em um estudo de análise da diversidade fitoquímica de trezes espécies de *Ficus* spp., onde os autores identificaram e caracterizaram trinta e cinco metabólitos das classes dos flavonoides, antocianinas, ácidos hidroxicinâmicos e derivados, como constituintes dos extratos etanólicos das plantas, (ELHAWARY *et al.*, 2018).

No estudo aqui realizado, a CLUE-EM-EM foi aplicada para identificação putativa dos constituintes do extrato orgânico e frações, majoritariamente, das folhas de *F. maxima*. O extrato e frações das cascas do caule encontram-se em fase de análise dos dados. Desse modo, alíquotas de CaxFls-EB e das partições CaxFls-Hex, CaxFls-DCM e CaxFls-AcOEt (aproximadamente 1 mg/mL cada) foram submetidas a análise por CLUE-EM-EM e ionização por eletrospray (IES), utilizando como fase móvel solução aquosa de ácido fórmico 0,1% (fase A) e solução de ACN:ácido fórmico 0,1% (fase B) e tempo de análise de 22 minutos.

O espectrômetro de massas foi configurado para varredura de *m/z* compreendida entre 100 e 1000 Da nos modos de ionização positivo (IES+) e modo de ionização negativo (IES-). Os cromatogramas foram gerados a partir da intensidade do pico base, que representa a intensidade dos picos mais intensos a cada ponto na análise.

O fluxo de trabalho *Molecular Library Search V2* da plataforma *Global Natural Products Social Networking* (GNPS) foi utilizado para auxiliar na anotação dos metabólitos presentes nas amostras através da pesquisa dos dados de EM-EM em bibliotecas espectrais públicas, permitindo a construção de uma tabela contendo todos os compostos anotados com suas respectivas razões *m/z*, tempo de retenção (TR), erro em PPM, e identificação (APÊNDICES E – K).

Um aspecto importante e que tem gerado resultados bastante satisfatórios é a combinação de ferramentas de bioinformática, como o uso das redes moleculares do GNPS, para visualização e anotação de dados adquiridos por CLUE-EM-EM (NOTHIAS *et al.*, 2020). Numa rede molecular, substâncias parecidas, mas com algumas transformações simples como, oxidação/redução, glicosilação etc., são relacionadas e agrupadas numa família molecular ou família espectral. Devido ao fato de que substâncias parecidas têm o mesmo padrão de fragmentação, elas acabam gerando espectros de massas relativamente parecidos. Baseado nisso, a construção de uma rede molecular facilita a identificação de similaridades e diferenças espectrais num dado conjunto de amostras (PILON *et al.*, 2021). A criação da rede molecular começa pelo tratamento dos dados espectrais, como a remoção de ruídos dos espectros de fragmentação EM sequencial (EM-EM ou EM$^2$), selecionando-se os $N$ picos mais intensos, com variação de +/- 50 unidades de *m/z* em todo o espectro. Em seguida, através do algoritmo do GNPS, o usuário define a tolerância de massa para agrupamento de espectros referentes ao mesmo íon precursor e íons produtos. O nível de similaridade entre os espectros dentro das amostras, ou entre as amostras e os espectros da biblioteca de referência, é definido por um valor de cosseno (variando de 0 a 1), sendo que 0 indica nenhuma similaridade, e 1 indica 100% de similaridade. Quando dois ou mais espectros comparados têm um valor de cosseno maior ou igual ao estabelecido pelo usuário, é criada uma conexão ou arco entre eles. Em seguida, é realizada uma etapa de agrupamentos de íons de uma mesma substância no chamado espectro consenso. Esse espectro consenso é representado na rede molecular por um nodo (um círculo). Neste nodo, ficam armazenados todos os espectros de uma mesma substância. Vários nodos podem ou não estar conectados pelos arcos, também chamados *edges* ou arestas. Quando conectados, a largura dos arcos representa o valor de similaridade entre os nodos, ou seja, nodos conectados entre si apresentam espectros de fragmentação semelhantes, indicando substâncias também semelhantes entre si. Vários nodos conectados representam, por sua vez, agrupamentos ou os chamados *clusters.*

O algoritmo comumente utilizado nas redes moleculares do GNPS, no entanto, não leva em consideração o tempo de retenção dos sinais dos experimentos de cromatografia líquida de alta eficiência (CLAE), fazendo com que isômeros e isóbaros sejam reunidos num mesmo nodo. Para contornar este problema, foi disponibilizado no GNPS um algoritmo avançado chamado de *Feature Based Molecular Networking* (FMBN)*,* que requer que os dados experimentais sejam adquiridos em técnicas associadas, como a CLUE-EM-EM, que leva em consideração o tempo de retenção das substâncias associado aos espectros. Neste caso, há uma etapa adicional de pré-processamento dos dados espectrais, onde as informações advindas das fontes de dados utilizadas são reunidas num conjunto de características chamado de *feature*.

Baseado nisso, no presente trabalho foram criadas redes moleculares do extrato orgânico das folhas de *F. maxima*, bem como das suas frações CaxFls-Hex, CaxFls-DCM e CaxFls-AcOEt, para melhor visualização e identificação dos metabólitos presentes, utilizando, para isso, o fluxo de trabalho FBMN disponível na plataforma de acesso aberto do GNPS. A etapa de pré-processamento dos dados de CLUE-EM-EM foi realizada no *software* MZmine 2.53. Os arquivos processados foram separados de acordo com o modo de ionização positivo ou negativo e, em seguida, carregados, juntamente com uma tabela de metadados, para o GNPS para a construção das redes moleculares. Os arquivos gerados no fluxo de trabalho foram exportados e as redes moleculares criadas foram visualizadas e editadas no *software* livre Cytoscape 3.10.

A rede molecular criada para os metabólitos presentes em CaxFls-EB, CaxFls-Hex, CaxFls-DCM e CaxFls-AcOEt no modo positivo pode ser visualizada na Figura 19.

Figura 19 - Rede molecular modo positivo das amostras CaxFls-EB, CaxFls-Hex, CaxFls-DCM e CaxFls-AcOEt editada no *software* Cytoscape 3.10.

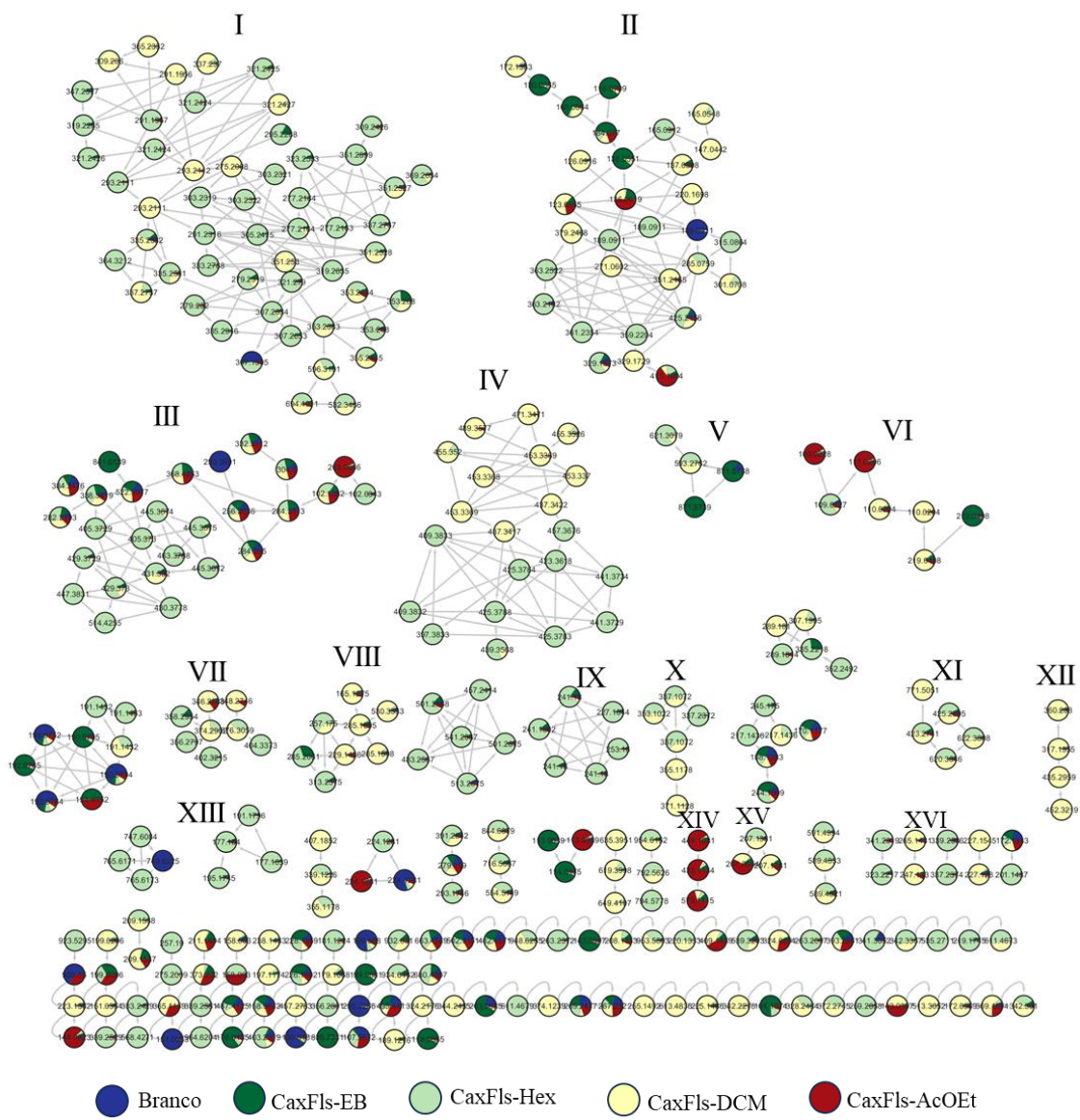

A rede molecular foi composta por 340 nodos e 527 arestas. Foram criados dezesseis agrupamentos ou *clusters* principais (I – XVI), os quais tiveram alguma anotação molecular com a biblioteca do GNPS, ou seja, substâncias identificadas putativamente nas amostras. Estão presentes pelo menos outros quinze agrupamentos que não geraram nenhuma anotação molecular e outros nodos que não tiveram similaridade entre si.

Os agrupamentos I, II e IV foram os que tiveram mais anotações moleculares, representando quase que exclusivamente substâncias presentes nas frações CaxFls-Hex e CaxFls-DCM, de caráter mais apolar, bem como algumas no extrato bruto de *F. maxima*.

Como observado na ampliação da rede molecular (Figura 20), a maioria das substâncias anotadas no agrupamento I são da classe dos ácidos graxos, sobretudo poliinsaturados presentes na fração hexânica das folhas de *F. maxima* (nodos em verde claro). Dentre eles, ésteres etílicos do ácido linoleico (**24**) (Tabela 4). Os ácido linoleico e linolênico, por exemplo, compreendem 75% ou mais de ácidos graxos encontrados nos óleos de sementes de espécies de *Ficus*, como *F. nota* e *F. septica* (KNOTHE; RAZON; CASTRO, 2019). Num estudo fitoquímico das folhas de cinco variedades de *F. carica*, a quantidade de ácidos graxos poliinsaturados variou de 53% a 71% no conteúdo total de ácidos graxos presente*,* com cadeias carbônicas contendo entre doze e vinte e quatro carbonos (SHIRAISHI *et al.*, 2023). As substâncias aqui anotadas e representadas estão de acordo com aquelas reportadas na literatura para espécies de *Ficus* spp.

Figura 20 - Ampliação do agrupamento I da rede molecular modo positivo das amostras CaxFls-EB, CaxFls-Hex, CaxFls-DCM e CaxFls-AcOEt.

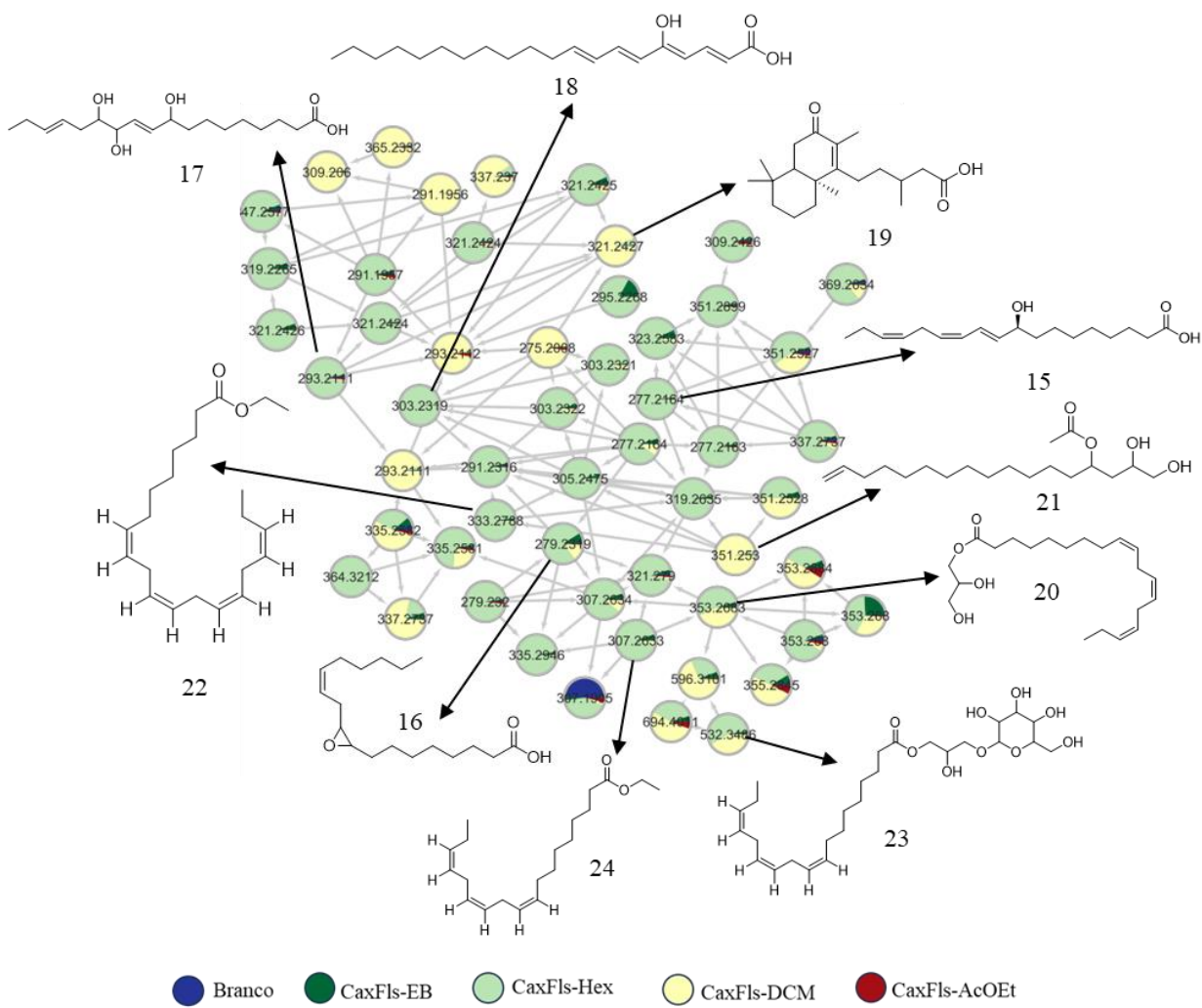

Fonte: Elaborado pelo autor.

Tabela 4 - Substâncias anotadas no agrupamento I.

| Entrada | Nome | *m/z* |
|---------|------|-------|
| 15 | Ác. 9-hidroxi-10,12,15-octadecatrienoico | 277,2164 |
| 16 | 9(10)-EpOME | 279,2319 |
| 17 | Ác. 9,12,13-trihidroxioctadeca-10,15-dienoico | 293,2111 |
| 18 | Ác. 5-hidroxieicosatetraenoico | 303,2319 |
| 19 | Ác. 5-[(8aS)-2,5,5,8a-tetrametil-3-oxo-4a,6,7,8-tetrahidro-4H-naftalen-1-il]-3-metilpentanoico | 321,2427 |
| 20 | Monolinolenina | 353,2683 |
| 21 | Acetato de 1,2-dihidroxiheptadec-16-en-4-il | 351,253 |
| 22 | Éster etilico do ácido ω-3-araquidônico | 333,2788 |
| 23 | Éster 3-(hexopiranosiloxi)-2-hidroxipropil do ác. 9, 12, 15-octadecatrienoico | 532,3486 |
| 24 | Éster etílico do ác. linolênico | 307,2623 |

O agrupamento II (Figura 21), por sua vez, anotou substâncias das classes dos aminoácidos e alguns metabólitos como flavonoides e cumarinas (Tabela 5), desta vez concentrando-se nas frações CaxFls-Hex e CaxFls-DCM. Substâncias fenólicas, como os flavonoides, cumarinas e outros ácidos fenólicos têm sido extraídas de todas as partes de espécies de *Ficus* estudadas mais recentemente (CRUZ *et al.*, 2022).

Figura 21 - Ampliação do agrupamento II da rede molecular modo positivo das amostras CaxFls-EB, CaxFls-Hex, CaxFls-DCM e CaxFls-AcOEt.

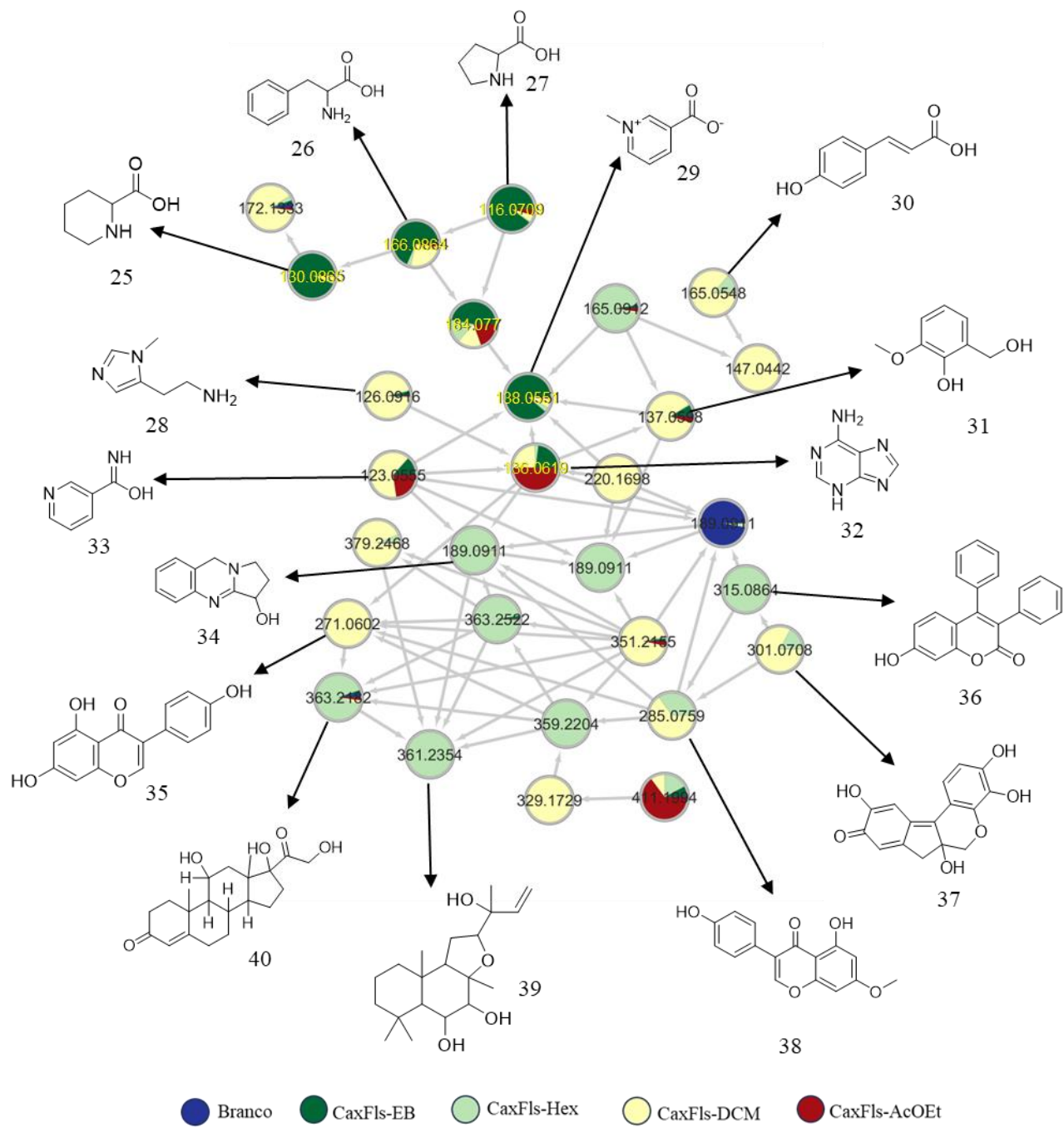

Tabela 5 - Substâncias anotadas no agrupamento II.

| Entrada | Nome | m/z |
|---------|------|-----|
| 25 | Pipecolato | 130,0865 |
| 26 | DL-fenilalanina | 166,0864 |
| 27 | Prolina | 116,0709 |
| 28 | 3-metilhistamina | 126,0916 |
| 29 | Trigonelina | 138,0551 |
| 30 | 4-cumarato | 165,0548 |
| 31 | Álcool vanílico | 137,0598 |
| 32 | Adenina | 136,0619 |
| 33 | Nicotinamida | 123,0555 |
| 34 | Vasicina | 189,0911 |
| 35 | Genisteína | 271,0602 |
| 36 | 3,4-difenil-7-hidroxicumarina | 315,0864 |
| 37 | Haemateína | 301,0708 |
| 38 | Prunetina | 285,0759 |
| 39 | 2-(2-hidroxibut-3-en-2-il)-3a,6,6,9a-tetrametil-2,4,5,5a,7,8,9,9b-octahidro-1H-benzo[E][1]benzofurano-4,5-diol | 361,2354 |
| 40 | (8S,9S,10R,11S,13S,14S,17S)-11,17-dihidroxi-17-(2-hidroxiacetil)-10,13-dimetil-2,6,7,8,9,11,12,14,15,16-decahidro-1H-ciclopenta[a]fenantren-3-ona | 363,2182 |

O agrupamento IV (Figura 22) mostrou-se um dos mais homogêneos, anotando substâncias apenas de uma classe metabólica específica, a dos triterpenos, em sua maioria, pentacíclicos (Tabela 6).

Figura 22 - Ampliação do agrupamento IV da rede molecular modo positivo das amostras CaxFls-EB, CaxFls-Hex, CaxFls-DCM e CaxFls-AcOEt.

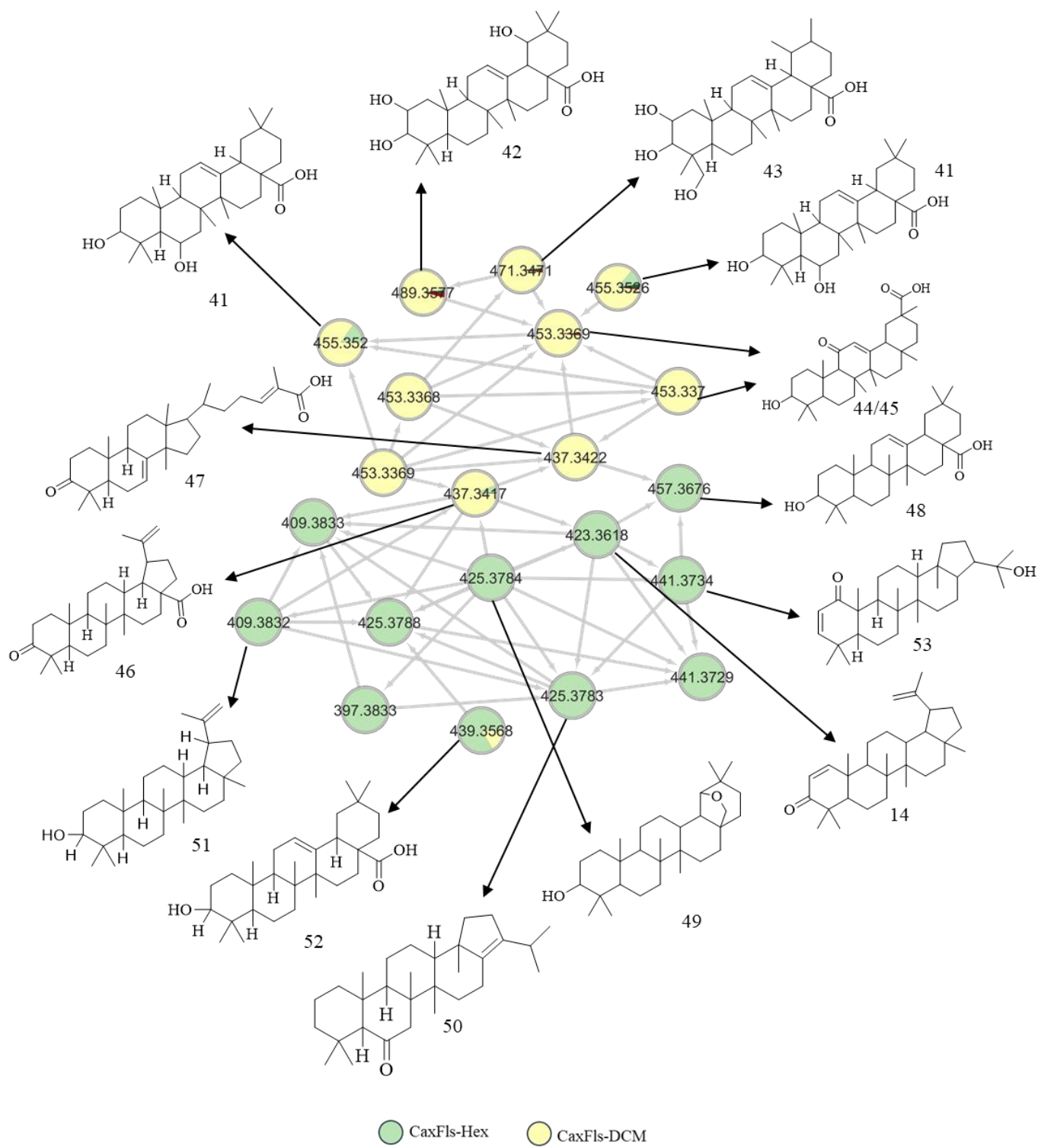

Fonte: Elaborado pelo autor.

Triterpenos pentacíclicos são amplamente distribuídos no reino vegetal. Representantes dessa classe possuem atividades biológicas descritas como anticancerígena (BISHAYEE *et al.*, 2011), anti-inflamatória (LIAW *et al.*, 2013), antioxidante (SILVA *et al.*, 2012), dentre outras. No gênero *Ficus*, triterpenos como o **51**, ursano e derivados como ácido ursólico e **48** isolados de *F. pseudopalma* foram avaliados quanto a sua atividade biológica e demonstraram atividade antioxidante e boa ação de eliminação de radicais livres (SANTIAGO *et al.*, 2014). Dentre os triterpenos anotados para *F. maxima*, destaca-se o **14**, substância anteriormente discutida na seção 4.3 e que possui atividade antinociceptiva registrada na literatura (KROGH *et al.*, 1999). Esta observação permite sugerir que o uso da espécie *F. maxima* para controle de dores pode estar relacionado a esta e/ou outras substâncias semelhantes.

Vale ressaltar que a maioria das espécies de *Ficus* estudadas reportam a classe dos flavonoides e polifenóis como constituintes majoritários de todas as partes da planta (NAWAZ; WAHEED; NAWAZ, 2019). As folhas de *F. maxima*, por sua vez, apresentaram uma grande quantidade de triterpenos pentacíclicos, como observado tanto no teste colorimétrico, quanto na análise por CLUE-EM-EM da rede molecular aqui representada (Tabela 6).

Tabela 6 - Substâncias anotadas no agrupamento IV.

| Entrada | Nome | *m/z* |
|---------|------|-------|
| 14 | Glochidona | 423,3618 |
| 41 | Ác. 8,10-dihidroxi-2,2,6a,6b,9,9,12a-heptametil-1,3,4,5,6,6a,6b,7,8,8a,9,10,11,12,12a,12b,13,14b-octadecahidropiceno-4a(2H)-carboxílico | 455,3526 |
| 42 | Ác. 2,3,19-trihidroxiolean-12-en-28-oico | 489,3577 |
| 43 | Ác. 2,3,23-trihidroxi-olean-12-en-28-oico | 471,3471 |
| 44 | Enoxolona | 453,3369 |
| 45 | Enoxolona (isômero) | 453,3370 |
| 46 | Ác. betulônico | 437,3417 |
| 47 | 3-hidroxi-13,28-epoxiurs-11-en-28-ona | 437,3422 |
| 48 | Ác. oleanólico | 457,3676 |

| Entrada | Nome | *m/z* |
|---------|------|-------|
| 49 | 4,5,9,9,13,20,20-heptametil-24-oxahexaciclo[17.3.2.0<1,18>.0<4,17>.0<5,14>.0<8,13>]tetracosan-10-ol | 425,3784 |
| 50 | 17(21)-hopen-6-ona | 425,3783 |
| 51 | Lupeol | 409,3832 |
| 52 | Ác. oleanólico ou isômero | 439,3568 |
| 53 | 22-hidoxi-2-hopen-1-ona | 441,3734 |

Nos outros agrupamentos menores VII, VIII, X e XII, representados em sequência na Figura 23, que não apresentaram similaridades com os demais, foi anotada apenas uma substância em cada: ácido traumático (**54**) (ácido dicarboxílico monoinsaturado), walleminona (**55**) (um sesquiterpeno do tipo cariofileno), carpacromeno (**56**) (uma flavona), e o bis[2-(2-butoxietoxi)etil] adipato (**57**), respectivamente (Tabela 7). De modo interessante, **55** representa um metabólito com núcleo cariofileno isolado do fungo toxigênico *Wallemia sebi* (FRANK *et al.*, 1999). Cariofilenos como o β-cariofileno e E-cariofileno estão presentes no óleo essencial de algumas plantas, como *Syzygium aromaticum* (GHELARDINI *et al.*, 2001) e *Cannabis sativa* (RIVERA *et al.*, 2023), respectivamente. Cariofilenos estão presentes também no óleo essencial de *F. benjamina* (OGUNWANDE *et al.*, 2012), corroborando para presença desse núcleo na amostra e primeira identificação da walleminona em *F. maxima*.

Figura 23 - Ampliação dos agrupamentos VII, VIII, X e XII da rede molecular modo
positivo das amostras CaxFls-EB, CaxFls-Hex, CaxFls-DCM e CaxFls-AcOEt.

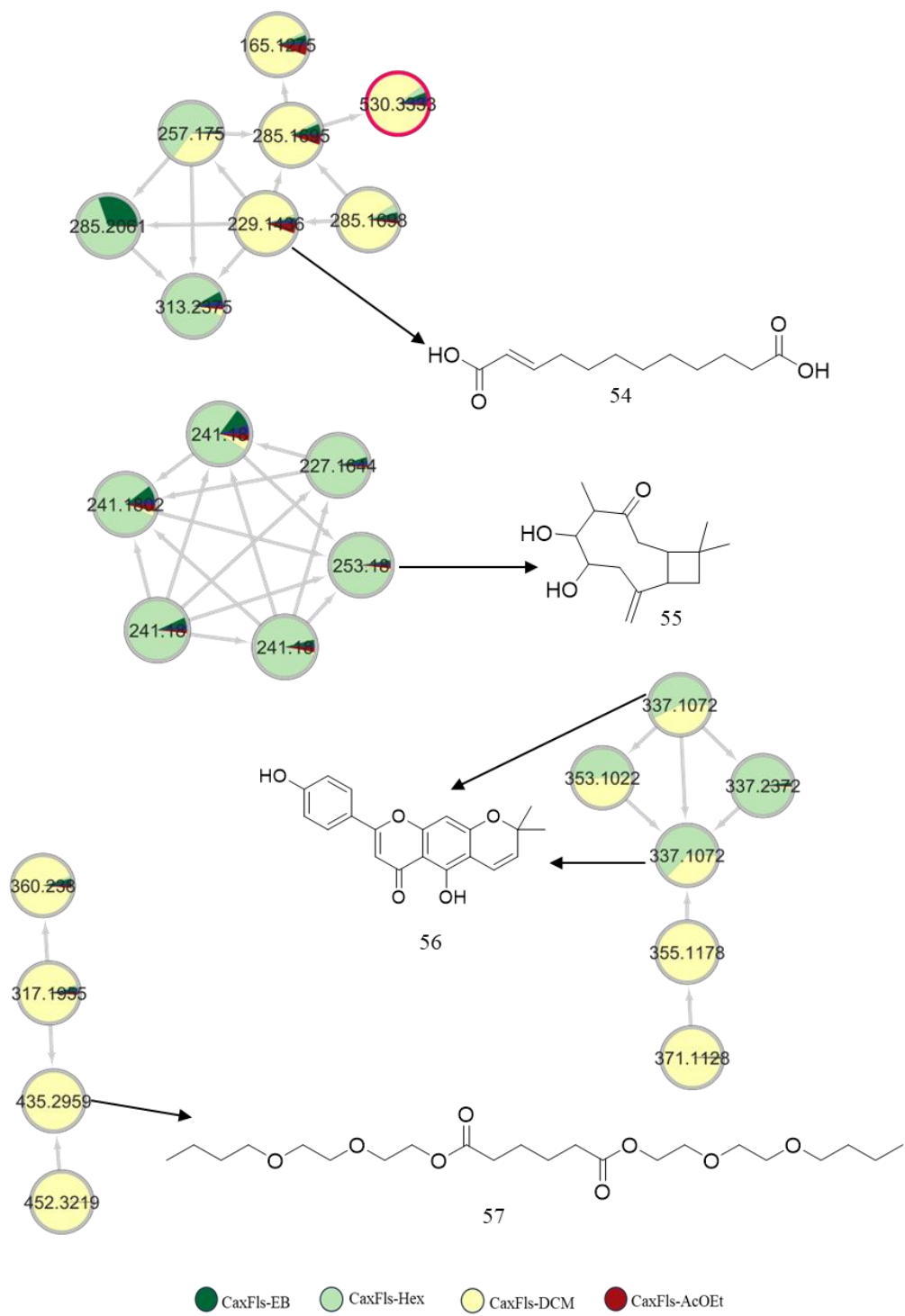

Fonte: Elaborado pelo autor.

Tabela 7 - Substâncias anotadas nos agrupamentos VII, VIII, X e XII.

| Entrada | Nome | m/z |
|---------|------|-----|
| 54 | Ác. traumático | 229,1496 |
| 55 | Walleminona | 253,179 |
| 56 | Carpacromeno | 337,1072 |
| 57 | bis[2-(2-butoxietoxi)etil] adipato | 435,2959 |

A rede molecular do modo negativo de CaxFls-EB, CaxFls-Hex, CaxFls-DCM e CaxFls-AcOEt está representada na Figura 24. Diferentemente do modo positivo, foram anotadas menos substâncias, devido à natureza dos constituintes em apresentarem mais sítios básicos que são mais favoráveis à ionização positiva. A rede foi composta de 205 nodos e 251 arestas. Foram criados seis agrupamentos ou *clusters* principais (XVII – XXII), três agrupamentos com apenas dois nodos cada (delimitados pelo retângulo) e outros dez nodos singulares (delimitados pela elipse), os quais tiveram alguma anotação molecular com a biblioteca do GNPS. Estão presentes também outros nodos que não tiveram nenhuma anotação ou correlação entre si.

Figura 24 - Rede molecular modo negativo das amostras CaxFls-EB, CaxFls-Hex, CaxFls-DCM e CaxFls-AcOEt editada no *software* Cytoscape 3.10.

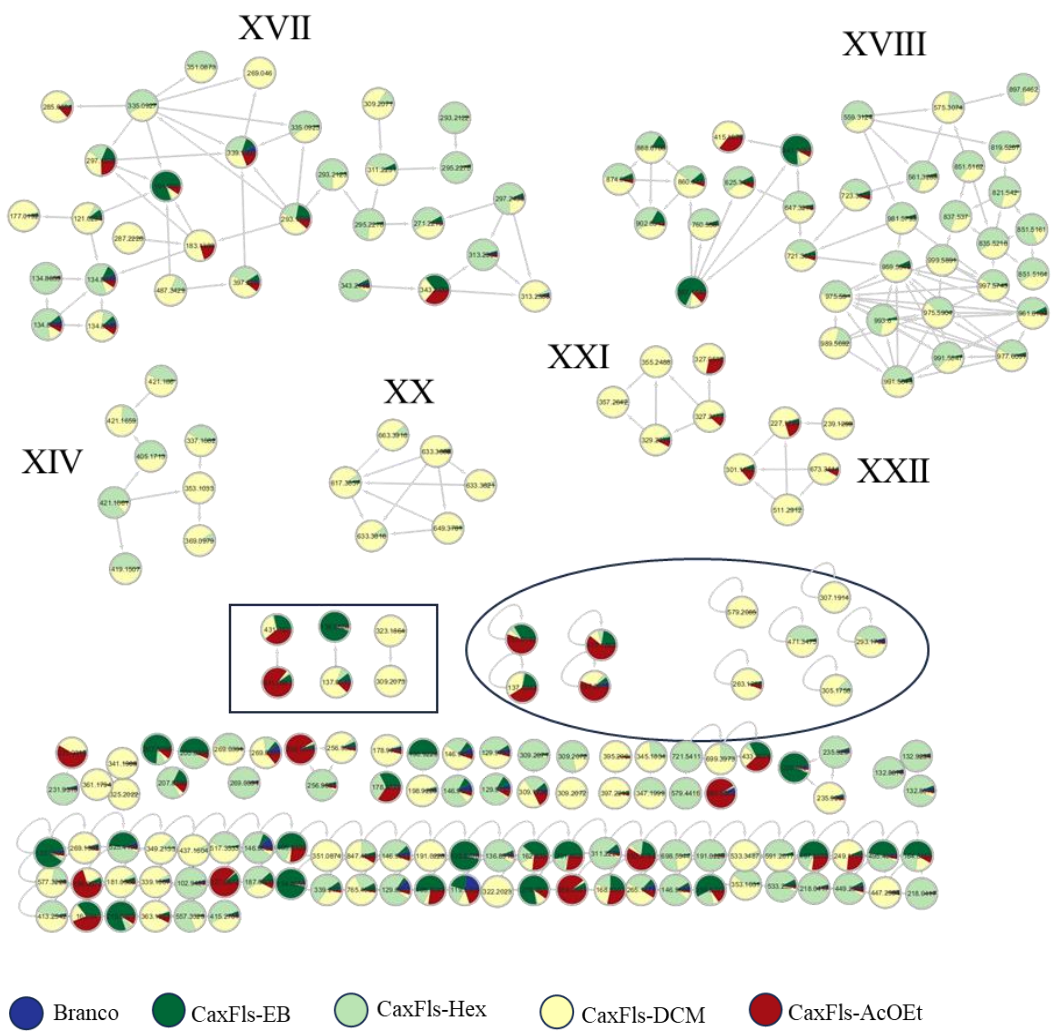

Fonte: Elaborado pelo autor.

O agrupamento XVII (Figura 25) anotou, além de alguns ácidos graxos poli-insaturados já detectados no modo positivo, flavona, isoflavona (SHAO *et al.*, 2022) uma cumarina (**65**) e um alcaloide de cinchona (**64**) (Tabela 8).

Figura 25 - Ampliação do agrupamento XVII da rede molecular modo negativo das amostras CaxFls-EB, CaxFls-Hex, CaxFls-DCM e CaxFls-AcOEt.

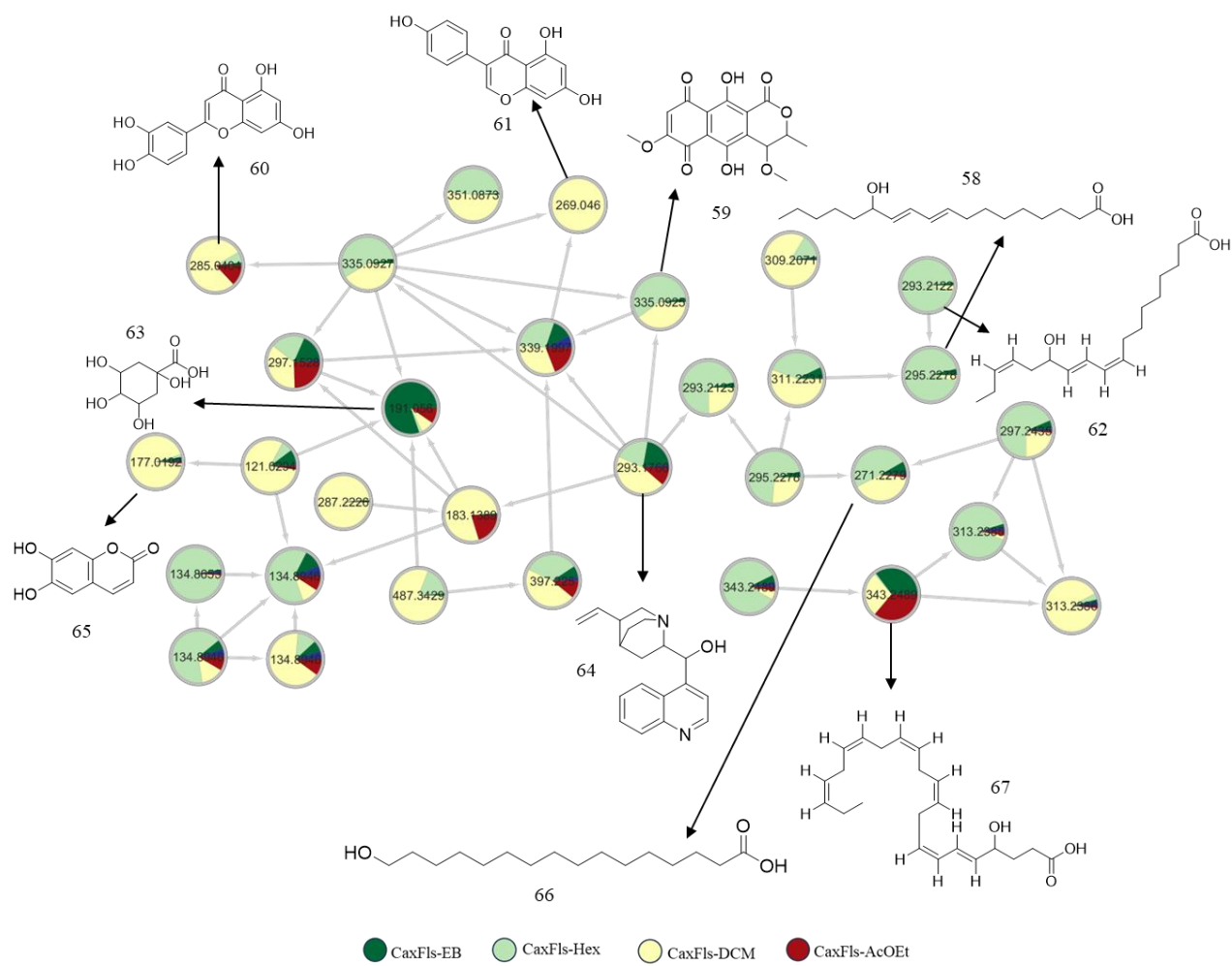

No agrupamento XVIII, foram anotados o éster 3-(hexopiranosiloxi)-2-hidroxipropil do ácido 9,12,15-octadecatrienoico (**68**) e o Glc-Glc-octadecatrienoil-sn-glicerol (**69**), em grande parte nas frações em Hex e DCM das folhas de *F. maxima* (Figura 26, Tabela 8).

Figura 26 - Ampliação do agrupamento XVIII da rede molecular modo negativo das amostras CaxFls-EB, CaxFls-Hex, CaxFls-DCM e CaxFls-AcOEt.

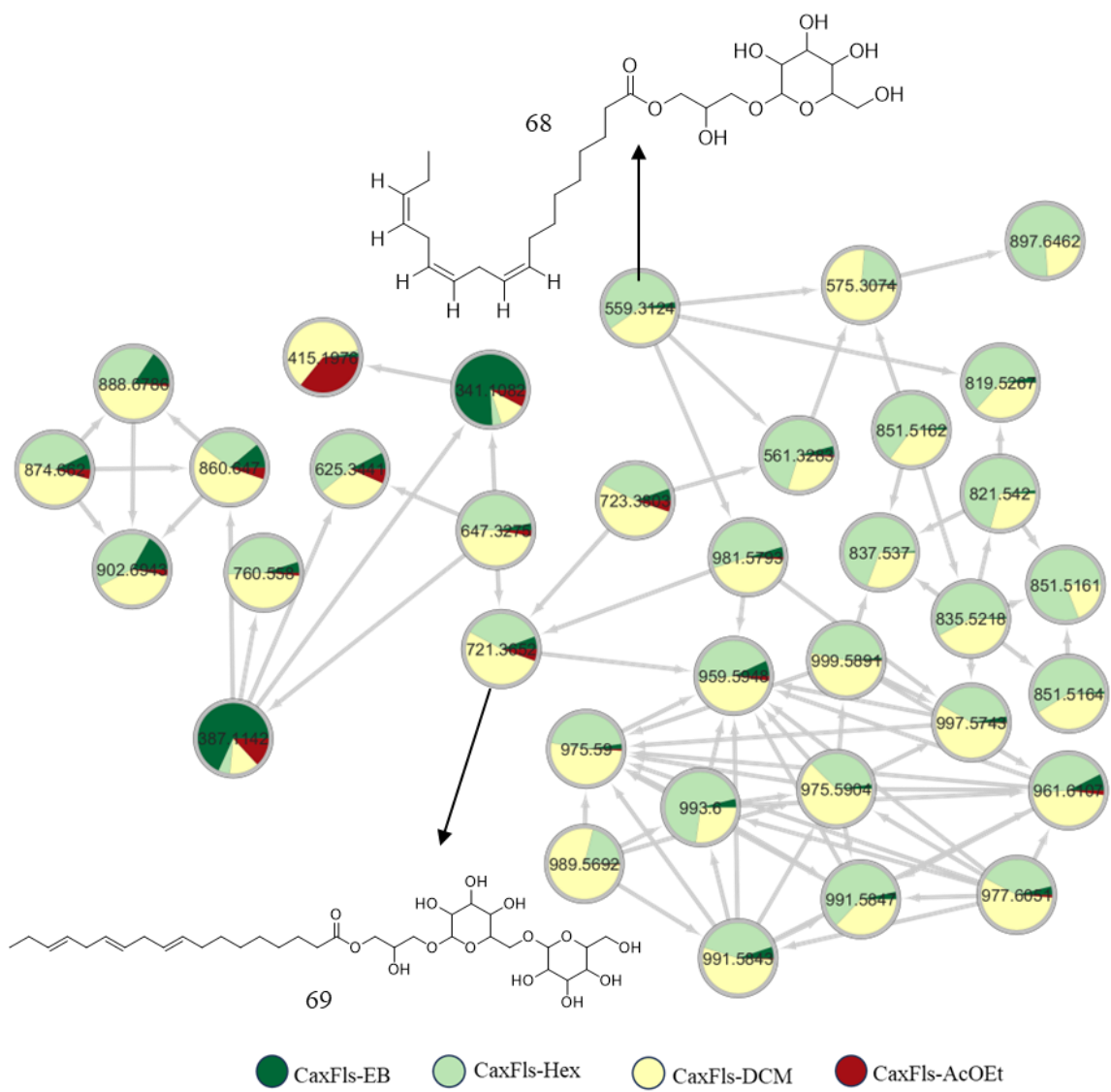

Fonte: Elaborado pelo autor.

Tabela 8 - Substâncias anotadas nos agrupamentos XVII e XVIII.

| Entrada | Nome | *m/z* |
|---------|------|-------|
| 58 | 13-HOTre | 295,2278 |
| 59 | Metoxihaemoventosina | 335,0946 |
| 60 | Luteolina | 285,0405 |
| 61 | Genisteína | 269,0459 |
| 62 | 13-OxoODE | 293,2122 |
| 63 | *D*-(-)-ácido quínico | 191,0559 |
| 64 | Cinchonina | 293,1770 |
| 65 | Esculetina | 177,0192 |
| 66 | Ác. 16-hidroxipalmítico | 271,2278 |
| 67 | Ác. (+/-)-4-hidroxi-5,7,10,13,16,19-docosahexaenoico | 343,2490 |
| 68 | Éster 3-(hexopiranosiloxi)-2-hidroxipropil do ác. 9,12,15-octadecatrienoico | 559,3126 |
| 69 | Glc-Glc-octadecatrienoil-sn-glicerol | 721,3618 |

No agrupamento XIV (Figura 27), foram anotados o ácido 2-*O*-*E*-*p*-coumaroil alftólico (**72**), o flavonoide prenilado 3-(3,4-dihidroxifenil)-5,7-dihidroxi-6,8-bis(3-metilbut-2-enil)cromen-4-ona (**70**) e seu isômero em CaxFls-Hex e CaxFls-DCM. Já nos agrupamentos XXI e XXII (Figura 28), foram anotados o ácido 9,12,13-triidroxioctadeca-10,15-dienoico (**73**) e o ácido traumático (**54**), presentes em maior concentração em CaxFls-DCM, e em menor concentração em CaxFls-AcOEt e no extrato bruto. Por sua vez, nos nodos delimitados pelo retângulo, foram anotados a vitexina (**75**), o aldeído protocatecuico (**76**) e mais um ácido octadecatrienoico.

Figura 27 - Ampliação dos agrupamentos XIV e XX  da rede molecular modo negativo das
amostras CaxFls-Hex e CaxFls-DCM.

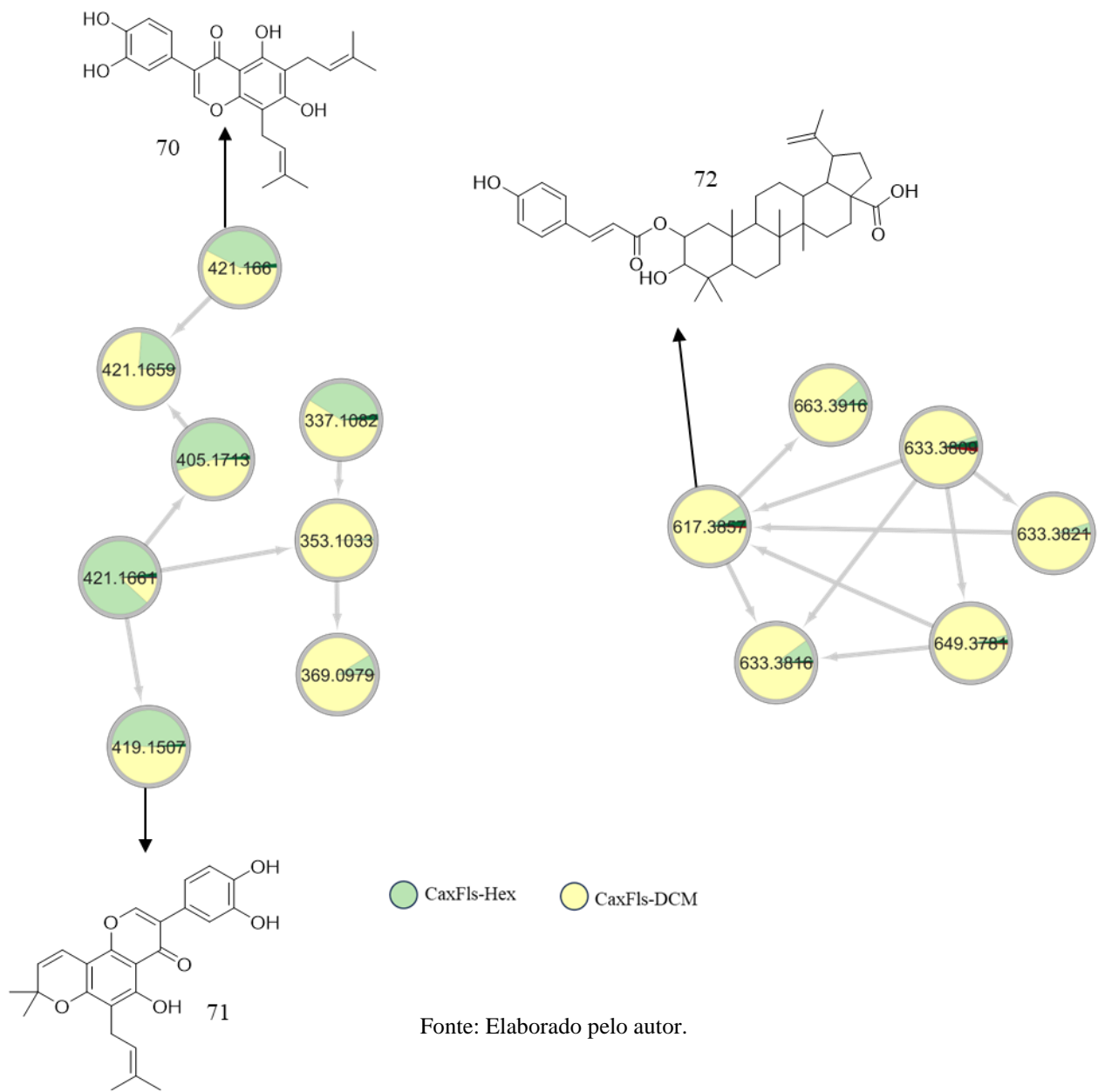

Fonte: Elaborado pelo autor.

Figura 28 - Ampliação dos agrupamentos XXI e XXII e dos nodos delimitados pelo retângulo da rede molecular do modo negativo de CaxFls-EB, CaxFls-Hex, CaxFls-DCM e CaxFls-AcOEt.

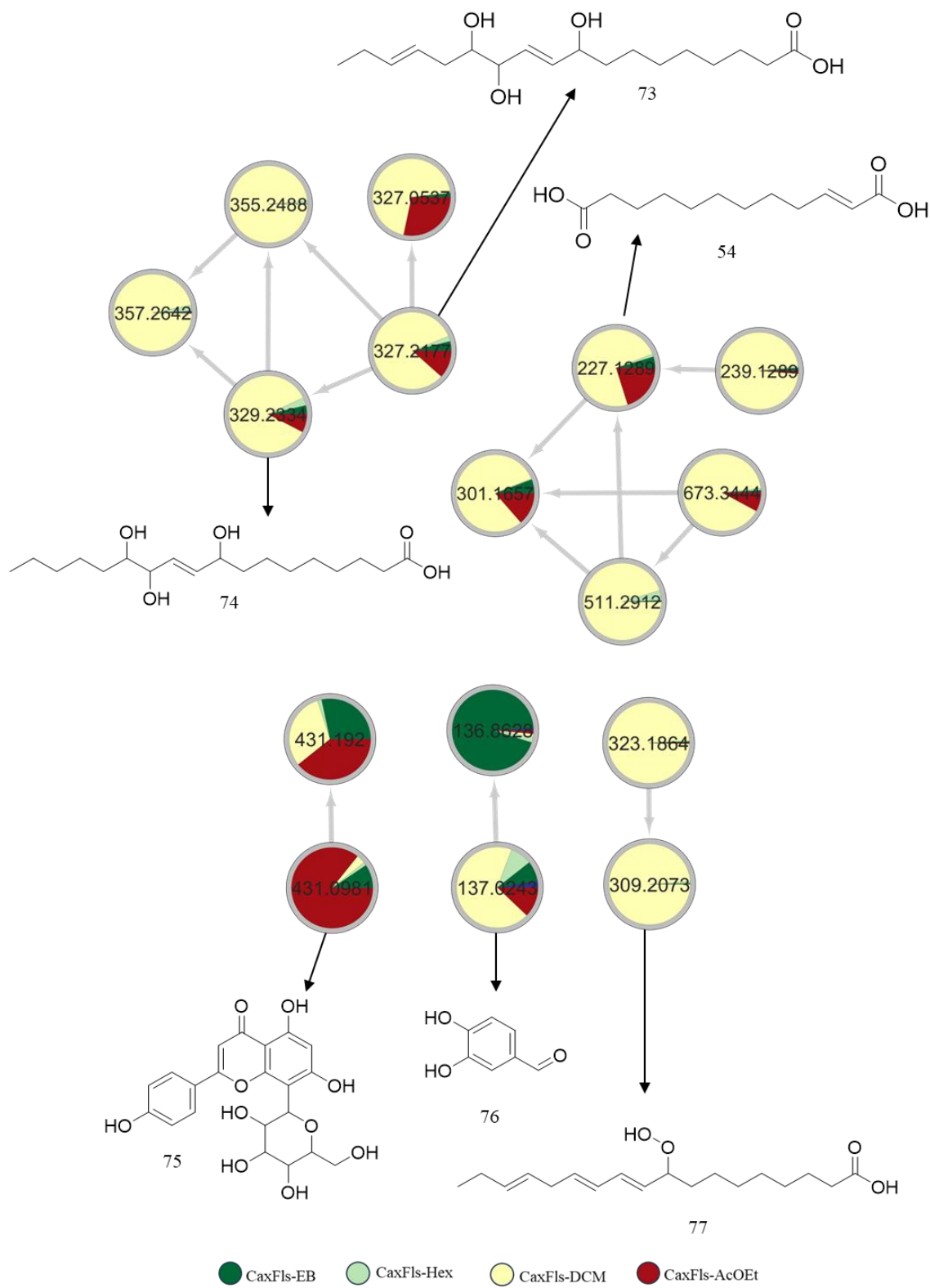

Fonte: Elaborado pelo autor.

Tabela 9 - Substâncias anotadas nos agrupamentos XXI, XXII e nodos delimitados pelo retângulo.

| Entrada | Nome | *m/z* |
|---------|------|-------|
| 70 | 3-(3,4-dihidroxifenil)-5,7-dihidroxi-6,8-bis(3-metilbut-2-enil)cromen-4-ona | 421,1657 |
| 71 | Pomiferina | 419,1507 |
| 72 | Ác. 2-*O*-*E*-*p*-coumaroil alftólico | 617,3847 |
| 73 | Ác. 9,12,13-trihidroxioctadeca-10,15-dienoico | 327,2177 |
| 74 | FA 18:1+3O | 329,2334 |
| 54 | Ác. traumático | 227,1289 |
| 75 | Vitexina | 431,0981 |
| 76 | 3-hidroxibenzoato | 137,0243 |
| 77 | FA 18:3+2O | 309,2073 |

Nos nodos delimitados pela elipse (Figura 29), foram anotados derivados do ácido benzoico como o 2,3-dihidroxibenzoato (**78**) e o ácido 3-hidroxibenzoico (**79**), uma

Figura 29 - Ampliação dos nodos delimitados pela elipse da rede molecular modo negativo de CaxFls-EB, CaxFls-Hex, CaxFls-DCM e CaxFls-AcOEt.

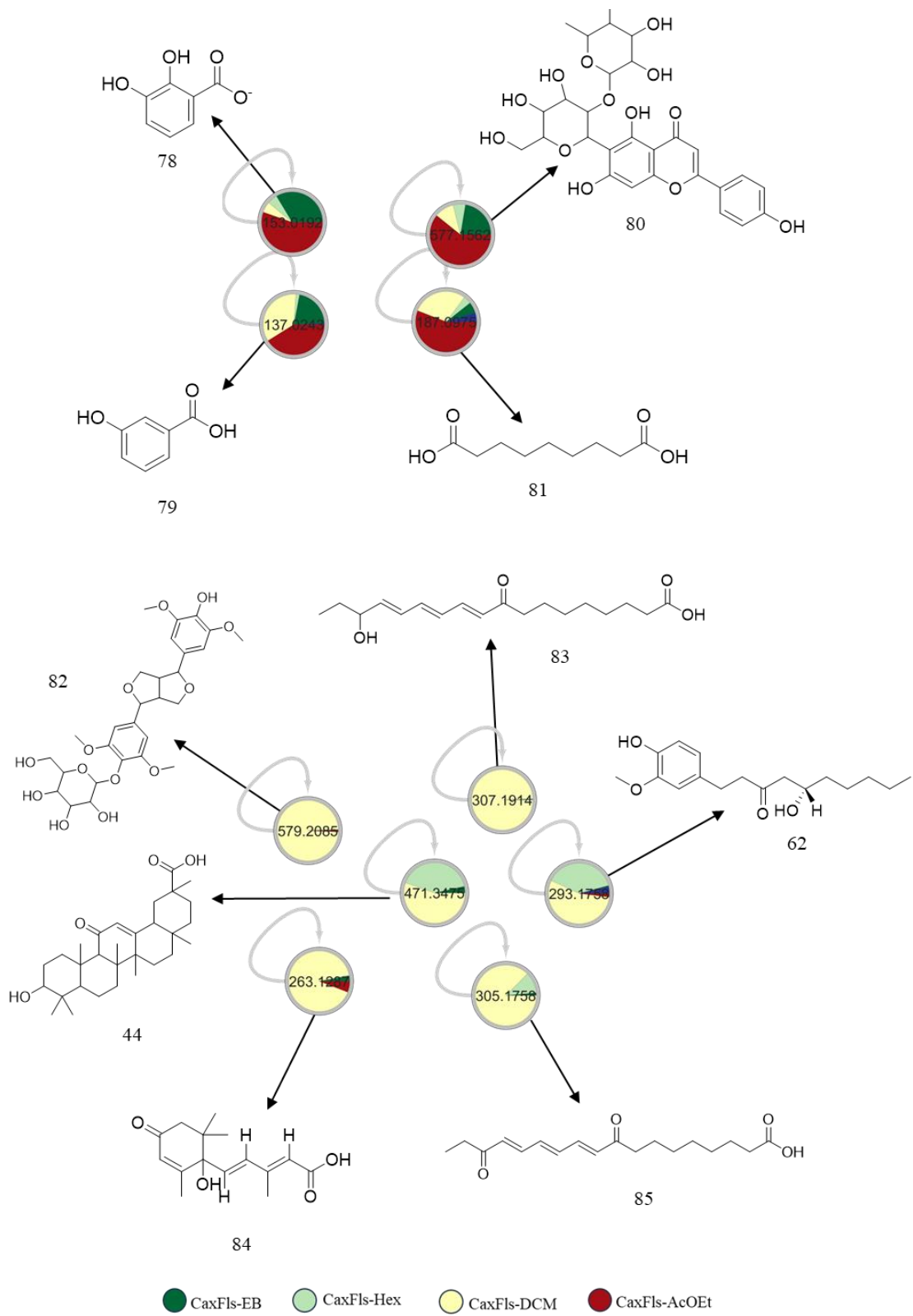

Fonte: Elaborado pelo autor.

isoflavona glicosilada (**80**), ácido azelaico (**81**), uma lignana (**82**), um triterpeno (**44**) e um sesquiterpeno (**84**), e o 6-gingerol (**62**) (Tabela 10).

Tabela 10 - Substâncias anotadas nos nodos delimitados pela elipse.

| Entrada | Nome | *m/z* |
|---------|------|-------|
| 78 | 2,3-dihidroxibenzoato | 153,0192 |
| 79 | Ác. 3-hidroxibenzoico | 137,0243 |
| 80 | 6-[4,5-dihidroxi-6-(hidroximetil)-3-[-3,4,5-trihidroxi-6-metiloxan-2-il]oxioxan-2-il]-5,7-dihidroxi-2-(4-hidroxifenil)cromen-4-ona | 577,1562 |
| 81 | Ác. azelaico | 187,0975 |
| 82 | 2-[4-[-3-(4-hidroxi-3,5-dimetoxifenil)-1,3,3a,4,6,6a-hexahidrofuro[3,4-c]furan-6-il]-2,6-dimetoxifenoxi]-6-(hidroximetil)oxano-3,4,5-triol | 579,2085 |
| 83 | FA 18:4+2O | 307,1914 |
| 44 | Enoxolona | 471,3475 |
| 62 | 6-gingerol | 293,1758 |
| 84 | Ác. abscísico | 263,1287 |
| 85 | FA 18:5+2O | 305,1758 |

A maioria das substâncias anotadas em ambos os modos de ionização foram pesquisados também nas bases de dados de espectrometria de massas do MassBank e PubChem para auxílio na correta identificação, a partir dos dados de EM-EM (EM$^2$). Desse modo, foi possível identificar alguns dos picos principais nos cromatogramas obtidos dos extratos e frações das folhas de *F. maxima*, representados nas Figura 30 - Figura 33. Outros picos não obtiveram anotação alguma nas bibliotecas do GNPS e nas demais pesquisadas, indicando, assim, substâncias possivelmente ainda não isoladas

Figura 30 - Cromatogramas de intensidade do pico base da amostra CaxFls-EB, modos positivo (superior) e negativo (inferior), com tempo total de análise de 22 min.

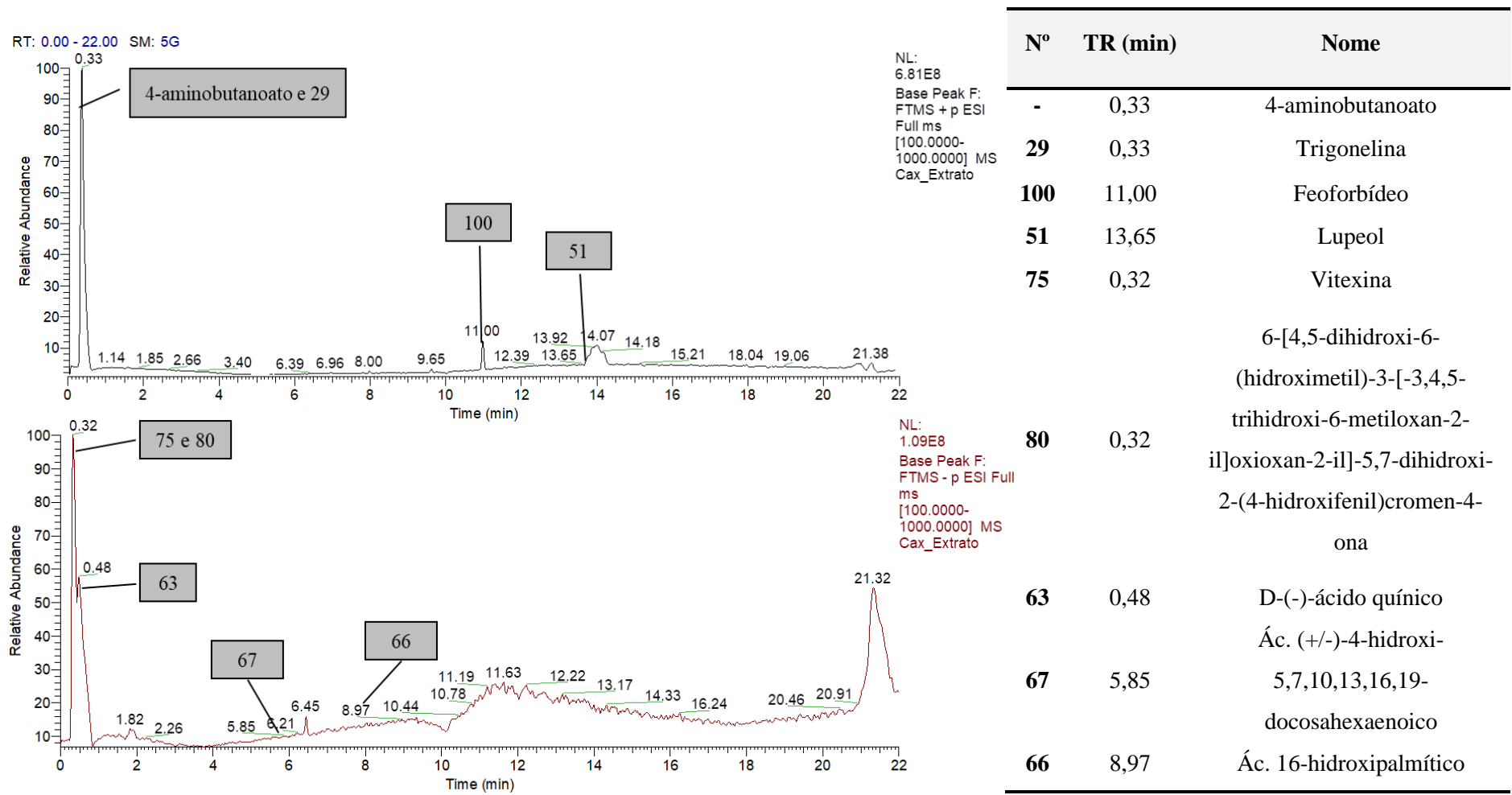

| Nº | TR (min) | Nome |
|---|---|---|
| - | 0,33 | 4-aminobutanoato |
| 29 | 0,33 | Trigonelina |
| 100 | 11,00 | Feoforbídeo |
| 51 | 13,65 | Lupeol |
| 75 | 0,32 | Vitexina |
| 80 | 0,32 | 6-[4,5-dihidroxi-6-(hidroximetil)-3-[-3,4,5-trihidroxi-6-metiloxan-2-il]oxioxan-2-il]-5,7-dihidroxi-2-(4-hidroxifenil)cromen-4-ona |
| 63 | 0,48 | D-(-)-ácido quínico |
| 67 | 5,85 | Ác. (+/-)-4-hidroxi-5,7,10,13,16,19-docosahexaenoico |
| 66 | 8,97 | Ác. 16-hidroxipalmítico |

Figura 31 - Cromatogramas de intensidade do pico base da amostra CaxFls-Hex, modos positivo (superior) e negativo (inferior), com tempo total de análise de 22 min.

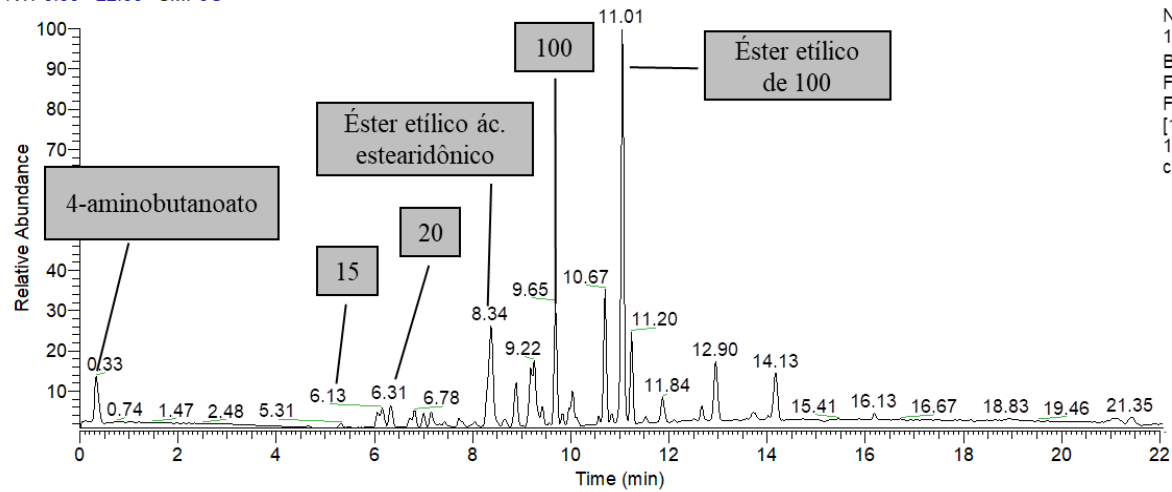

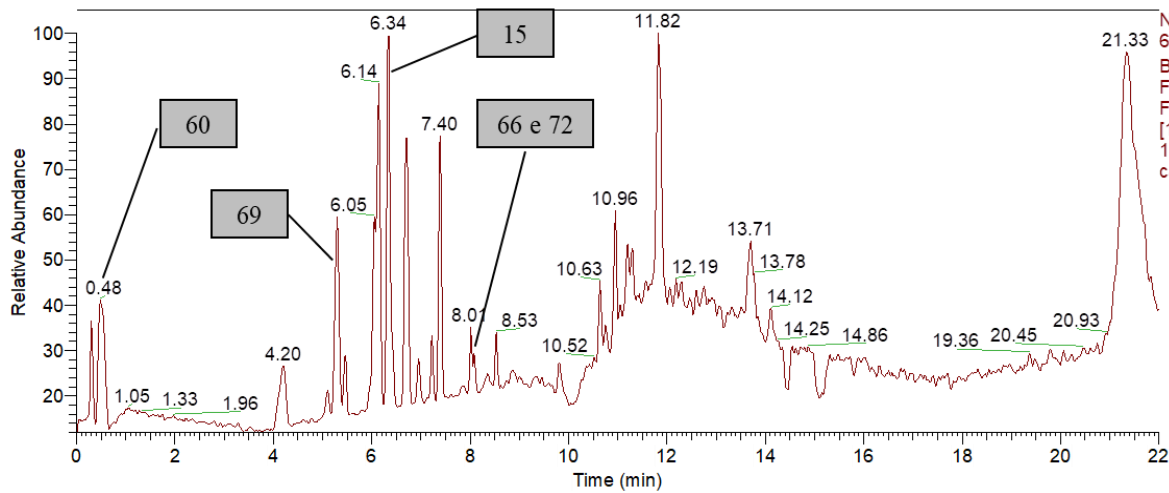

| Nº | TR (min) | Nome |
|---|---|---|
| - | 0,33 | 4-aminobutanoato |
| 15 | 6,13 | Ác. 9-hidroxi-10, 12, 15-octadecatrienoico |
| 20 | 6,31 | Monolinolenina (9c, 12c, 15c) |
| - | 8,34 | Éster etílico do ác. esteparidônico |
| 100 | 9,65 | Feoforbídeo |
| - | 11,01 | Éster etílico do feoforbídeo |
| 60 | 0,48 | Luteolina |
| 69 | 5,47 | Glc-Glc-octadecatrienoil-sn-glicerol |
| 15 | 6,34 | Ác. 9, 12, 15-octadecatrienoico |
| 66 | 8,01 | Ác. 16-hidroxipalmítico |
| 72 | | 2-O-E-p-cumaroil alftólico |

Figura 32 - Cromatogramas de intensidade do pico base da amostra CaxFls-DCM, modos positivo (superior) e negativo (inferior), com tempo total de análise de 22 min.

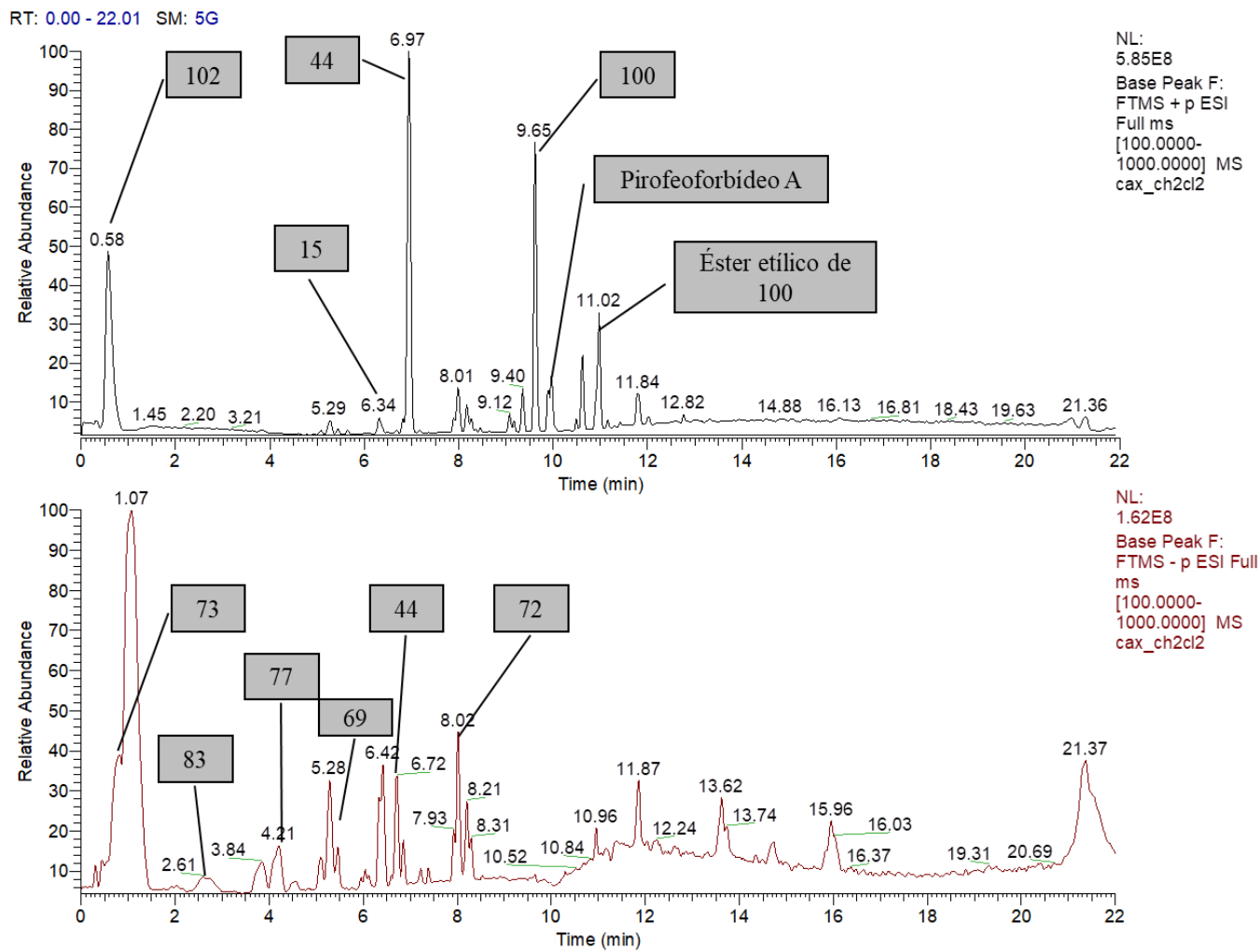

| Nº | TR (min) | Nome |
|---|---|---|
| **102** | 0,58 | Loliolídeo |
| **15** | 6,34 | Ác. 9, 12, 15-octadecatrienoico |
| **44** | 6,97 | Enoxolona |
| **100** | 9,65 | Feoforbídeo A |
| **-** | 10,01 | Pirofeoforbídeo A |
| **-** | 11,02 | Éster etílico do feoforbídeo |
| **73** | 0,87 | Ác. 9,12,13-trihidroxioctadeca-10,15-dienoico |
| **83** | 2,61 | FA 18:4+2O |
| **77** | 4,21 | FA 18:3+2O |
| **69** | 5,46 | Glc-Glc-octadecatrienoil-sn-glicerol |
| **42** | 6,72 | Enoxolona |
| **72** | 8,02 | 2-O-E-*p*-cumaroil alftólico |

Figura 33 - Cromatogramas de intensidade do pico base da amostra CaxFls-AcOEt, modos positivo (superior) e negativo (inferior), com tempo total de análise de 22 min.

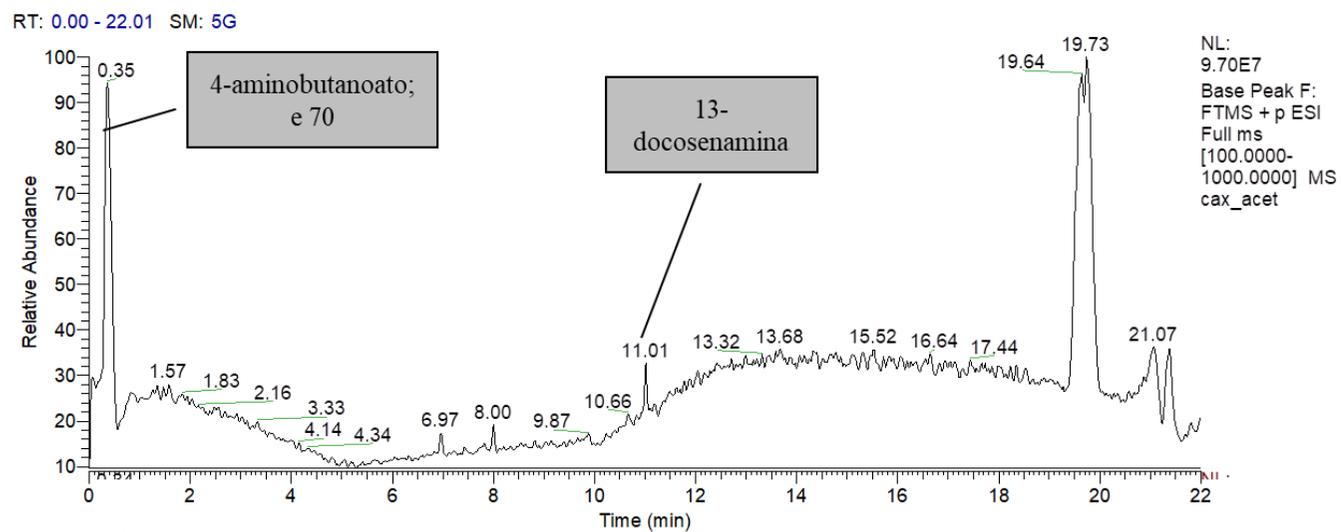

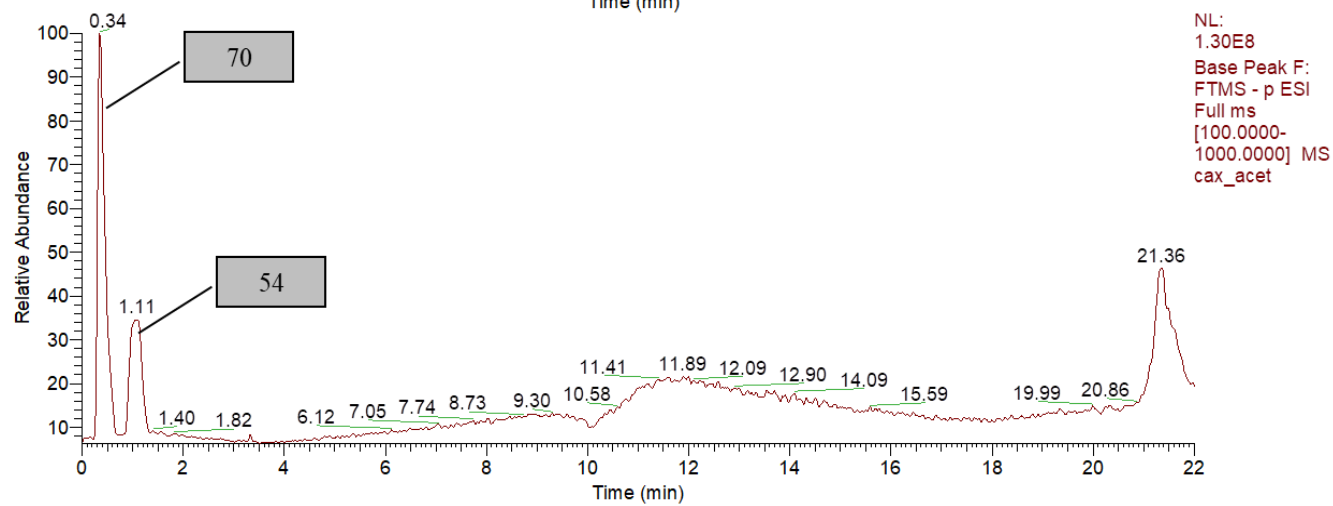

| Nº | Nome | TR (min) |
|---|---|---|
| - | 4-aminobutanoato | 0,35 |
| 70 | Isovitexina | 0,35 |
| - | 13-docosenamida | 11,01 |
| 70 | Isovitexina | 0,34 |
| 54 | Ác. traumático | 1,11 |

A comparação das anotações realizadas pelo GNPS e dos espectros de EM-EM das substâncias identificas foi realizada com auxílio da ferramenta *Metabolomics Spectrum Resolver*. Essa ferramenta, integrada à plataforma do GPNS, está disponível no site https://metabolomics-usi.gnps2.org/ e permite aos usuários construir o espectro de EM experimental e espelhar correspondências entre dois espectros de EM-EM mostrando sua similaridade (BITTREMIUEX *et al.*, 2020). Nesse caso, os espectros experimentais de EM-EM de alguns dos principais picos mais intensos das amostras são discutidos a seguir e estão mostrados em preto, enquanto os da biblioteca do GNPS são mostrados em verde. O valor de "*cosine similarity*" indica a similaridade entre os dois espectros comparados diretamente. Do mesmo modo que esse parâmetro foi definido para a criação das redes moleculares, valores próximos a 1,0 indicam 100% de similaridade.

O espectro de EM-EM comparado correspondente ao 4-aminobutanoato, com TR de 0,33 está representado na Figura 34. Por se tratar de uma substância pequena, a IES não provoca muita fragmentação, o que é nítido pela presença de apenas quatro picos principais no espectro. É possível observar que o pico do íon molecular de *m/z* 104 ($C_4H_9NO_2$) dos dois espectros, representando um aduto [M+H]+ são semelhantes, com a mesma intensidade observada em ambos. A proposta de fragmentação (Esquema 1) sugere que o segundo pico mais intenso é justificado por uma quebra heterolítica entre os carbonos α e β para gerar o fragmento de *m/z* 60. É observado também uma perda de $H_2O$ para gerar o íon de *m/z* 87.

Figura 34 - Espectro de EM-EM (IES+) experimental (superior) do 4-aminobutanoato comparado ao da biblioteca do GNPS (inferior).

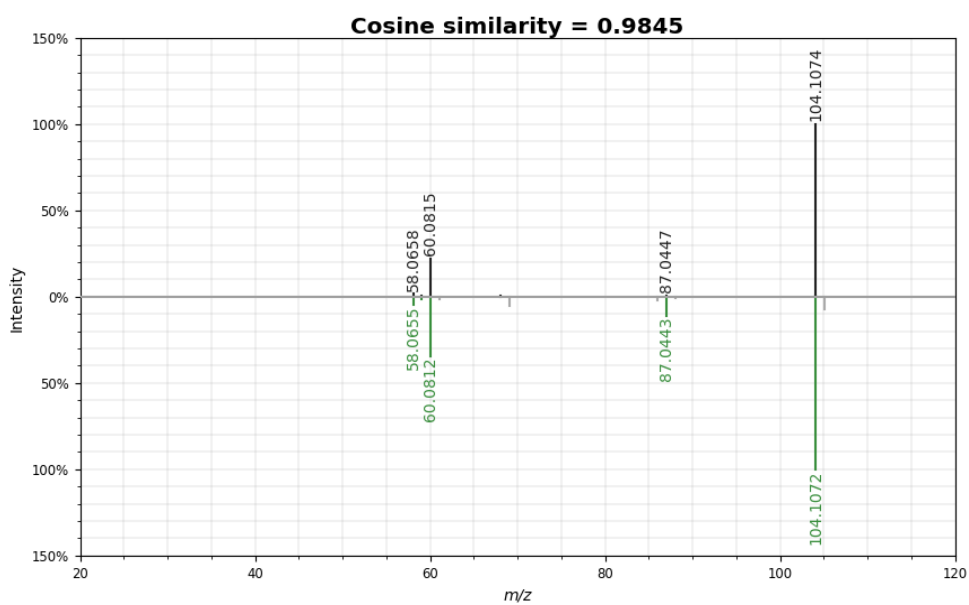

Esquema 1 - Proposta de fragmentação do 4-aminobutanoato.

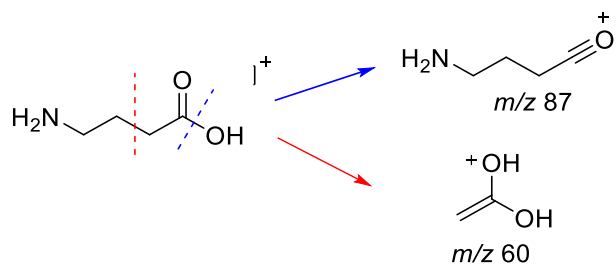

Fonte: Elaborado pelo autor.

O espectro de EM-EM comparado correspondente ao ácido 9-hidroxi-10, 12, 15-octadecatrienoico, com TR 6,13 está representado na Figura 35.

Figura 35 - Espectro de EM-EM (IES+) experimental (superior) do ác. 9-hidroxi-10, 12, 15-octadecatrienoico comparado ao da biblioteca do GNPS (inferior).

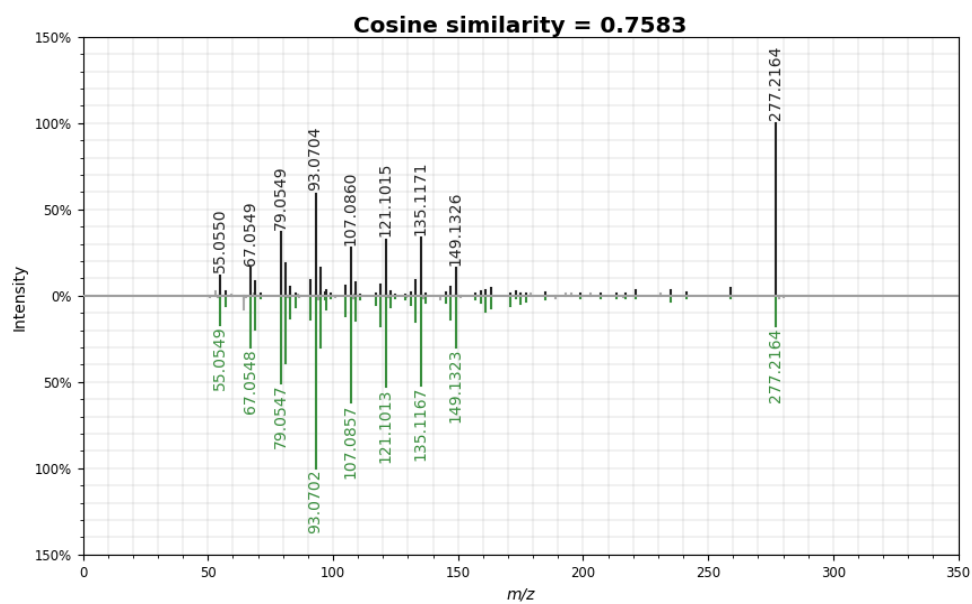

O ácido 9-hidroxi-10, 12, 15-octadecatrienoico é um ácido graxo policonjugado. O valor de cosseno um pouco mais baixo (0,75) pode ser justificado para as diferenças nas intensidades observadas para alguns picos, inclusive o do íon molecular, cuja intensidade é maior no espectro experimental. O íon molecular apresentou razão *m/z* de 277 ($C_{18}H_{30}O_3$), justificado por uma perda de $H_2O$ [M+H-H2O]+. As fragmentações de ácidos graxos semelhantes na literatura são geralmente produzidas por impacto de elétrons (IE), que tem uma energia de colisão superior à IES, gerando mais fragmentos (YOSHIKAWA *et al.*, 1998). Um dos picos mais observados por IE para esses compostos é o gerado pela clivagem da ligação C9-C10 próximo à hidroxila, que pode justificar a presença do íon de *m/z* 121 (Esquema 2) (DONG; ODA; HIROTA, 2000; TRAPP *et al.*, 2015).

Esquema 2 - Proposta de fragmentação para o ác. 9-hidroxi-10, 12, 15-octadecatrienoico

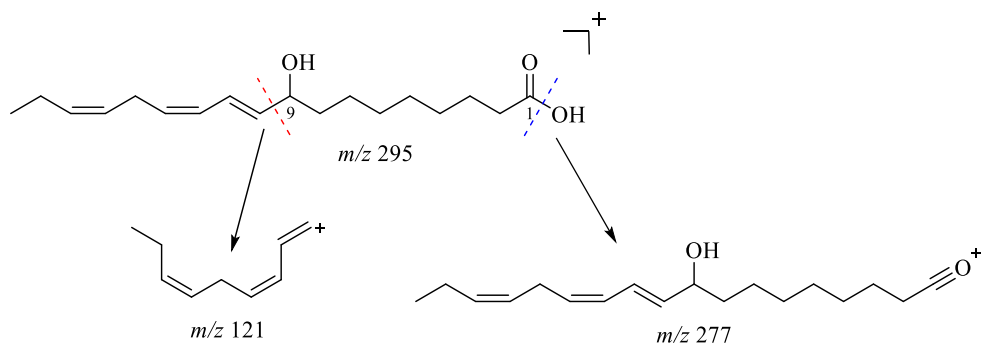

Fonte: Elaborado pelo autor.

O espectro de EM-EM comparado correspondente à trigonelina (**29**), com TR 0,33 está representado na Figura 36:

Figura 36 - Espectro de EM-EM (IES+) experimental (superior) de **29** comparado ao da biblioteca do GNPS (inferior).

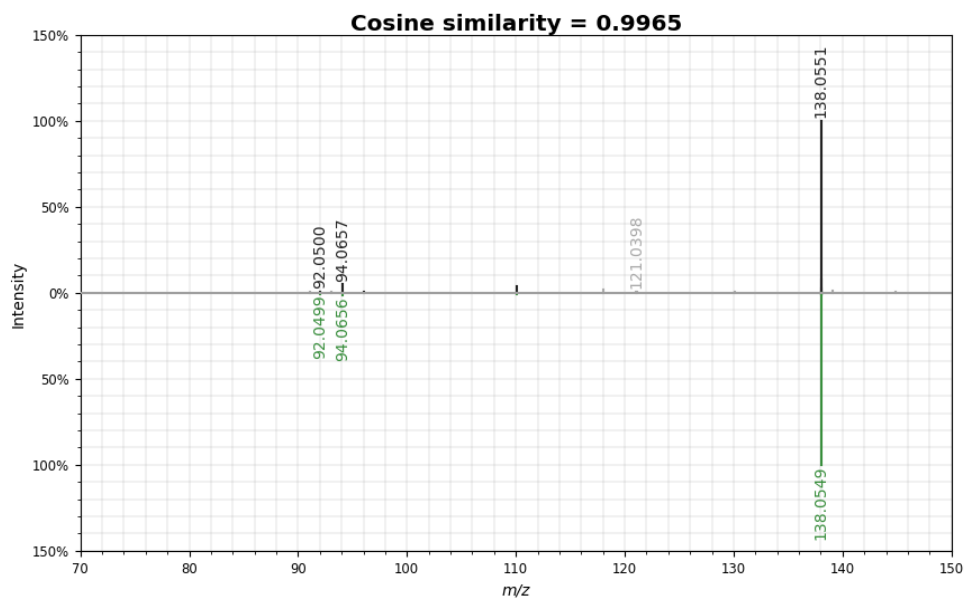

O íon molecular [M+H]+ apresentou *m/z* 138 ($C_7H_7NO_2$), com intensidades semelhantes em ambos os espectros. A substância **29** por se tratar de alcaloide relativamente pequeno, também não produz muitas fragmentações por IES. Um outro pico que tem uma intensidade relativa na faixa de 10%, mas que é imperceptível no espectro, é o gerado pela perda de $H_2O$, com *m/z* 121. O pico de *m/z* 94 pode ser justificado pela perda de $CO_2$, que, consequentemente, pela perda de $H_2$ leva à formação de um íon estável em *m/z* 92, conforme proposta de fragmentação (Esquema 3) (HAU *et al.*, 2001).

Esquema 3 - Proposta de fragmentação para **29**.

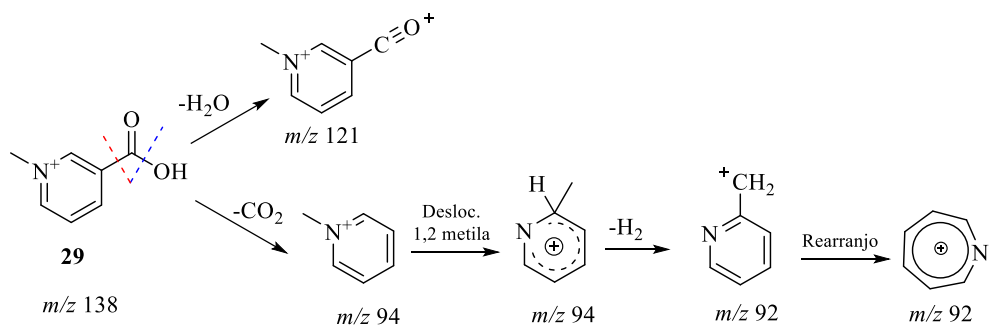

Fonte: Elaborado pelo autor.

A trigonelina foi pela primeira vez isolada das sementes de feno-grego (*Trigonella foenum-graecum* L.), da família Fabaceae. Na literatura, são relatadas algumas atividades biológicas para a substância, como efeitos hipoglicemiantes e doenças do SNC (ZHOU; CHAN; ZHOU, 2012) o que corrobora para sua presença no extrato de *F. maxima*, já que espécies de *Ficus* também são utilizadas popularmente para tratamento de diabetes.

O espectro de EM-EM comparado correspondente ao feoforbídeo, com TR 11,00 está representado na Figura 37.

Figura 37 - Espectro de EM-EM (IES+) experimental (superior) do feoforbídeo comparado ao da biblioteca do GNPS (inferior).

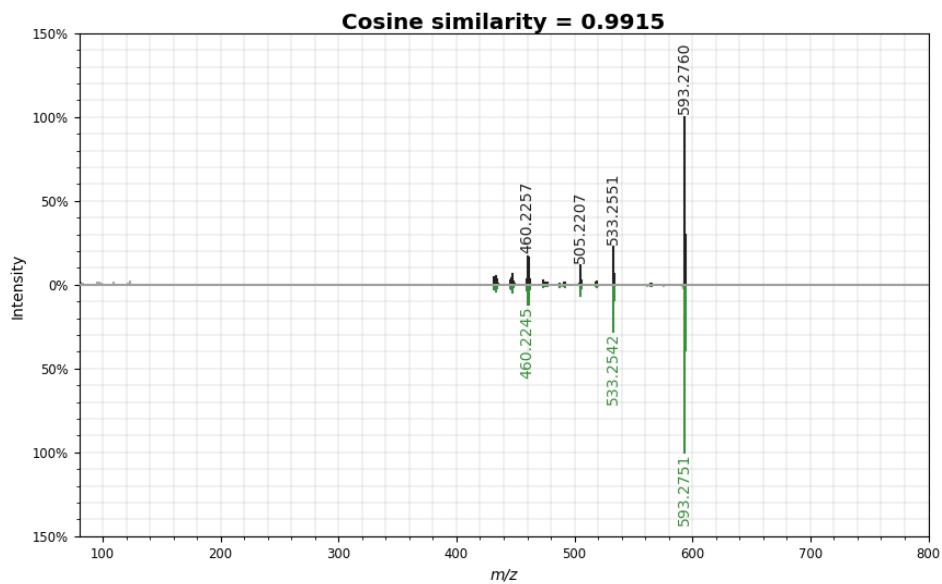

Como observado na comparação dos espectros, existe uma alta similaridade entre os picos e suas intensidades, justificando a presença do feoforbídeo. O feoforbídeo é um produto derivado da clorofila, normalmente por decomposição dessa. Na literatura, derivados da clorofila como o feoforbídeo A e feofitinas têm sido isolados de folhas de *F. exasperata* mostrando atividades no tratamento de desordens uterinas (BAFOR; ROWAN; EDRADA-EBEL, 2018). A literatura também relata que análises por EM desses derivados de clorofila apresentam o íon molecular [M+H]+ ($C_{35}H_{36}N_4O_5$) como sendo mais abundante do que outros íons fragmentos, o que também é observado nos espectros comparados (BREEMEN; CANJURA; SCHWARTZ, 1991).

O espectro de EM-EM comparado correspondente ao lupeol (**51**), com TR 13,65, um triterpeno pentacíclico, está representado na Figura 38:

Figura 38 - Espectro de EM-EM (IES+) experimental (superior) de **51** comparado ao da biblioteca do GNPS (inferior).

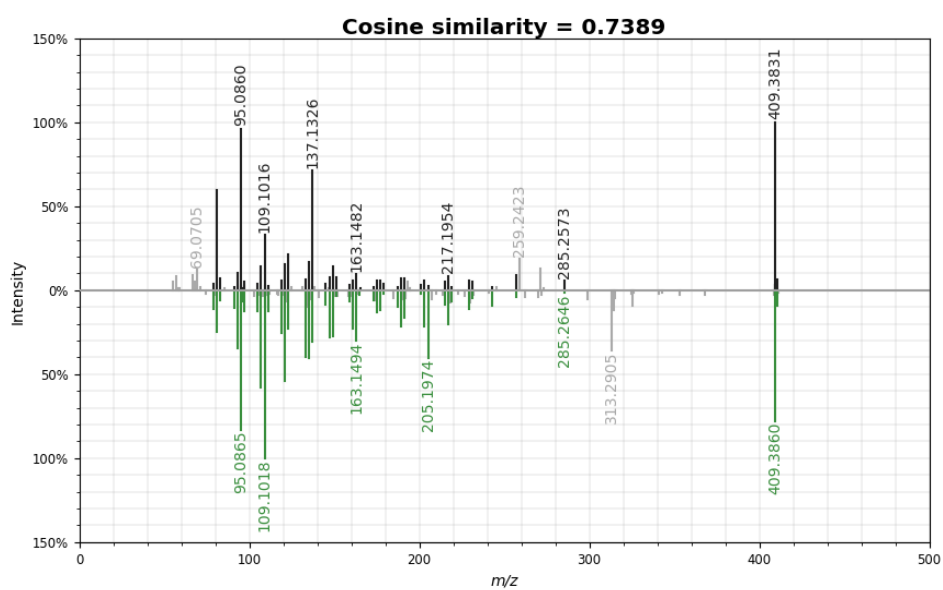

O baixo valor de similaridade (0,73) pode ser justificado pelas diferenças de intensidade observadas para alguns íons fragmentos, dentre eles o de *m/z* 205 que, no espectro experimental, apresenta uma intensidade muito baixa. A fragmentação mais observada para **51** é a perda de $H_2O$ que gera o íon *m/z* 409 ($C_{30}H_{50}O$), como representado na proposta abaixo (Esquema 4).

Esquema 4 - Proposta de fragmentação para **51**.

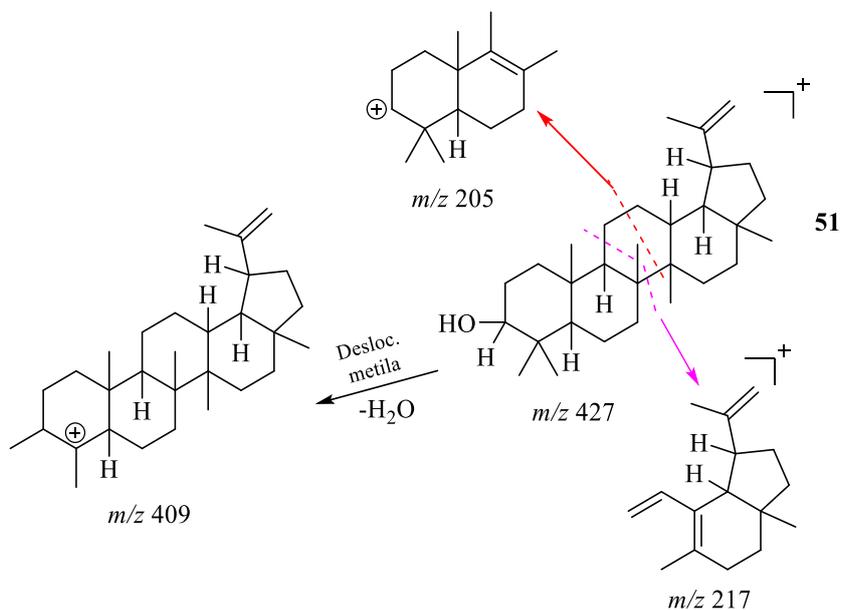

Fonte: Elaborado pelo autor.

O espectro de EM-EM correspondendo à enoxolona (**44**), com TR 6,987, outro triterpeno pentacíclico, está representado na Figura 39:

Figura 39 - Espectros de EM-EM (IES+) experimental (superior) de **44** comparado ao da biblioteca do GNPS (inferior).

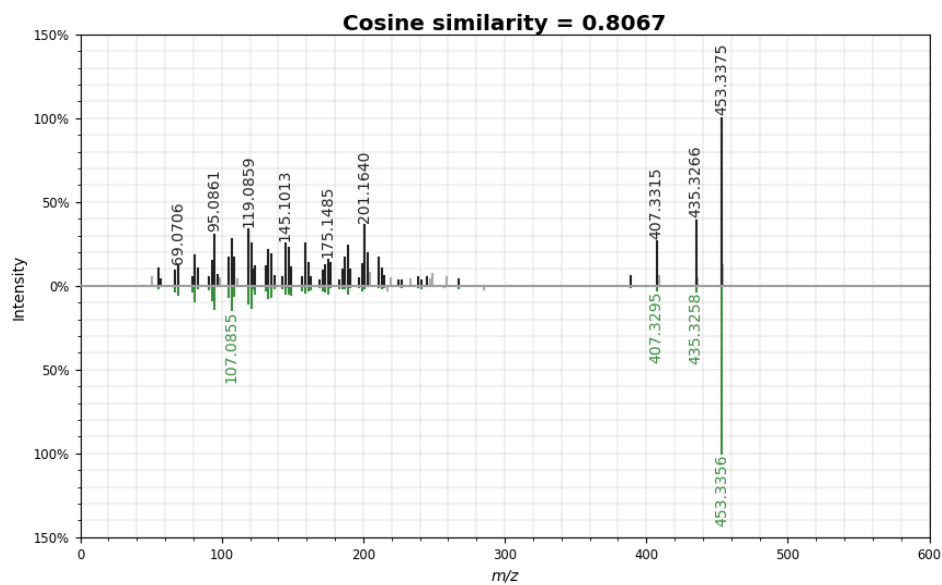

Os picos principais de **44** indicam a perda de H$_2$O (-18 Da), que gera o íon de *m/z* 453 (C$_{30}$H$_{46}$O$_4$) representando o aduto [M+H-H2O]+ e mais uma perda -18 Da, para gerar o íon de *m/z* 435. Em seguida, uma diferença de 46 Da indicando perda de HCOOH da molécula, gera o íon fragmento de *m/z* 407, conforme proposta de fragmentação abaixo (Esquema 5) (MUSHARRAF; KANWAL; ARFEEN, 2013).

Esquema 5 - Proposta de fragmentação para **44**.

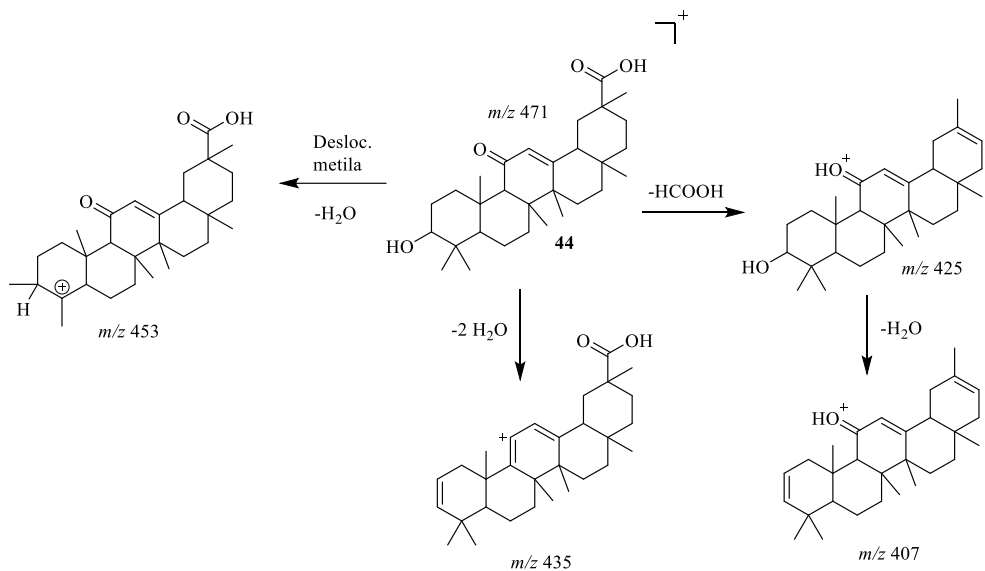

O espectro de EM-EM comparado correspondente à isovitexina (**70**), com TR 0,35 está representado na Figura 40:

Figura 40 - Espectro de EM-EM (IES-) experimental (superior) de **70** comparado ao da biblioteca do GNPS (inferior).

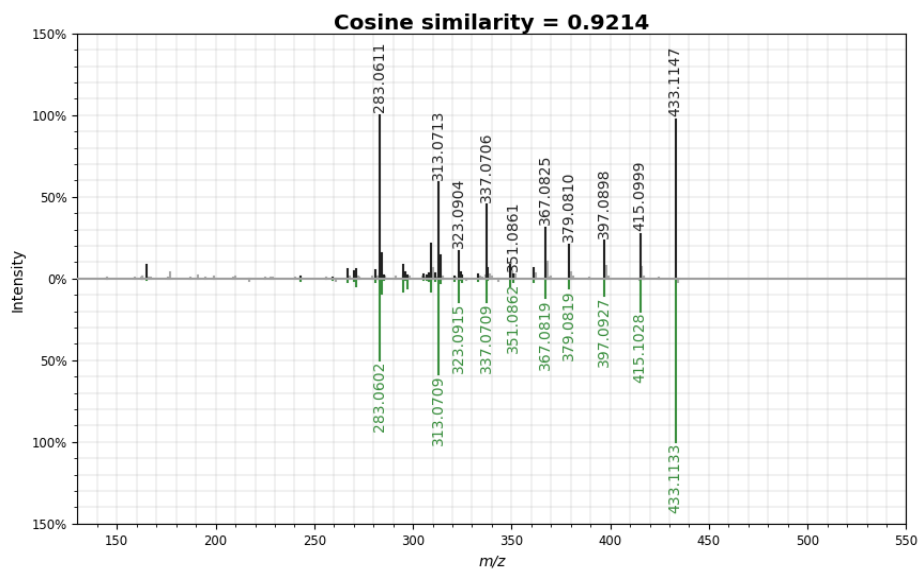

O pico do íon molecular [M-H]- de $m/z$ 433 ($C_{21}H_{20}O_{10}$) apresenta intensidade próxima a 100% em ambos os espectros de EM-EM. A substância **70** é um flavonoide 6-*C*-glicosilado. O segundo pico mais intenso para a isovitexina é o de $m/z$ 283, que pode ser justificado pela clivagem da molécula de glicose na ligação O-C1. Em seguida, o segundo pico mais intenso de $m/z$ 313 também pode ser justificado pela clivagem da molécula de glicose na ligação O-C2. A perda de H₂O (-18 Da) também é observado para a substância, gerando o íon *m/z* 415 (Esquema 6) (PEREIRA; YARIWAKE; MCCULLAGH, 2005; SUN *et al.*, 2013).

Esquema 6 - Proposta de fragmentação para **70**.

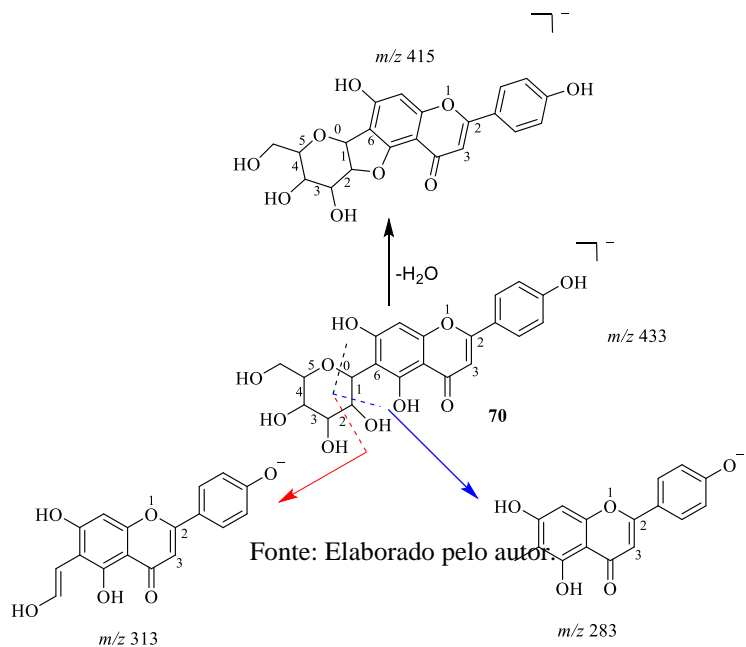

Fonte: Elaborado pelo autor.

Os cromatogramas adquiridos das análises do extrato etanólico e das frações Hex, DCM e AcOEt das folhas de *F. maxima* encontram-se também em anexo para melhor visualização.

## 4.5    Análise das subfrações CaxFls-DCM R3+4, CaxFls-DCM R5+6 e CaxFls-DCM R7 por CLUE-EM-EM e construção de redes moleculares

As frações discutidas na seção 4.2  foram submetidas a análise por CLUE-EM-EM e ionização por eletrospray (IES), com a mesma metodologia supracitada, contudo utilizando como fase móvel solução aquosa de ácido fórmico 0,1% (fase A) e solução de MeOH:ácido fórmico 0,1% (fase B) e tempo de análise de 20 minutos. O espectrômetro de massas foi configurado da mesma maneira que as análises anteriores.

Os cromatogramas obtidos encontram-se representados a seguir (Figura 41). É possível observar pelos cromatogramas que as três amostras possuem bastante similaridade no perfil químico, onde os constituintes majoritários apresentam sinais na região de baixa polaridade, com o mesmo tempo de retenção entre 14 e 16 min no modo positivo. Exceto por CaxFls-DCM R7 que apresentou um pico de média intensidade em 14,06 min, indicando concentração alta de substância não detectada na mesma concentração nas outras amostras.

Figura 41 - Cromatogramas de intensidade do pico base das subfrações CaxFls-DCM R3+4, R5+6 e R7 no modo positivo, tempo total de análise de 20 min.

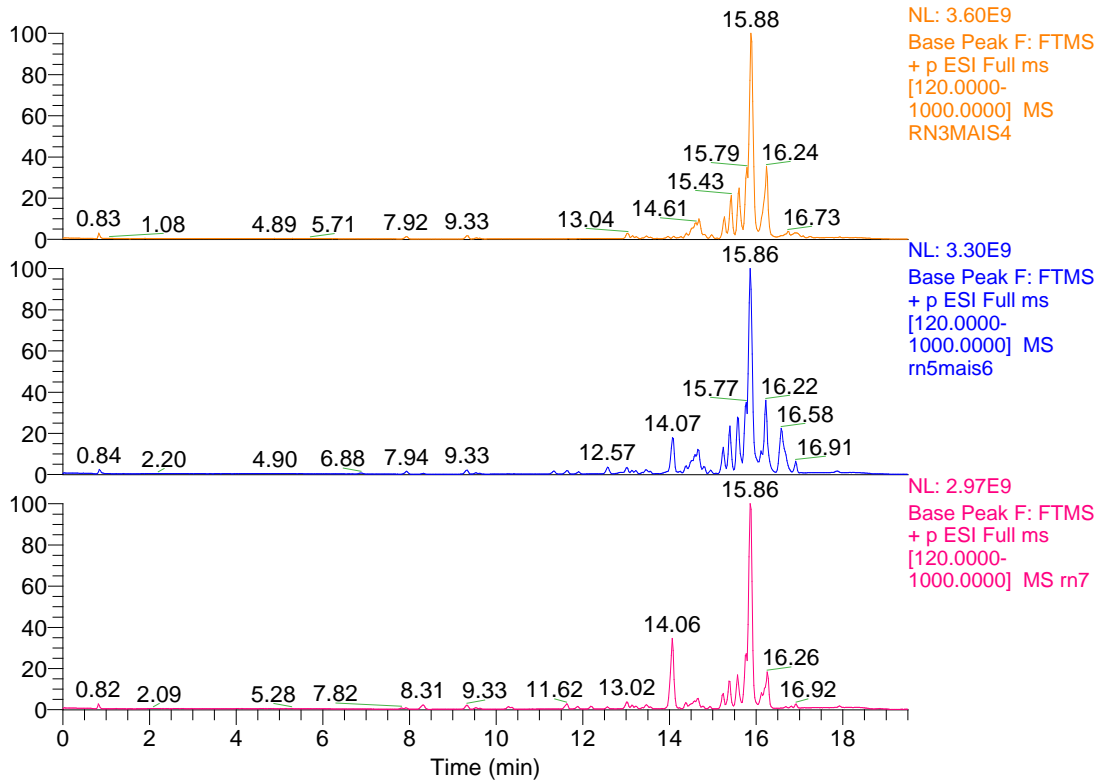

Fonte: Elaborado pelo autor.

No modo negativo (Figura 42), as subfrações apresentam perfis ligeiramente distintos entre si, exceto pelo sinal em 7,93 min que aparece praticamente nas três análises.

Figura 42 - Cromatogramas de intensidade do pico base das subfrações CaxFls-DCM R3+4, R5+6 e R7 no modo negativo, tempo total de análise de 20 min.

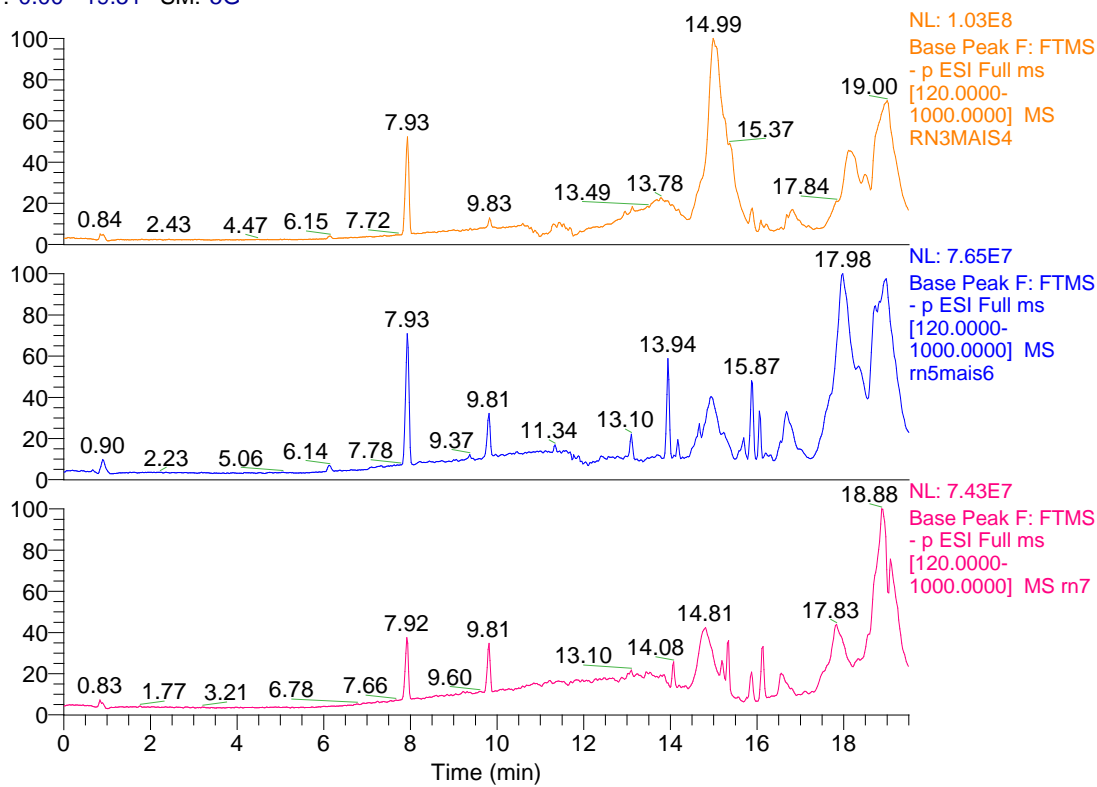

Fonte: Elaborado pelo autor.

As redes moleculares para as subfrações supracitadas foram criadas da mesma maneira que as anteriores. A rede molecular do modo positivo encontra-se na Figura 43. Foram criados sete agrupamentos principais (XXIII – XXIX), os quais tiveram mais de uma anotação molecular com a biblioteca do GNPS. Os agrupamentos delimitados pelo retângulo, tiveram ao menos uma anotação em cada. Há ainda a presença de alguns nodos singulares que obtiveram anotação molecular, porém não apresentaram similaridade com os demais através do valor de cosseno definido e, portanto, não foram agrupados entre si.

Figura 43 - Rede molecular das subfrações CaxFls-DCM R3+4, R5+6 e R7, modo positivo, editada no *software* Cytoscape 3.10.

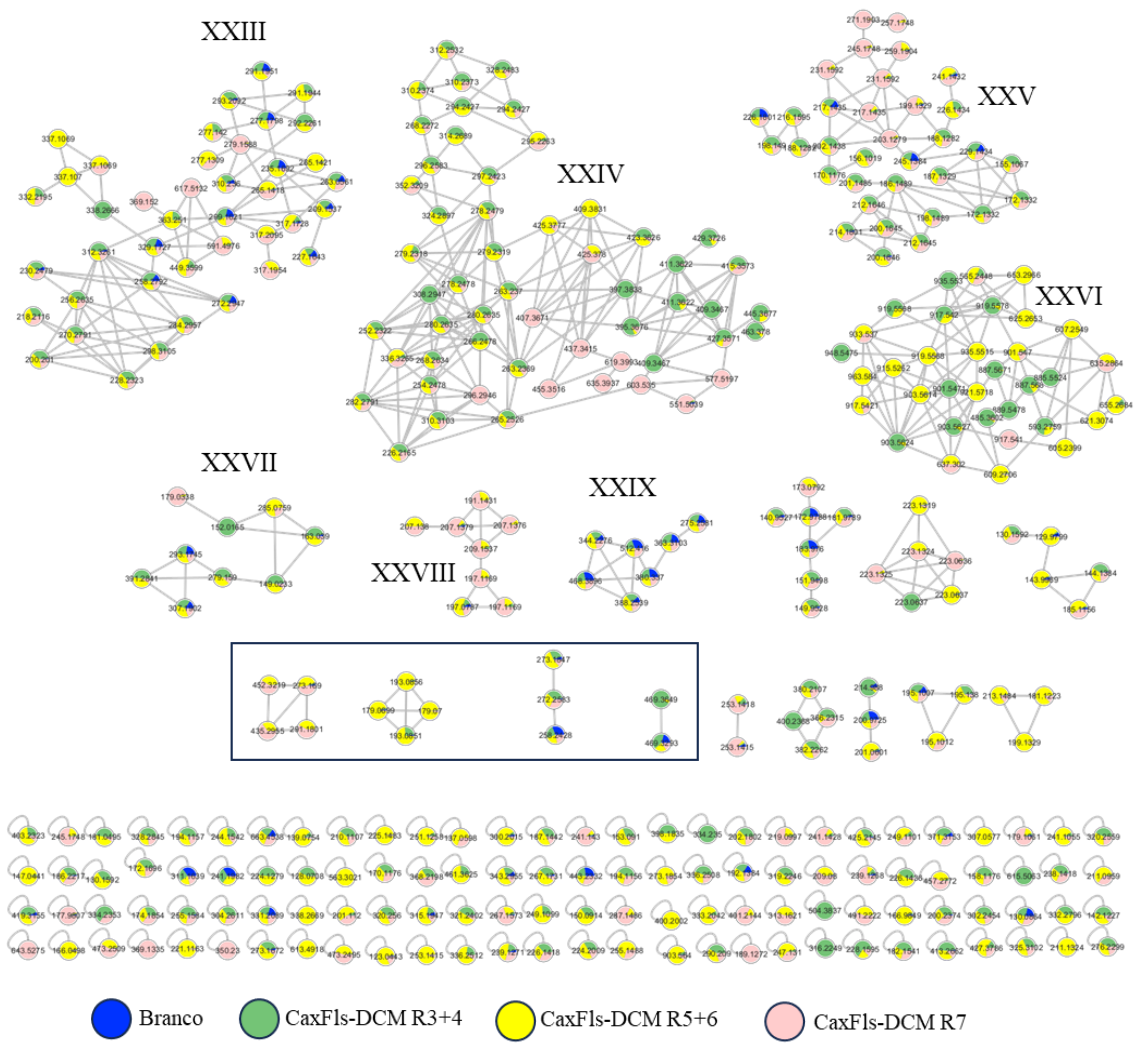

Fonte: Elaborado pelo autor.

O agrupamento XXIII (Figura 44) anotou substâncias da classe das amidas graxas, fenilpropanoides, flavonoides e benzofuranos. Algumas dessas substâncias, como as amidas graxas (**87** e **90**), estavam presentes nas três subfrações analisadas. O flavonoide **60**, contudo, ficou restrito à subfração CaxFls-DCM R5+6.

Figura 44 - Ampliação do agrupamento XXIII da rede molecular modo positivo; subfrações CaxFls-DCM R3+4, 5+6 e R7.

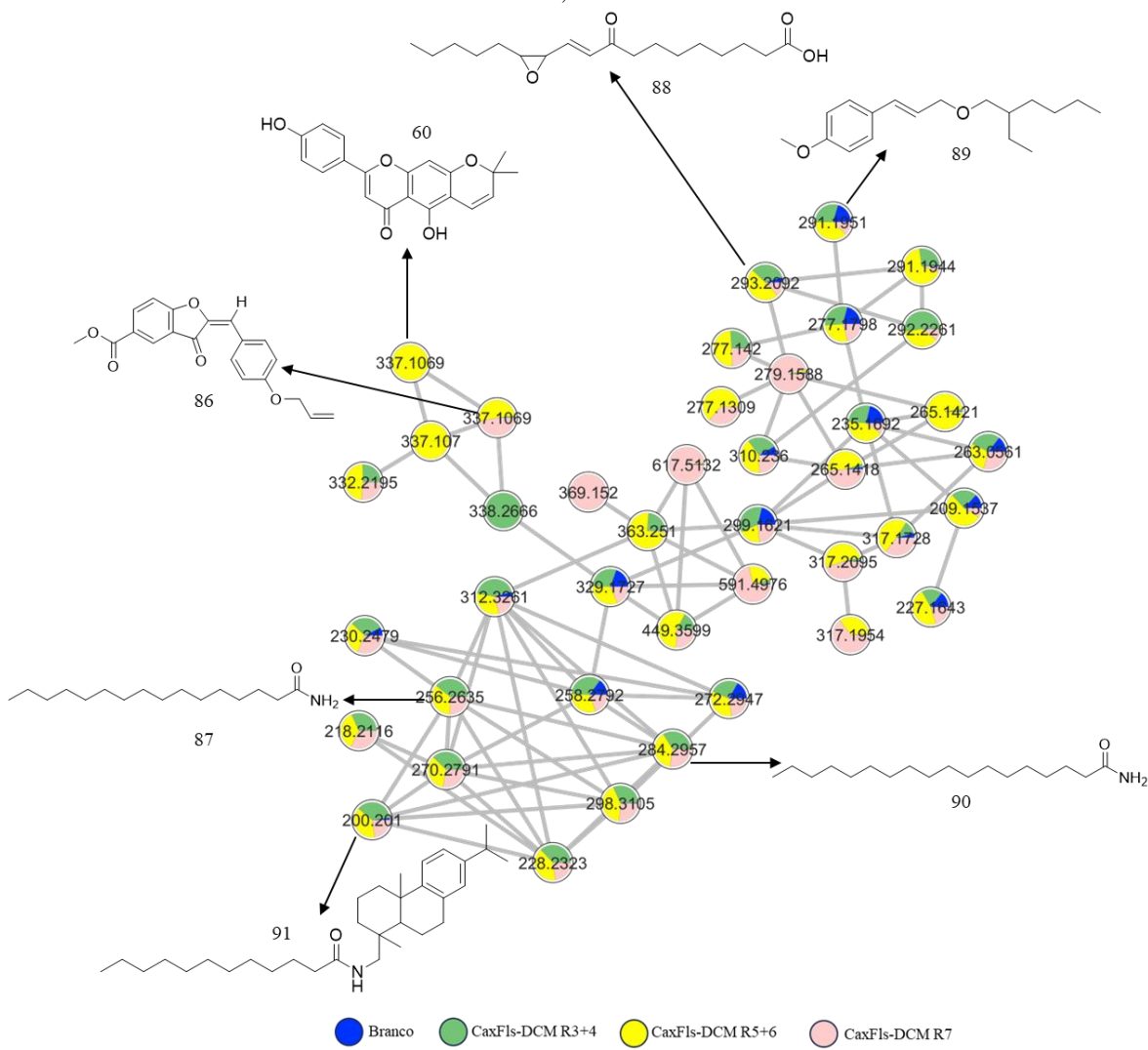

Fonte: Elaborado pelo autor.

Tabela 11 - Substâncias anotadas no agrupamento XXIII.

| Entrada | Nome | *m/z* |
|---|---|---|
| 86 | Metil-3-oxo-2-[((4-prop-2-eniloxifenil)metileno]benzo[b]furano-5-carboxilato | 337,1069 |
| 60 | Carpacromeno | 337,1069 |
| 87 | Palmitamídeo | 256,2635 |
| 88 | trans-EKODE-(E) | 293,2092 |
| 89 | Octinoxato | 291,1951 |

| Entrada | Nome | *m/z* |
|---------|------|-------|
| 90 | Octadecanamida | 284,2957 |
| 91 | Ác. laurico leelamida | 200,201 |

No agrupamento XXIV (Figura 45) foram anotados alguns triterpenos já identificados nas frações Hex e DCM anteriores, como **14**, **50** e **51** (Tabela 12).

Figura 45 - Ampliação do agrupamento XXIV da rede molecular modo positivo; subfrações CaxFls-DCM R3+4, 5+6 e R7.

Fonte: Elaborado pelo autor.

Tabela 12 - Substâncias anotadas no agrupamento XXIV.

| Entrada | Nome | *m/z* |
|---------|------|-------|
| 92 | Ác. oleico | 265,2526 |
| 93 | Ác. sumaresinólico | 455,3516 |
| 46 | Ác. betulônico | 437,3415 |
| 94 | Ác. linoleico conjugado | 263,2369 |
| 50 | 17(21)-hopen-6-ona | 425,378 |
| 51 | Lupeol | 409,3834 |
| 95 | Betulina | 425,3777 |
| 96 | Ác. 8-hidroxi-8-(3-octiloxiran-2-il)octanoico | 297,2423 |
| 97 | 1-palmitoil-2-oleoil-sn-glicerol | 557,5197 |
| 14 | Glochidona | 423,3626 |

Nos agrupamentos XXV e XXVI (Figura 46) foram anotados o dimetilsebacato (**98**), a vitamina K (**99**), o feoforbídeo (**100**) e um terpeno glicosilado (**101**).

Figura 46 - Ampliação dos agrupamentos XXV e XXVI da rede molecular modo positivo; subfrações CaxFls-DCM R3+4, 5+6 e R7.

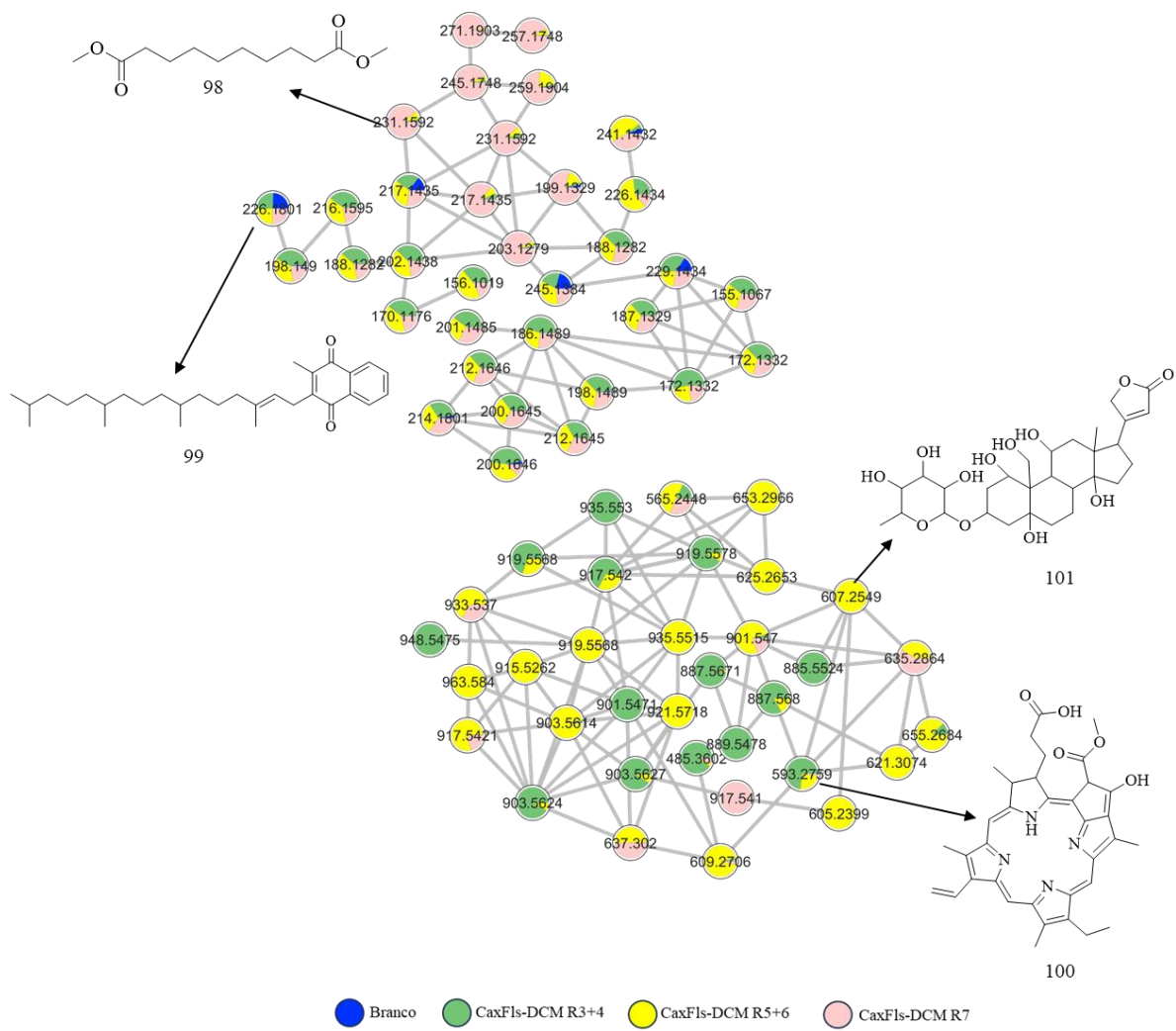

Fonte: Elaborado pelo autor.

Tabela 13 - Substâncias anotadas nos agrupamentos XXV e XXVI.

| Entrada | Nome | m/z |
|---|---|---|
| 98 | Dimetilsebacato | 231,1592 |
| 99 | Vitamina K | 226,1801 |
| 100 | Feoforbídeo A | 593,2759 |
| 101 | (3-[(6-desoxi-β-D-talopiranosil)oxi]-1,5,11,14,19-pentahidroxicard-20(22)-enolideo | 607,2549 |

Nos agrupamentos XXVII e XXVIII (Figura 47) foram anotados **65**, **38** e o triterpeno **44**, presentes nas redes anteriores; além do loliolídeo (**102**) e coniferaldeído (**104**).

Figura 47 - Ampliação dos agrupamentos XXVII, XXVIII e nodos delimitados pelo retângulo da rede molecular modo positivo; subfrações CaxFls-DCM R3+4, 5+6 e R7.

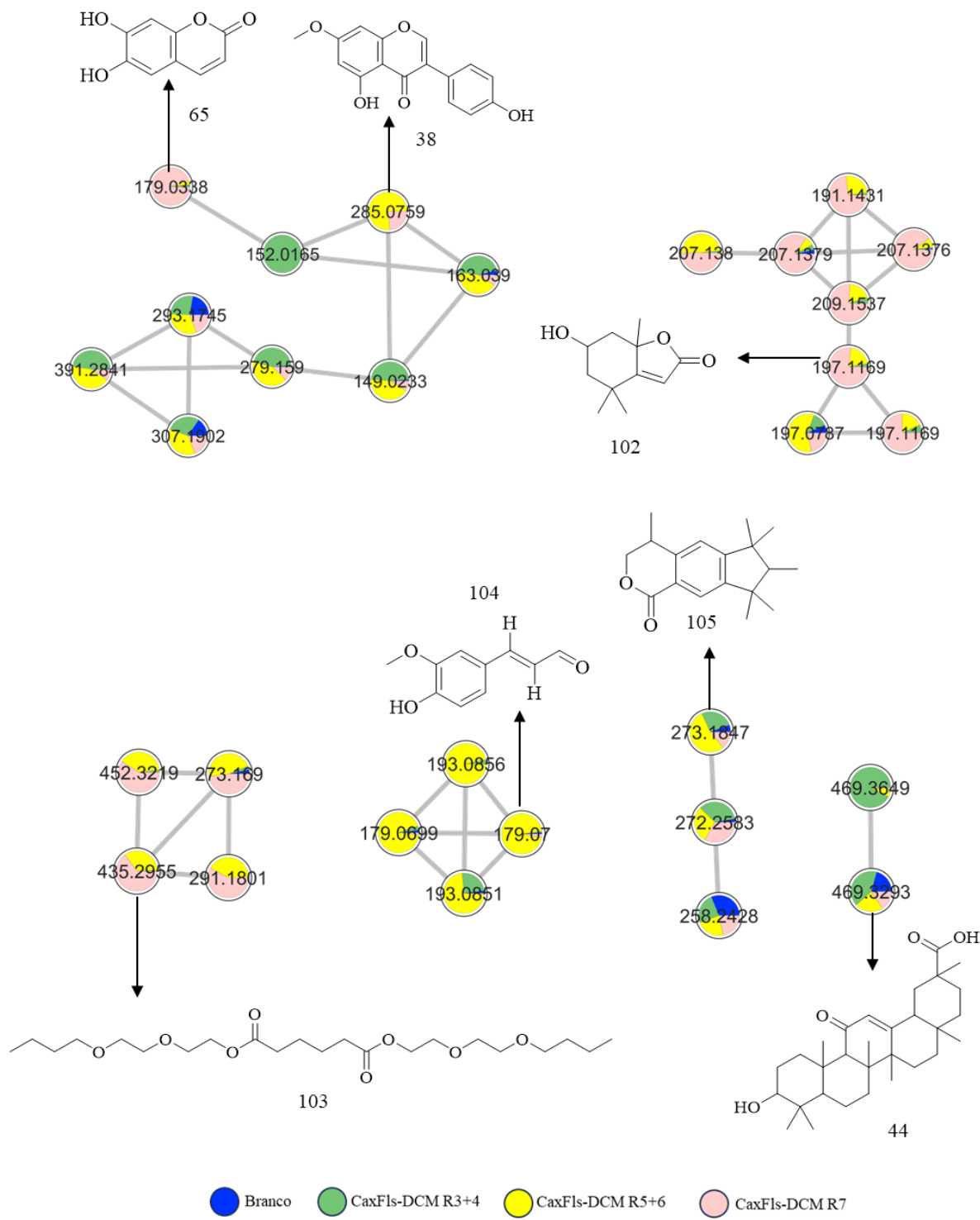

Fonte: Elaborado pelo autor.

Tabela 14 - Substâncias anotadas XXVII, XXVIII e nodos delimitados pelo retângulo

| Entrada | Nome | m/z |
|---------|------|-----|
| 65 | Esculetina | 179,0338 |
| 38 | Prunetina | 285,0759 |
| 102 | Loliolídeo | 197,1169 |
| 103 | Bis[2-(2-butoxietoxi)etil]hexadioato | 435,2955 |
| 104 | Coniferaldeído | 179,07 |
| 105 | Galaxolidona | 273,1847 |
| 44 | Enoxolona | 469,3293 |

A rede molecular do modo negativo (Figura 48), por sua vez, formou um único agrupamento, o qual foram anotadas cinco substâncias, dentre elas, algumas já anotadas no agrupamento XVII do modo positivo, como o alcaloide cinchonina, como mostrado na Figura 49.

Figura 48 - Rede molecular modo negativo das subfrações CaxFls-DCM R3+4, R5+6 e R7, editada no *software* Cytoscape 3.10.

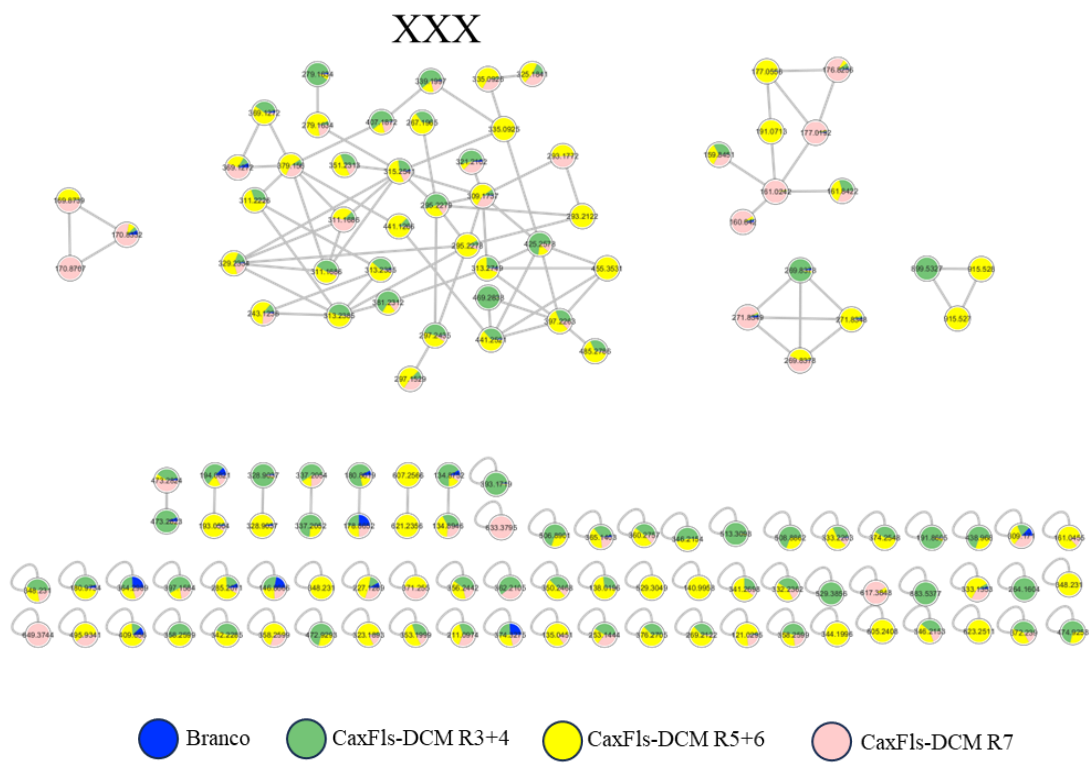

Fonte: Elaborado pelo autor.

Figura 49 - Ampliação do agrupamento XXX da rede molecular modo negativo; subfrações CaxFls-DCM R3+4, R5+6 e R7.

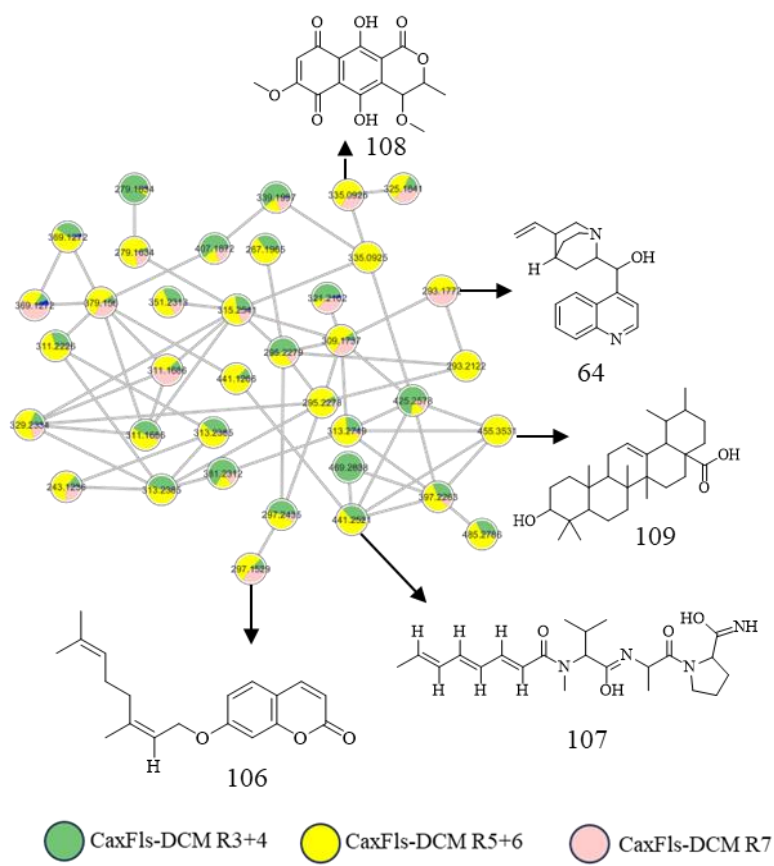

Fonte: Elaborado pelo autor.

Tabela 15 - Substâncias anotadas no agrupamento XXX.

| Entrada | Nome | m/z |
|---------|------|-----|
| 106 | Aurapteno | 297,1529 |
| 107 | N-metil-N-[(2E,4E,6E)-2,4,6-octatrienoil]valilalanilprolinamida | 441,2521 |
| 64 | Cinchonina | 293,1772 |
| 108 | Metoxihaementosina | 335,0925 |
| 109 | Ác. ursólico | 455,3531 |

Algumas das substâncias anotadas para as subfrações da fração em DCM das folhas de *F. maxima* já foram identificadas no extrato bruto e nas frações CaxFls-Hex e CaxFls-DCM, como os triterpenos **50** e **51**, que possuem o núcleo do lupano (**13**) e o triterpeno **14**. Uma das substâncias majoritárias que foi anotada para as subfrações em DCM foi o loliolídeo (**102**), uma monoterpenolactona, com TR 7,93 min. A comparação do espectro de EM-EM correspondente a **102** com a biblioteca do GNPS mostrou uma similaridade com valor de cosseno 0,98, com picos e intensidades relativas semelhantes, indicando a presença da substância (Figura 50).

Figura 50 – Espectro de EM-EM (IES+) experimental (superior) de **102** comparado ao da biblioteca do GNPS (inferior).

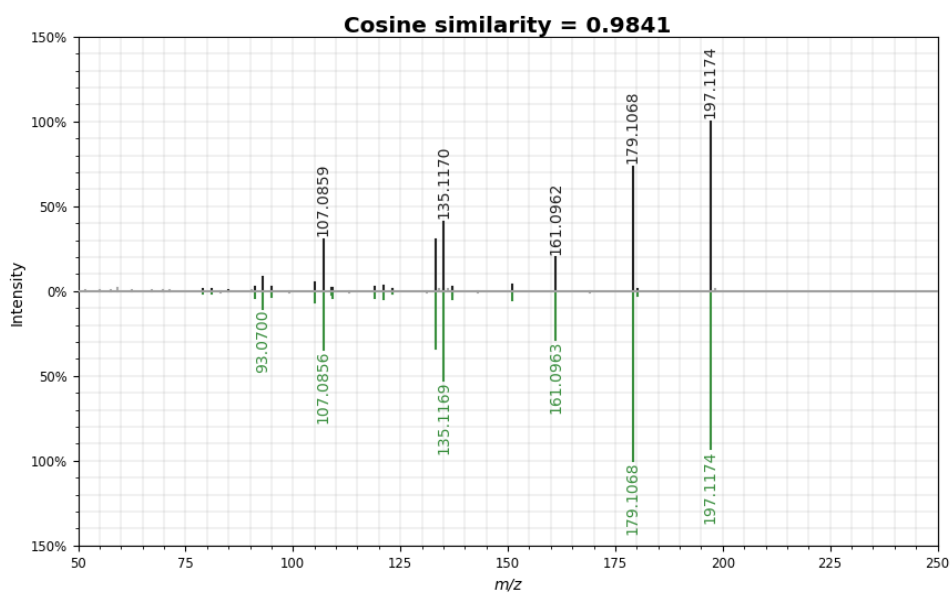

A proposta de fragmentação de **102** abaixo (Esquema 7) justifica os dois principais picos mais intensos por uma desidratação, gerando o íon fragmento de $m/z$ 179, e posterior perda de $CO_2$ para gerar o íon de $m/z$ 135.

Esquema 7 - Proposta de fragmentação para **102**.

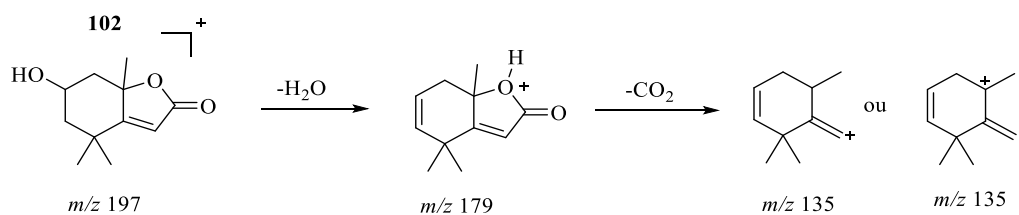

Fonte: Elaborado pelo autor.

**102** já foi encontrado nas folhas da espécie *Morus alba* L., também da família Moraceae. Na medicina popular, as folhas de *M. alba,* cujo componente majoritário é **102**, são utilizadas para baixar o nível de açúcar no sangue. Embora os estudos não tenham comprovado atividade direta de **102** no consumo de glicose *in vitro*, a presença desse componente nas folhas de *M. alba* pode indicar atividade hipoglicemiante por outras vias ainda não completamente estudadas (HUNYADI *et al.*, 2012). Mais recentemente, outro trabalho mostrou que **102** isolado da alga verde *Codium tomentosum* possui atividades neuroprotetoras e anti-inflamatórias, mostrando ser um potencial agente terapêutico neuroprotetor (SILVA *et al.*, 2021).

## 4.6    Ensaio de atividade antinociceptiva via teste de formalina

Para investigação e validação do uso popular de *F. maxima* como agente anti-inflamatório, foi realizado o teste de atividade antinociceptiva das folhas e cascas do caule via teste da formalina.

O teste da formalina é um método de avaliação comportamental utilizado para medir a efetividade de agentes antinociceptivos, com a vantagem, sobre outros métodos semelhantes, de medir dois tipos diferentes de dor em duas fases (fase neurogênica e fase inflamatória) ao longo do tempo.

O teste emprega um estímulo de nocicepção química, através da injeção i.p. de formalina 2,5% (preparada a partir de solução de 37% (p/p) de formaldeído em dimetilsulfóxido (DMSO), que elicia uma resposta espontânea indicativa de dor (HUNSKAAR; HOLE, 1987; TJOLSEN *et al.*, 1992). Enquanto ocorre inflamação no sítio de injeção da formalina, é possível entender o papel da inflamação nas respostas espontâneas nas duas fases.

A primeira fase (fase neurogênica), que se estende pelos primeiros cinco minutos, é caracterizada pela ação direta da formalina sobre os nociceptores, gerando dor neurogênica ou aguda, estimulando diretamente os nociceptores das fibras C. Nessa fase, são liberados aminoácidos excitatórios e substância P (neuropeptídeo). Substâncias como opioides, que atuam no sistema nervoso central, são os principais agentes a atuarem nessa fase (TJOLSEN *et al.*, 1992)

A segunda fase (fase inflamatória), que ocorre entre quinze e trinta minutos após a injeção de formalina, está relacionada com a liberação de vários mediadores pró-inflamatórios, como bradicinina, prostaglandinas, histamina, entre outros. Os agentes que atuam nessa fase são anti-inflamatórios não-esteroidais (AINEs), como o ácido acetilsalicílico (AAS), por exemplo.

Os resultados envolvendo as respostas neurogênica e inflamatória resultantes da ação da formalina foram observados logo após a administração i.p. do veículo (DMSO), AAS (150mg/kg i.p.), morfina (1 mg/kg i.p.) e das amostras CaxCsc-EB e CaxFls-EB, nas concentrações de 3, 10 e 30 mg/kg i.p..

O tratamento com a amostra CaxCsc-EB produziu os seguintes resultados:

Na primeira fase, houve decréscimo da reatividade à formalina de 124,3 ± 25,98 s para 84,7 ± 7,53 s, 49,8 ± 4,7 s e 9,9 ± 4,4, referentes ao veículo e os controles AAS e morfina, respectivamente. A amostra CaxCsc-EB reduziu para 57,3 ± 8,8 s; 64,9 ± 6,9 s; e 40,8 ± 7,4 s, nas concentrações de 3, 10 e 30 mg/kg, respectivamente. Na segunda fase, a redução foi de 124,3 ±25,98 s para 49,8 ± 4,7 s e 9,9 ± 4,4 s, referente aos controles AAS e morfina, respectivamente. A amostra CaxCsc-EB alterou para 136, ± 6,5 s; 86,8 ± 16,4 s; e 32,5 ± 8,5 s, nas concentrações de 3, 10 e 30 mg/kg, respectivamente (Gráfico 1).

Gráfico 1 - Tempo de reatividade à formalina após administração i.p. do veículo (DMSO), dos controles (morfina e AAS) e do extrato bruto da casca do caule de *F. maxima* (CaxCsc-EB) (3, 10 e 30 mg/kg).

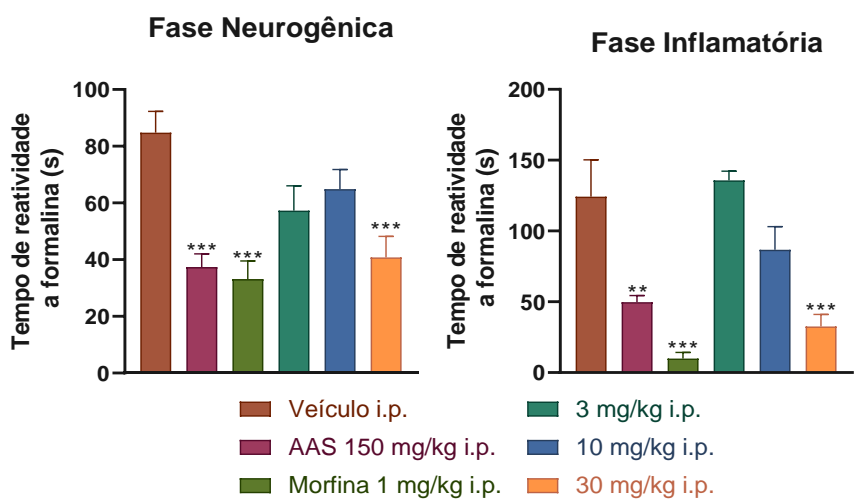

*Os dados são expressos em média ± SEM (n = 6 – 8 ). Valores significantes quando \*p < 0,05 e \*\*\* p < 0,001 comparados ao veículo. Para a análise estatística da comparação, foi utilizado o One-way ANOVA Dunnett's multiple comparisons.*

O tratamento com a amostra CaxFls-EB produziu os seguintes resultados (Gráfico 2):

Na primeira fase, houve decréscimo da reatividade à formalina de 124,3 ± 25,98 s para 87,4 ± 7,53 s, 37,4 ± 4,7 s e 33,1 ± 6,4 s para o veículo e controles AAS e morfina, respectivamente. A amostra CaxFls-EB variou para 47,5 ± 6,3 s, 43,5 ± 3,9 s, 53,6 ± 2,00 s, nas concentrações de 3, 10 e 30 mg/kg, respectivamente.

Na segunda fase, a redução foi de 124,3 ±25,98 s para 49,8 ± 4,7 s e 9,9 ± 4,4 s, para os controles AAS e morfina, respectivamente. A amostra CaxFls-EB alterou para 72,1 ± 17,3 s; 59,8 ± 12,8 s; e 61,8 ± 18,4 s, nas concentrações de 3, 10 e 30 mg/kg, respectivamente.

Gráfico 2 - Tempo de reatividade à formalina após administração i.p. do veículo (DMSO), dos controles (morfina e AAS) e do extrato bruto das folhas de *F. maxima* (CaxFls-EB) (3, 10 e 30 mg/kg).

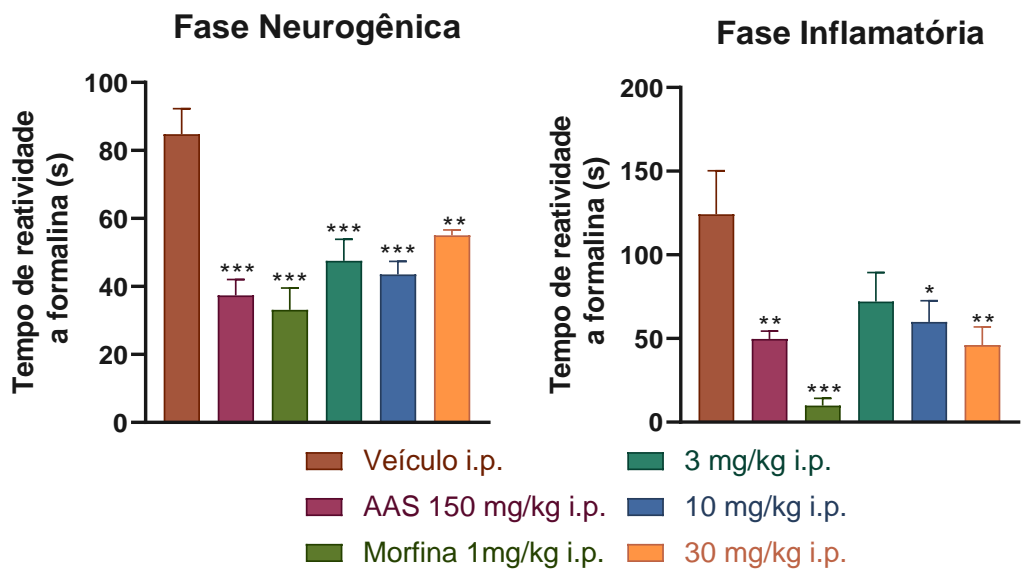

*Os dados são expressos em média ± SEM (n = 8 ). Valores significantes quando < 0,05 e \*\*\* p < 0,001 comparados ao veículo. Para a análise estatística da comparação, foi utilizado o One-way ANOVA Dunnett's multiple comparisons.*

Assim, foi observado o efeito de dose-resposta para ambos os extratos brutos de *F. maxima*, e, principalmente, para a amostra CaxCsc-EB na fase inflamatória, na concentração de 30 mg/kg. Sugerindo, desse modo, inibição dos mediadores químicos de processo inflamatório. Na fase neurogênica, os extratos também reduziram, ainda que com ligeira variação, a nocicepção causada pela injeção de formalina.

As frações em Hex, DCM e AcOEt das folhas de *F. maxima* também foram testadas quanto a sua atividade antinociceptiva, na concentração de 30 mg/kg i.p. em comparação com os extratos. Os resultados obtidos e representados no Gráfico 3 foram os seguintes:

O tempo de resposta à formalina na primeira fase foi reduzido para 37,4 ± 4,7 s e 33,1 ± 6,4 s, referentes aos controles AAS e morfina, respectivamente. As amostras CaxCsc-EB e CaxFls-EB, reduziram para 40,8 ± 7,4 s e 53,6 ± 2,0 s, respectivamente. Já as frações CaxFls-Hex, CaxFls-DCM e CaxFls-AcOEt, reduziram para 49,7 ± 4,2 s, 46,8 ± 8,1 s e 52,4 ± 5,3 s, respectivamente.

Na segunda fase, a resposta à formalina foi reduzida de 124,3 ± 25,98 s para 49,8 ± 4,7 s referente ao controle AAS, e 9,9 ± 4,4 s referente ao controle morfina. A amostra CaxCsc-EB reduziu para 32,5 ± 8,5 s, e a amostra CaxFls-EB reduziu para 61,8 ± 18,4 s. Já as frações

CaxFls-Hex, CaxFls-DCM e CaxFls-AcOEt, reduziram para 54,4 ± 7,4 s, 100,2 ± 17,9 s e 118,4 ± 35,3 s, respectivamente.

Gráfico 3 - Tempo de reatividade à formalina após administração i.p. do veículo (DMSO), dos controles (morfina e AAS), do extrato bruto das folhas (CaxFls-EB), do extrato bruto das cascas do caule (CaxCsc-EB) e das frações em Hex, DCM e AcOEt de *F. maxima* (30 mg/kg).

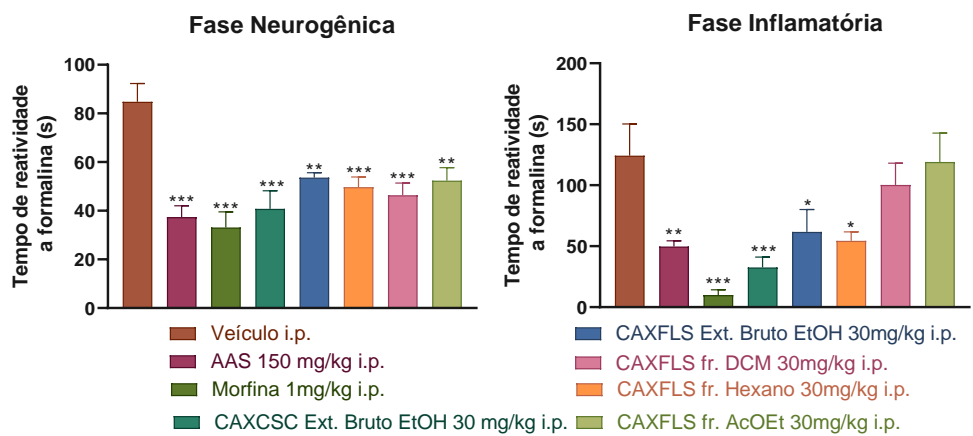

*Os dados são expressos em média ± SEM (n = 6 – 8). Valores significantes quando < 0,05 e *** p < 0,001 comparados ao veículo. Para a análise estatística da comparação, foi utilizado o One-way ANOVA Dunnett's multiple comparisons.*

Desse modo, foi possível observar que o extrato orgânico das cascas do caule de *F. maxima* apresentou melhor atividade na fase inflamatória, na concentração de 30 mg/kg em comparação com o extrato das folhas. Embora a fração em hexano das folhas também tenha apresentado boa atividade em comparação com o extrato bruto, mas ainda inferior àquela observada para as cascas.

A literatura relata que triterpenos pentacíclicos como o lupeol, presente no extrato de *F. carica*, por exemplo, betulina, ácido betulínico e derivados, possuem atividades anti-inflamatórias comprovadas em diversos ensaios com modelos murinos (PATOCKA, 2003; SALEEM, 2009; WOODE *et al.*, 2009; HOWLADER *et al.*, 2017). Essas substâncias também identificadas putativamente no extrato das folhas de *F. maxima* pode justificar a atividade antinociceptiva observada.

Para investigar os mecanismos de ação envolvidos nas atividades antinociceptivas observadas, a amostra CaxCsc-EB foi testada na concentração de 30 mg/kg i.p. e após o pré-tratamento junto com um antagonista do receptor muscarínico (atropina 1 mg/kg i.p.). O mecanismo de ação de CaxCsc-EB não mostrou reversão significativa na fase neurogênica. O tempo de reatividade à formalina reduziu para 40,8 ± 7,4 s referente a CaxCsc-EB, e 47,5 ± 6,5 s referente a CaxFls-EB + atropina. Na fase inflamatória, a reversão observada foi mais significativa. O tempo de reatividade foi reduzido para 32,5 ± 8,5 s referente à amostra Cax Csc-EB e 47,5 ± 6,5 s referente a CaxFls-EB + atropina, sugerindo a possibilidade de ação do extrato das cascas nas vias muscarínicas (Gráfico 4).

Gráfico 4 - Tempo de reatividade à formalina após administração i.p. do veículo (DMSO), tratamento com o extrato bruto etanólico da casca do caule de *F. maxima* (30 mg/kg i.p.) e o pré-tratamento com atropina (1 mg/kg i.p.).

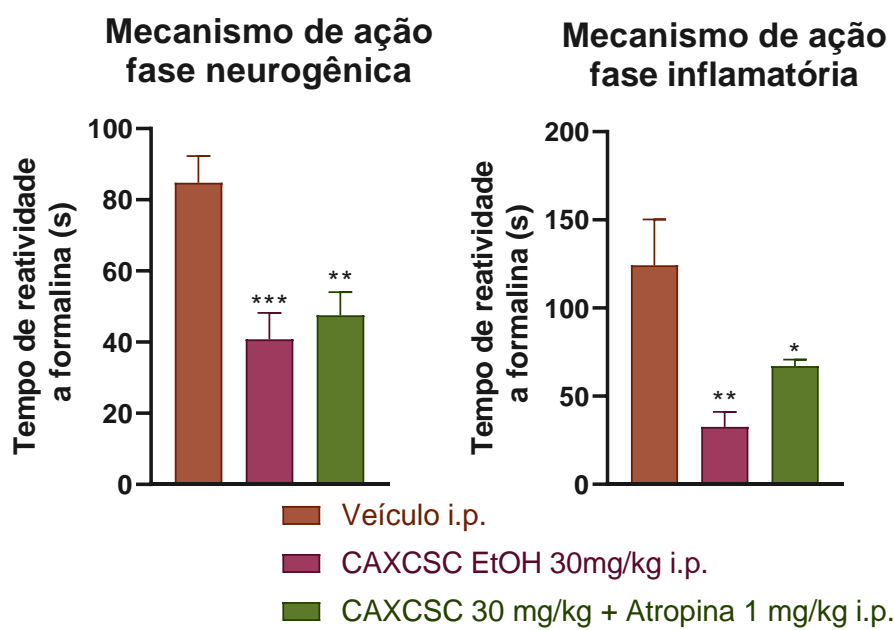

*Os dados são expressos em média ± SEM (n = 6 – 8). Valores significantes quando < 0,05 e *** p < 0,001 comparados ao veículo. Para a análise estatística da comparação, foi utilizado o One-way ANOVA Dunnett's multiple comparisons.*

A amostra CaxFls-EB também foi testada na concentração de 30 mg/kg i.p. após o pré-tratamento com um antagonista do receptor opioide não-seletivo (naloxona 1 mg/kg i.p.). Os resultados estão representados no Gráfico 5. A reversão foi observada somente na fase inflamatória, reduzindo o tempo de reatividade à formalina para 61,7 ± 18,4 s referente a CaxFls-EB, e 101,3 ± 5,4 s referente a CaxFls-EB + naloxona.

Gráfico 5 - Tempo de reatividade à formalina após administração i.p. do veículo (DMSO), do tratamento com o extrato bruto etanólico das folhas de *Ficus maxima* (30 mg/kg i.p.) e o pré-tratamento com naloxona (1 mg/kg i.p.).

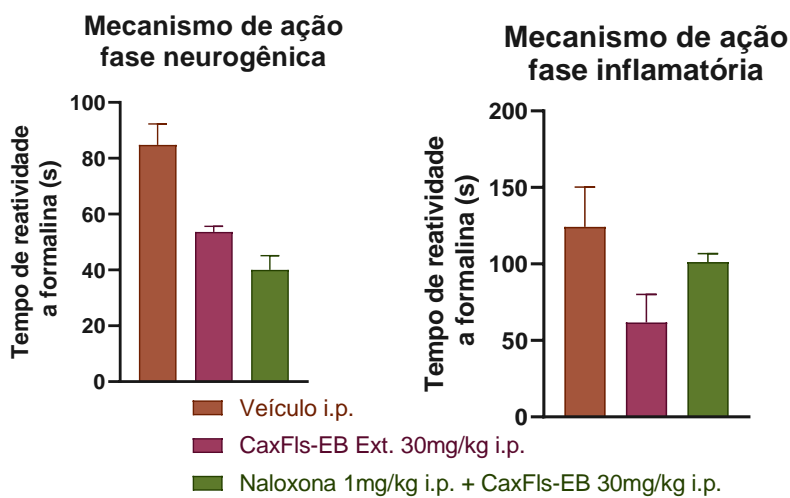

*Os dados são expressos em média ± SEM (n = 6 – 8 ). Para a análise estatística da comparação, foi utilizado o One-way ANOVA Dunnett's multiple comparisons.*

## 4.7 Ensaio de atividade antinociceptiva em placa quente

A fim de se avaliar a eficácia da atividade antinociceptiva observada para o extrato bruto da casca de *F. maxima* foi realizado o ensaio em placa quente. O teste se baseia no princípio em que os roedores quando são colocados sob uma superfície quente (52 ± 0,5 °C) irão inicialmente demonstrar os efeitos aversivos ao estímulo térmico, indicando resposta nociceptiva, através de lambidas das patas e/ou saltos para escapar do ambiente (Eddy e Leimback, 1953). A latência observada para os primeiros sinais de lambida de pata ou salto é então cronometrada. Substâncias que atuam inibindo a nocicepção aumentam a latência para a lambida/salto, caracterizando um efeito analgésico mediado por mecanismos do SNC.

A amostra CaxCsc-EB foi então injetada intraperitonealmente nas concentrações de 3, 10 e 30 mg/kg, e os efeitos foram observados na janela de 5 a 150 min após a administração das doses. A maior porcentagem de atividade antinociceptiva (%AA) para CaxCsc-EB na concentração de 3 mg/kg foi de 37,7 ± 10,7 % após 40 min, enquanto para o grupo veículo foi de 3,13 ±1,7%. Para CaxCsc-EB na concentração de 10 mg/kg, a maior %AA foi 39,3 ± 9,5% após 60 min, enquanto para o controle foi 0,9 ± 0,9%, no mesmo tempo. Na concentração de 30 mg/kg, a maior %AA para CaxCsc-EB foi de 62,6 ± 9,2% no tempo de 90 após administração, comparado ao veículo (%AA=0,7 ± 0,7%) no mesmo tempo.

Os resultados demonstraram que CaxCsc-EB apresentou atividade antinociceptiva moderada e mais constante nas concentrações de 3 e 10 mg/kg (Gráfico 6). O aumento da dose para 30 mg/kg mostrou atividade antinociceptiva mais significante quando comparada ao veículo.

Desse modo, os resultados obtidos pelo teste da placa quente indicam substâncias presentes no extrato que atuam nas vias dos opioides.

Gráfico 6 - Porcentagem de atividade antinociceptiva avaliada pelo ensaio da placa quente após administração do veículo, da morfina (10 mg/kg i.p.) e da amostra CaxCsc-EB (3, 10 e 30 mg/kg i.p).

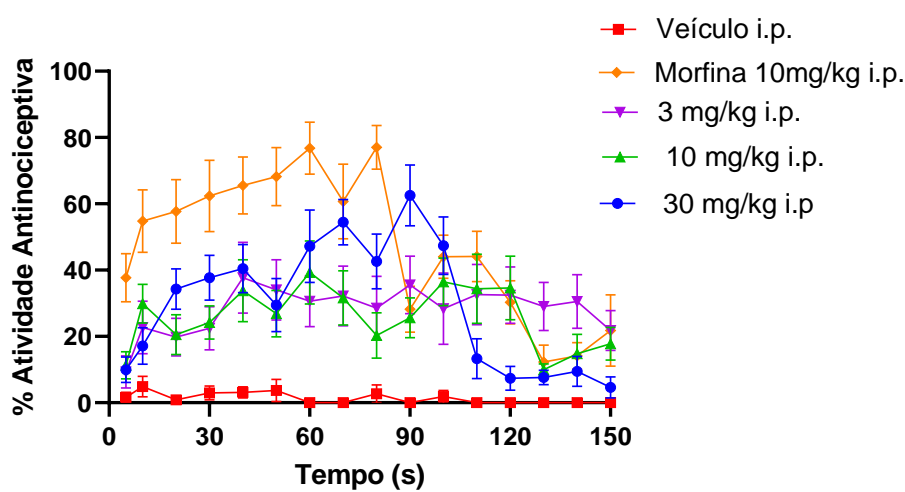

## 4.8    Ensaio de atividade inibitória de enzimas do sistema citocromo P450

Enzimas do sistema citocromo P450 (CYP450) são o grupo mais importante de enzimas metabolizadoras de fármacos. CYP450 é uma superfamília de proteínas monooxigenases contendo o grupo heme que catalisam diversas reações oxidativas, que, quando se ligam ao monóxido de carbono, atingem absorbância máxima em 450 nm (COOK *et al.*, 2016).

Muitos medicamentos são primeiramente metabolizados através de mecanismos de oxidação por diversas enzimas CYP450 nas chamadas Reações de Fase I (LYNCH; PRYCE, 2007). O processo de metabolismo é uma transformação irreversível de fármacos administrados em moléculas menores ou metabólitos, funcionando como um primeiro mecanismo de defesa contra substâncias danosas ou xenobióticos (SAHA, 2018). O principal objetivo do metabolismo é converter essas substâncias em compostos mais hidrofílicos ou mesmo inativos, a fim de se facilitar sua excreção. As enzimas CYP450 são responsáveis pela maioria dessas reações da fase I catalisadas no fígado humano (BLUMER, 2011).

Existem mais de 50 enzimas individuais da CYP450, contudo, há seis subfamílias mais importantes que podem metabolizar de 75-90% de todos os fármacos: CYP1A2, CYP2C9, CYP2D6, CYP2C19, CYP3A4 e CYP3A5 (LYNCH; PRYCE, 2007). A CYP1A2 faz parte da subfamília CYP1A que contém também a isoforma CYP1A1, dois genes funcionais altamente conservados entre espécies (MARTIGNONI; GROOTHUIS; KANTER, 2006).

Reações catalisadas pela CYP1A incluem hidroxilação e oxidação de compostos aromáticos, onde a CYP1A1 está principalmente envolvida no metabolismo de hidrocarbonetos aromáticos, enquanto a CYP1A2 está envolvida preferencialmente no metabolismo de aminas aromáticas e compostos heterocíclicos (LU *et al.*, 2020).

Assim substâncias indutoras de CYP podem aumentar sua atividade enzimática e acelerar o metabolismo de fármacos, levando a uma redução na eficácia desses. Por outro lado, os inibidores de CYP podem reduzir o metabolismo de fármacos e gerar aumento da eficácia ou mesmo toxicidade (SAHA, 2018).

A CYP1A é conhecida pelo metabolismo de algumas substâncias, especialmente os hormônios sexuais. Alterações nas atividades metabólicas das enzimas CYP1A podem geram um desbalanceamento na concentração de hormônios no corpo, o que pode aumentar a sensibilidade a determinadas doenças, sobretudo carcinogênicas. Por exemplo, estudos mostraram que a CYP1A e sua isoforma CYP1A2 aumentam a susceptibilidade de câncer de próstata (DING *et al.*, 2013), câncer cervical (KLEINE *et al.*, 2015), e perda recorrente de gravidez em humanos (LI *et al.*, 2017).

Diante do exposto, fica evidente a importância de estudos acerca de substâncias inibidoras de CYP1A. Diferentes espécies de *Ficus* estudadas na literatura como *F. religiosa* (MOUSA *et al.*, 1994)*, F. benghalensis* (SHARMA *et al.*, 2007)*, F. racemosa* (KHAN; SULTANA, 2005) têm diversas atividades biológicas, incluindo atividades anticancerígenas, associadas aos seus constituintes extraídos de todas as partes da planta. No entanto, nenhum desses estudos foram realizados em enzimas CYP450. Assim, foi conduzida uma avaliação de atividade inibitória de CYP1A, sobretudo da subfamília CYP1A1, a partir dos extratos brutos etanólicos das cascas do caule e folhas, bem como das frações de *F. maxima*.

O protocolo utilizado para o ensaio foi o definido por Santos *et al*. (2021), através do ensaio de etoxiresorufina-*O*-desetilasae (EROD), que descreve a taxa de desetilação do substrato 7-etoxiresorufina (7-ER) mediada pela enzima CYP1A1 para formação do produto resorufina, segundo Esquema 8:

Esquema 8 - Reação de *O*-desetilação da 7-etoxiresorufina (7-ER), mediada pela enzima CYP1A1, para produção da resorufina.

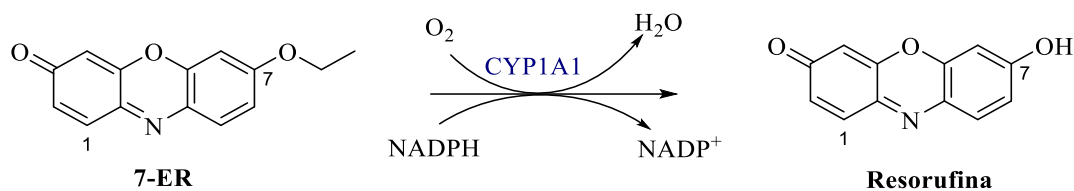

**7-ER**      **Resorufina**

Fonte: Elaborado pelo autor.

A concentração de resorufina produzida pode ser medida pela intensidade da fluorescência avaliada por um espectrofluorímetro (PARKINSON, 2001; NERURKAR *et al.*, 1993). Uma inibição da enzima produz quantidades mínimas do produto que é representada pela concentração de resorufina por mg de proteína por minuto (picomol/mg ptn/ min).

O controle positivo utilizado para o ensaio foi a α-naftoflavona (ANF), um potente inibidor já descrito para CYP1A1/2, enquanto a β-naftoflavona (BNF) foi utilizado como controle negativo (SANTOS *et al.*, 2021).

Assim, as amostras CaxFls-EB e suas subfrações, e CaxCsc-EB foram testados, na concentração de 100 μg/mL, para avalição da potencial atividade inibitória de CYP1A e os resultados obtidos foram representados no Gráfico 7.

Gráfico 7 - Atividade de CYP1A avaliada pela reação de etoxiresorufina-*O*-desetilase

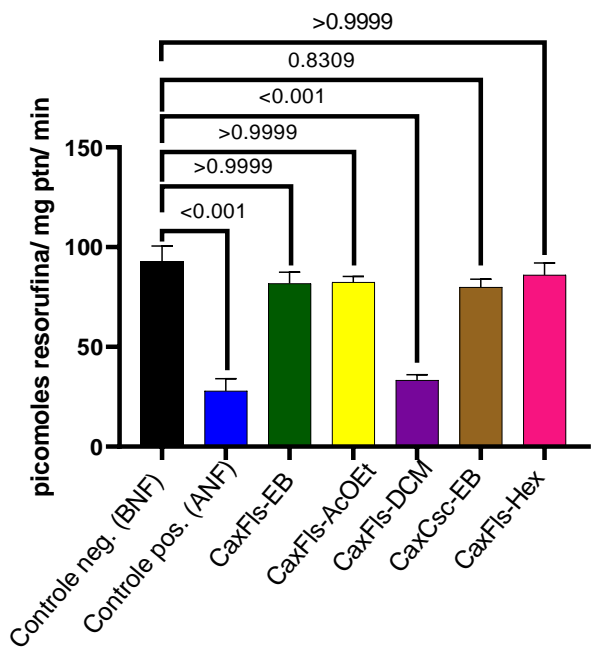

*Dados analisados pelo teste one-way ANOVA, pós teste Bonferroni. Valores significantes quando p<0.05, comparados ao controle.*

O controle positivo para CYP1A1 resultou uma média de 27,97 ± 6,06 picomoles/mg ptn/min. O controle negativo resultou uma média de 92,74 ± 7,80 picomoles/mg ptn/min. As frações CaxFls-EB, CaxFls-AcOEt, CaxCsc-EB e CaxFls-Hex resultaram em 81,83 ± 5,58, 82,45 ± 2,81, 79,90 ± 4,07 e 86,07 ± 5,98 picomoles/mg ptn/min, respectivamente. Esses valores estão acima da média registrada para o controle positivo e próximas o suficiente do controle negativo, mostrando que essas frações não apresentaram inibição de CYP1A. A fração CaxFls-DCM, contudo, resultou uma média de 33,36 ± 2,67 picomoles/mg ptn/ min, mostrando atividade inibitória para CYP1A tão potente quanto o inibidor ANF.

A presença de substâncias da classe dos flavonoides anotadas principalmente na fração CaxFls-DCM pode ser sugerida como uma justificativa para a atividade inibitória observada de *F. maxima*. Flavonoides são reportados como uma das principais classes a interagir com enzimas CYP1 (SOUSA *et al.*, 2013; MIRON *et al.*, 2017). O flavonoide quercetina, por exemplo, isolada do extrato etanólico de *Piper rivinoides* inibiu 80% da enzima num estudo mais recente (SANTOS *et al.*, 2021). Nesse mesmo trabalho, estudos de modelagens moleculares mostraram a importância da planaridade e possibilidade de ligações de hidrogênio das substâncias testadas para uma boa atividade de inibição. Num outro estudo envolvendo a interação de flavonoides com a expressão de CYP1A, foi demonstrado que a luteolina (**60**), flavonoide também encontrado em CaxFls-DCM, agiu como inibidor de CYP1A. O mesmo trabalho revelou que algumas combinações de inibidores de CYP1A com indutores como as flavonas, acabam interagindo de forma sinergística aumentando a indução de CYP1A após um período de até 24 h, o que pode explicar o efeito observado para o extrato bruto de *F. maxima*, onde foram anotadas algumas flavonas (CHATUPHONPRASERT *et al.*, 2010).

Os resultados observados estão de acordo com aqueles reportados na literatura para os inibidores de CYP1A. Contudo, como a atividade inibitória de CYP1A1 aqui relatada foi observada somente para a fração em DCM, o que dificilmente pode ser recriado no uso popular da espécie *F. maxima*, são necessários mais estudos envolvendo a quantificação, mecanismos de ação e métodos de extração dos flavonoides presentes para justificar o uso da espécie como agente anticancerígeno.

## 5 CONCLUSÕES

Ao menos 45 substâncias foram identificadas putativamente no extrato bruto etanólico das folhas de *F. maxima* por CLUE-EM-EM com auxílio da base de dados do GNPS e da FBMN. A análise fitoquímica revelou a presença de triterpenos pentacíclicos, ácidos graxos poliinsaturados e seus ésteres, aminoácidos e derivados de clorofila como constituintes majoritários nas folhas. Há presença de flavonoides, flavonas (na forma de agliconas e nas formas glicosiladas), alcaloides quinolínicos, lignanas, cumarinas e ácidos fenólicos em menor quantidade, e alguns traços de açúcares livres; esses últimos sobretudo na fração em AcOEt. As frações em Hex e DCM detêm boa parte dos constituintes de baixa polaridade.

As análises por CLUE-EM-EM das cascas não foram finalizadas a tempo para discussão dos seus constituintes.

Os ensaios *in vitro* de atividades inibitória enzimática dos extratos brutos e frações revelou que a fração em DCM das folhas de *F. maxima* é um potente inibidor de enzimas CYP1A1, justificado pela presença de flavonoides e cumarinas nessa fração.

Os extratos brutos das folhas e cascas do caule de *F. maxima* apresentaram atividade antinociceptiva *in vivo* com efeito dose-resposta na concentração de 30 mg/kg i.p. na fase inflamatória do teste da formalina. A presença de triterpenos com atividade anti-inflamatória, como **51**, identificados putativamente nas folhas sugere relação direta com a atividade observada para esse extrato. O mecanismo de ação envolvido na atividade observada para o extrato das cascas, sugere substâncias atuantes nas vias muscarínicas.

O extrato bruto das cascas do caule apresentou potente atividade antinociceptiva, de 62,6 ± 9,2% na concentração de 30 mg/kg i.p., *in* vivo, via teste de placa quente. O mecanismo de ação estudado sugere a presença de substâncias no extrato das cascas que são atuantes na via dos opióides.

O estudo fitoquímico e de atividades biológicas de *F. maxima* aqui realizado mostrou a eficácia do uso popular da espécie para tratamentos de inflamações e dores. Contudo, mais estudos acerca de outras atividades biológicas reportadas e quantificação de metabólitos devem ser realizados para garantir maior segurança no uso da espécie, visto que há relatos de toxicidade associados à presença dos alcaloides. Assim, a espécie *F. maxima* representa uma importante e potencial fonte terapêutica alternativa na medicina popular.

APÊNDICE A – Cromatograma de intensidade do pico base do extrato etanólico das folhas de *F. maxima*

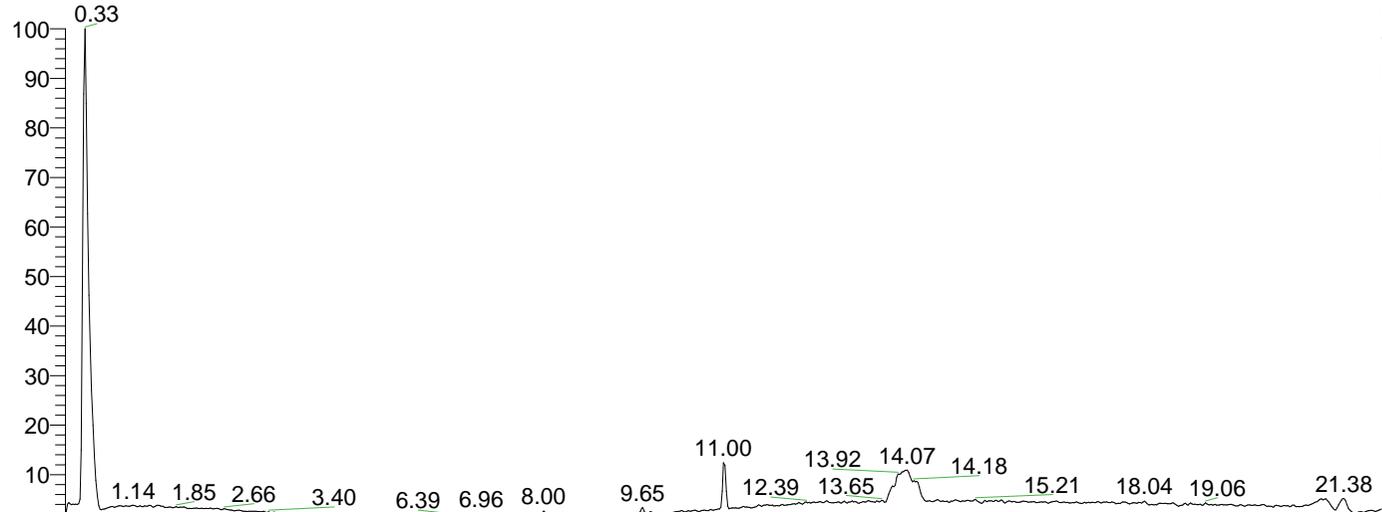

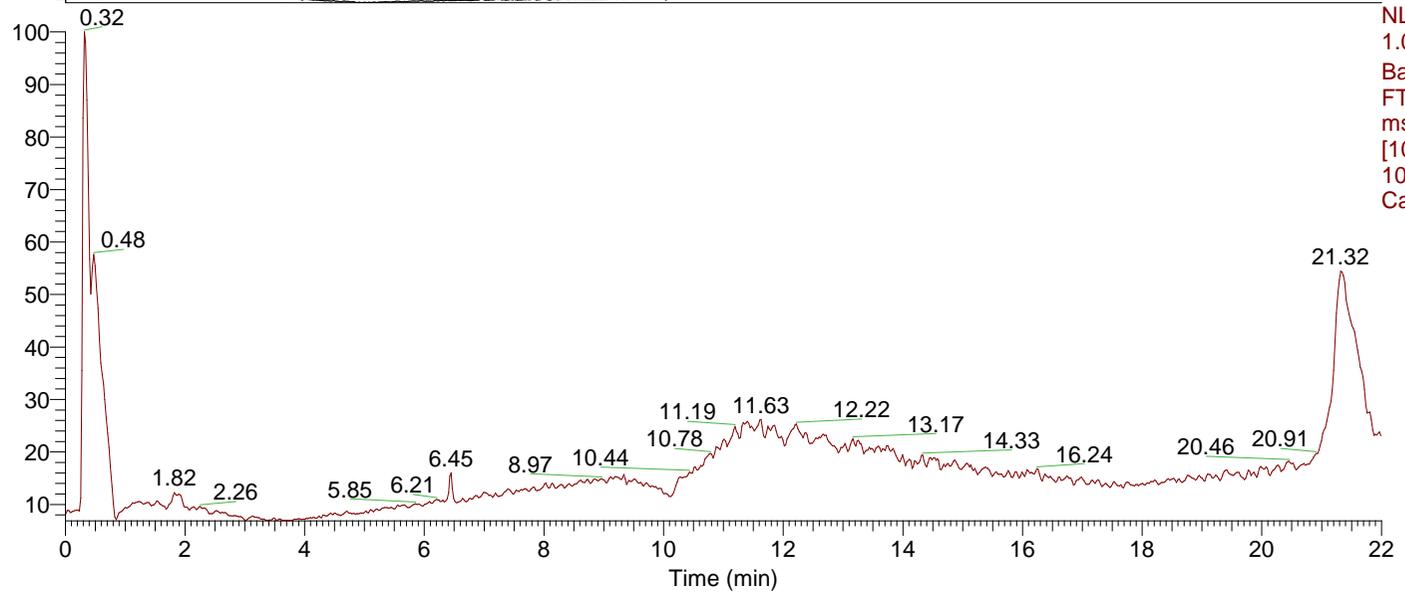

APÊNDICE B – Cromatograma de intensidade do pico base da fração em hexano das folhas de *F. maxima*.

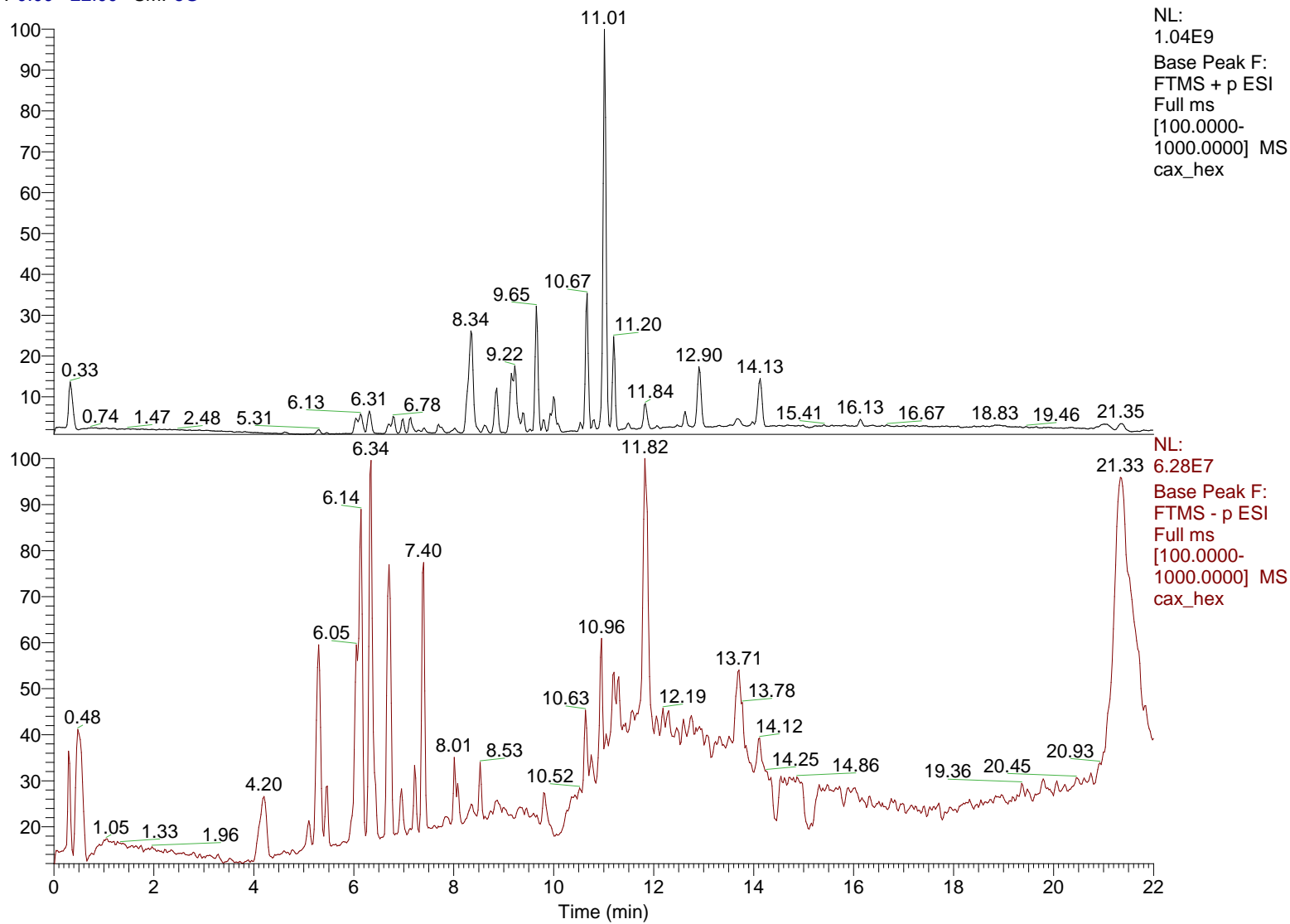

APÊNDICE C – Cromatograma de intensidade do pico base da fração em diclorometano das folhas de *F. maxima*

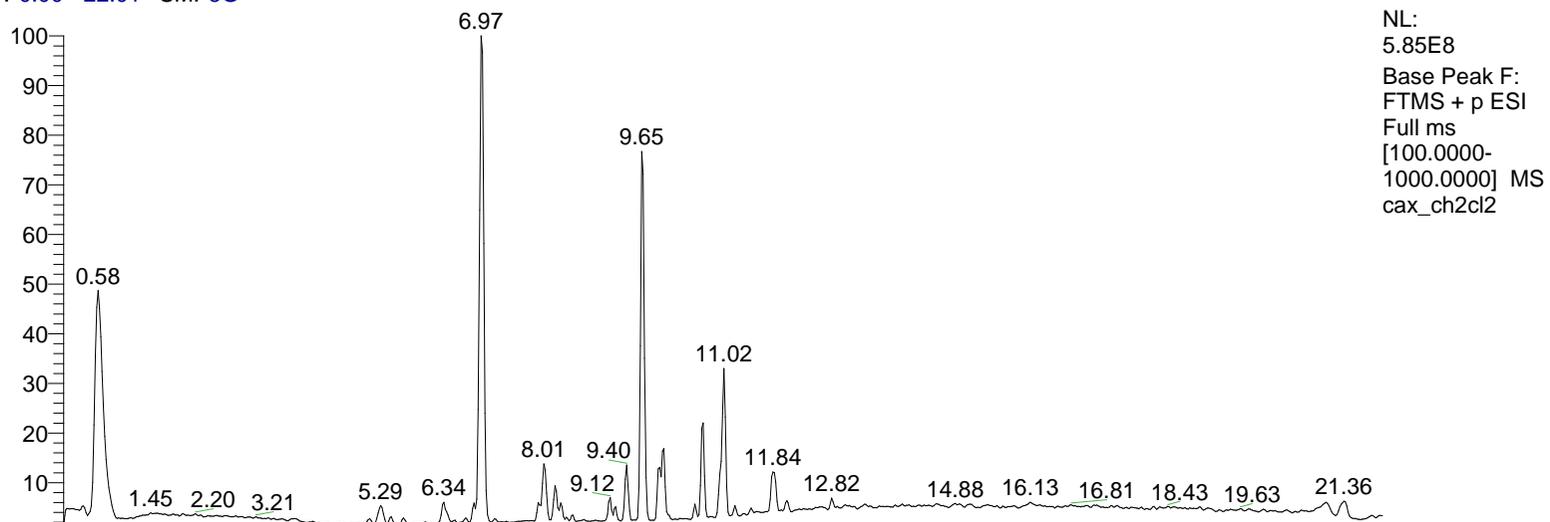

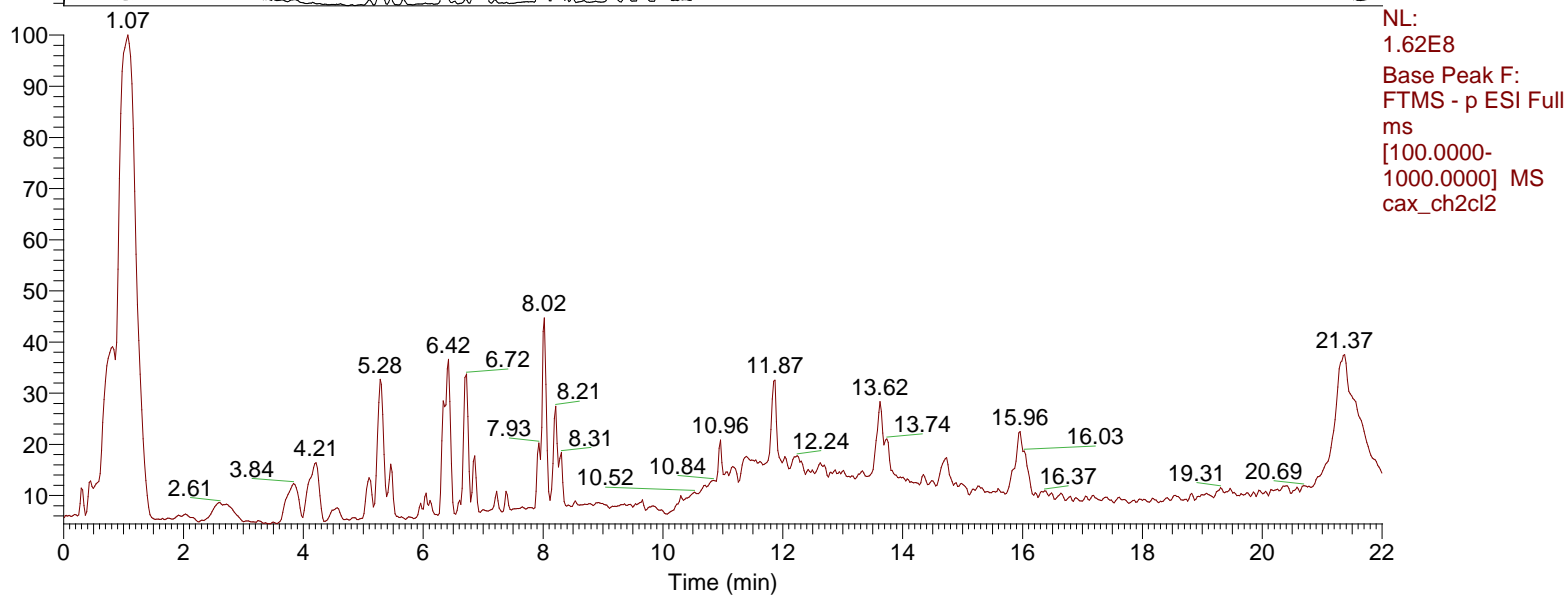

APÊNDICE D – Cromatograma de intensidade do pico base da fração em acetato de etila das folhas de *F. maxima*.

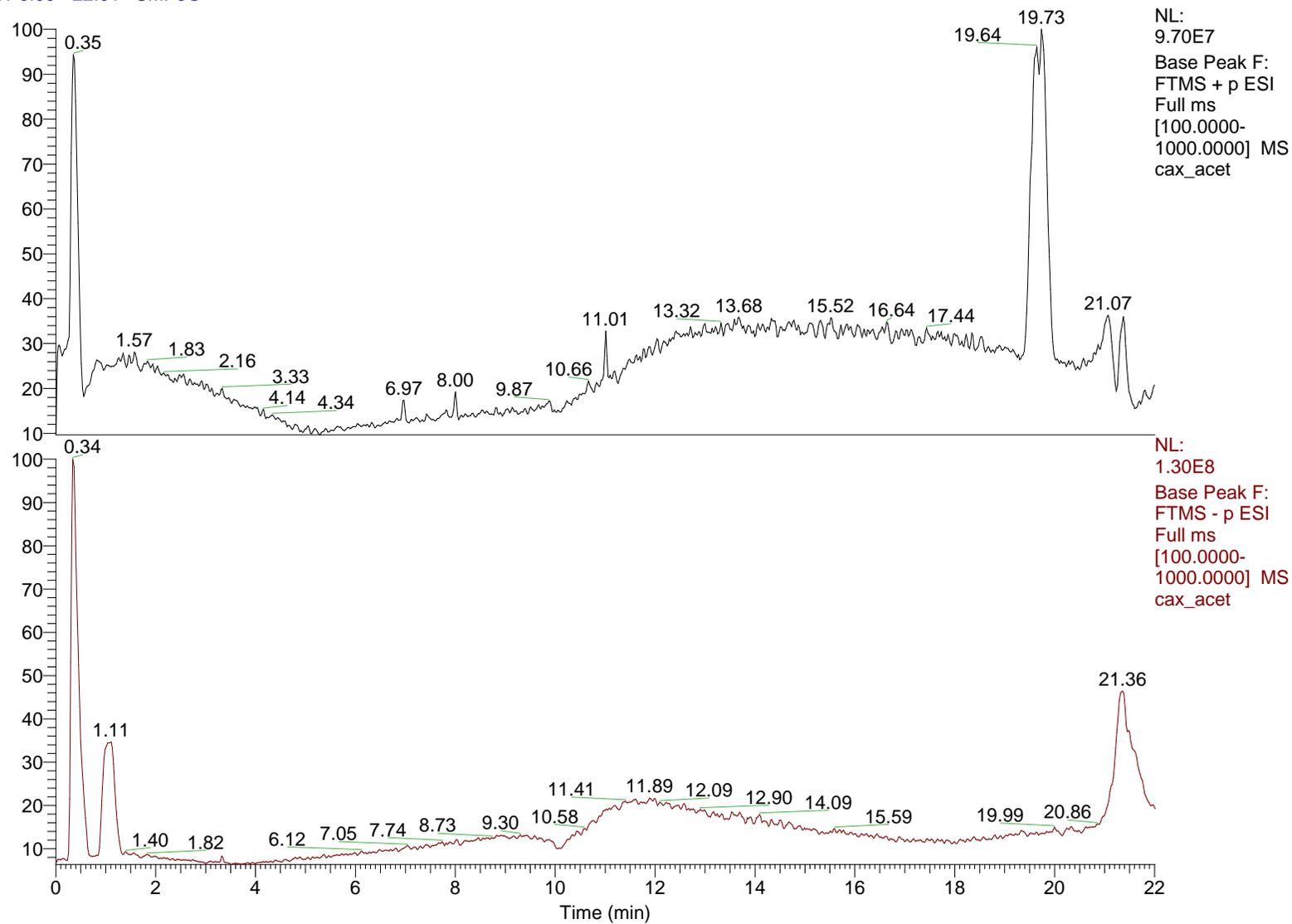

APÊNDICE E - Substâncias anotadas no extrato etanólico das folhas de *F. maxima*

| Nome | TR (min) | Aduto | m/z bibliot. | m/z Exp. | Erro ppm | Cosseno | Modo Ionização | Classe Química |
|---|---|---|---|---|---|---|---|---|
| 4-Aminobutanoato | 0,33 | [M+H]⁺ | 104,107 | 104,1074 | 3,0 | 0,90 | POS (+) | Aminoácido |
| Prolina | 0,33 | [M+H]⁺ | 116,071 | 116,0709 | 0,0 | 0,89 | POS (+) | Aminoácido |
| L-Valina | 0,31 | [M+H]⁺ | 118,086 | 118,0865 | 4,0 | 0,95 | POS (+) | Aminoácido |
| Nicotinamida | 0,36 | [M+H]⁺ | 123,055 | 123,0556 | 4,0 | 0,97 | POS (+) | Alcaloide de piridina |
| Pipecolato | 0,31 | [M+H]⁺ | 130,086 | 130,0864 | 3,0 | 0,96 | POS (+) | Aminoácido |
| Adenina | 0,38 | [M+H]⁺ | 136,062 | 136,0619 | 0,0 | 0,96 | POS (+) | Alcaloide de pteridina |
| 3-hidroxibenzoato | 0,71 | [M-H]⁻ | 137,024 | 137,0243 | 0,0 | 0,84 | NEG (-) | Fenilpropanoide |
| Aldeído protocatecuico | 0,71 | [M-H]⁻ | 137,024 | 137,0243 | 0,0 | 0,72 | NEG (-) | Fenilpropanoide |
| Trigonellina | 0,33 | [M+H]⁺ | 138,055 | 138,0551 | 0,0 | 0,94 | POS (+) | Alcaloide |
| 2,3-dihidroxibenzoato | 0,41 | [M-H]⁻ | 153,019 | 153,0192 | 0,0 | 0,95 | NEG (-) | Fenilpropanoide |
| 4-Cumarato | 7,92 | [M+H]⁺ | 165,055 | 165,0548 | 1,0 | 0,95 | POS (+) | Ác. cinâmico e derivados |

APÊNDICE E - Substâncias anotadas no extrato etanólico das folhas de *F. maxima*

| Nome | TR (min) | Aduto | m/z bibliot. | m/z Exp. | Erro ppm | Cosseno | Modo Ionização | Classe Química |
|---|---|---|---|---|---|---|---|---|
| D-fenilalanina | 0,33 | [M+H]$^+$ | 166,086 | 166,0864 | 2,0 | 0,93 | POS (+) | Aminoácido |
| Esculetina | 0,39 | [M-H]$^-$ | 177,019 | 177,0192 | 0,0 | 0,93 | NEG (-) | Cumarina |
| 4-(2,6,6-Trimethil-2-Ciclohexen-1-il)2-Butanona | 7,12 | [M+H-H2O]$^+$ | 177,164 | 177,1639 | 0,0 | 0,94 | POS (+) | Sesquiterpeno |
| Ác. Azelaico | 0,48 | [M-H]$^-$ | 187,098 | 187,0974 | 2,0 | 0,91 | NEG (-) | Ácido graxo |
| D-(-)-ácido quínico | 0,41 | [M-H]$^-$ | 191,056 | 191,0559 | 0,0 | 0,95 | NEG (-) | Ác. carboxílico |
| Loliolídeo | 0,62 | [M+H]$^+$ | 197,117 | 197,1174 | 2,0 | 0,95 | POS (+) | Apocarotenóide |
| Ác. abscísico | 0,60 | [M+H-H2O]$^+$ | 247,133 | 247,1331 | 0,0 | 0,94 | POS (+) | Apocarotenóide |
| Palmitamideo | 8,73 | [M+H]$^+$ | 256,262 | 256,2635 | 0,0 | 0,94 | POS (+) | Amida graxa |
| Ác. 16-hidroxipalmítico | 8,53 | [M-H]$^-$ | 271,227 | 271,2278 | 3,0 | 0,78 | NEG (-) | Ác. graxo |
| Ác. 9-hidroxi-10,12,15-octadecatrienoico | 6,14 | [M+H-H2O]$^+$ | 277,216 | 277,2164 | 1,0 | 0,89 | POS (+) | Ácido graxo |

APÊNDICE E - Substâncias anotadas no extrato etanólico das folhas de *F. maxima*

| Nome | TR (min) | Aduto | m/z bibliot. | m/z Exp. | Erro ppm | Cosseno | Modo Ionização | Classe Química |
|---|---|---|---|---|---|---|---|---|
| 9(10)-Epome | 8,64 | [M+H-H2O]⁺ | 279,231 | 279,2320 | 3,0 | 0,91 | POS (+) | ND |
| Luteolina | 0,54 | [M-H]⁻ | 285,040 | 285,0405 | 1,0 | 0,88 | NEG (-) | Flavonoide |
| Prunetina | 3,41 | [M+H]⁺ | 285,076 | 285,0759 | 0,0 | 0,94 | POS (+) | Isoflavona |
| FA 18:5+2O | 3,84 | [M-H]⁻ | 305,175 | 305,1758 | 2,0 | 0,76 | NEG (-) | Octadecanóide |
| Éster etílico do ác. linolênico | 8,80 | [M+H]⁺ | 307,262 | 307,2632 | 4,0 | 0,88 | POS (+) | Monoéster |
| 3,4-Difenil-7-hidroxicumarina | 4,38 | [M+H]⁺ | 315,102 | 315,0864 | 49,0 | 0,77 | POS (+) | Neoflavonoide |
| Ác. 9,12,13-trihidroxioctadeca-10,15-dienoico | 0,79 | [M-H]⁻ | 327,218 | 327,2176 | 0,0 | 0,88 | NEG (-) | Octadecanóide |
| Carpacromeno | 6,56 | [M+H]⁺ | 337,107 | 337,1071 | 0,0 | 0,72 | POS (+) | Flavonoide |
| Sucrose | 0,30 | [M-H]⁻ | 341,109 | 341,1085 | 2,0 | 0,96 | NEG (-) | Sacarídeo |
| Ác. (+/-)-4-hidroxi-5,7,10,13,16,19-docosahexaenoico | 5,73 | [M-H]⁻ | 343,230 | 343,2490 | 55,0 | 0,77 | NEG (-) | Docosanoide |

APÊNDICE E - Substâncias anotadas no extrato etanólico das folhas de *F. maxima*

| Nome | TR (min) | Aduto | m/z bibliot. | m/z Exp. | Erro ppm | Cosseno | Modo Ionização | Classe Química |
|---|---|---|---|---|---|---|---|---|
| (4-hidroxi-3,5-dimetoxifenil)metil-β-D-Glicopiranosideo | | [M+Na] | 369,116 | 369,0980 | 49,0 | 0,72 | POS (+) | ND |
| 4-(2,6,6-Trimetil-4-oxo-2-ciclohexen-1-il)-2-butanil β-D-glicopiranosídeo | 0,37 | [M+H]$^+$ | 373,222 | 373,2225 | 0,0 | 0,77 | POS (+) | Apocarotenóide |
| α,α-Trehalose | 0,30 | N/D | 387,115 | 387,1143 | 0,0 | 0,92 | POS (+) | Sacarídeo |
| Lupeol | 13,63 | [M-H2O+H]$^+$ | 409,382 | 409,3832 | 2,0 | 0,74 | POS (+) | Triterpeno |
| 3-(3,4-dihidroxifenil)-5,7-dihidroxi-6,8-bis(3-metilbut-2-enil)cromen-4-ona | 5,98 | [M-H]$^-$ | 421,166 | 421,1657 | 0,0 | 0,82 | NEG (-) | ND |
| Vitexina | 0,34 | [M-H]$^-$ | 431,098 | 431,0980 | 0,0 | 0,91 | NEG (-) | Flavonoide |
| Isoorientina | 0,40 | [M+H]$^+$ | 449,108 | 449,1084 | 0,0 | 0,85 | POS (+) | Flavona |
| Enoxolona | 6,22 | [M+H-H2O]$^+$ | 453,336 | 453,3371 | 2,0 | 0,82 | POS (+) | Triterpeno |
| Enoxolona | 6,70 | [M+H]$^+$ | 471,347 | 471,3476 | 2,0 | 0,92 | POS (+) | Triterpeno |

APÊNDICE E - Substâncias anotadas no extrato etanólico das folhas de *F. maxima*

| Nome | TR (min) | Aduto | m/z bibliot. | m/z Exp. | Erro ppm | Cosseno | Modo Ionização | Classe Química |
|---|---|---|---|---|---|---|---|---|
| Ác. 2,3,19-trihidroxiolean-12-en-28-oico | 4,87 | [M+H]$^+$ | 489,357 | 489,3578 | 1,0 | 0,72 | POS (+) | Triterpeno |
| Ác. -3-(hexopiranosiloxi)-2-hidroxipropil ester-9,12,15-Octadecatrienoico | 6,34 | M+FA-H | 559,312 | 559,3126 | 1,0 | 0,89 | POS (+) | Glicerolipídeo |
| 6-[4,5-dihidroxi-6-(hidroximetil)-3-[-3,4,5-trihidroxi-6-metiloxan-2-il]oxioxan-2-il]-5,7-dihidroxi-2-(4-hidroxifenil)cromen-4-ona | 0,34 | [M-H]$^-$ | 577,156 | 577,1559 | 0,0 | 0,77 | NEG (-) | ND |

APÊNDICE E - Substâncias anotadas no extrato etanólico das folhas de *F. maxima*

| Nome | TR (min) | Aduto | m/z bibliot. | m/z Exp. | Erro ppm | Cosseno | Modo Ionização | Classe Química |
|---|---|---|---|---|---|---|---|---|
| 6-[4,5-dihidroxi-6-(hidroximetil)-3-[-3,4,5-trihidroxi-6-metiloxan-2-il]oxioxan-2-il]-5,7-dihidroxi-2-(4-hidroxifenil)cromen-4-ona | 0,33 | [M+H]$^+$ | 579,171 | 579,1716 | 0,0 | 0,80 | POS (+) | Flavona |

APÊNDICE F – Tabela das substâncias anotadas na fração em hexano das folhas de *F. maxima*

| Nome | TR (min) | Aduto | *m/z* bibliot. | *m/z* Exp. | Erro ppm | Cosseno | Modo Ionização | Classe Química |
|---|---|---|---|---|---|---|---|---|
| 4-Aminobutanoato | 0,33 | [M+H]⁺ | 104,107 | 104,1074 | 3,0 | 0,90 | POS (+) | Aminoácido |
| Adenina | 0,38 | [M+H]⁺ | 136,062 | 136,0619 | 0,0 | 0,96 | POS (+) | Alcaloide de pteridina |
| Ác. p-cumárico | 7,95 | [M+H-H2O]⁺ | 147,044 | 147,0442 | 0,0 | 0,94 | POS (+) | Ác. cinâmico e derivados |
| 4-Cumarato | 7,92 | [M+H]⁺ | 165,055 | 165,0548 | 1,0 | 0,95 | POS (+) | Ác. cinâmico e derivados |
| 4-(2,6,6-Trimethil-2-Ciclohexen-1-il)2-Butanona | 7,12 | [M+H-H2O]⁺ | 177,164 | 177,1639 | 0,0 | 0,94 | POS (+) | ND |
| Vasicina | 14,56 | [M+H]⁺ | 189,102 | 189,0912 | 57 | 0,76 | POS (+) | Alcaloide quinazolínico |
| 2,4,7,9-tetrametil-5-decino-4,7-diol | 9,54 | [M+H-2H2O]⁺ | 191,179 | 191,1797 | 3,0 | 0,85 | POS (+) | ND |
| Loliolídeo | 0,62 | [M+H]⁺ | 197,117 | 197,1174 | 2,0 | 0,95 | POS (+) | Apocarotenóide |
| 4-hidroxi-4-(3-hidroxibutil)-3,5,5-trimetil-2-ciclohexen-1-one | 0,79 | [M-H2O+H]⁺ | 209,153 | 209,1538 | 3,0 | 0,70 | POS (+) | Apocarotenóide |
| (+)-Nootkatone | 5,87 | [M+H]⁺ | 219,174 | 219,1746 | 2,0 | 0,87 | POS (+) | Sesquiterpeno |
| Azelato de dietila | 6,14 | [M+H]⁺ | 245,174 | 245,1749 | 4,0 | 0,85 | POS (+) | Monoéster |
| Walleminona | 8,46 | [M+H]⁺ | 253,179 | 253,1799 | 3,0 | 0,76 | POS (+) | Sesquiterpeno |
| Palmitamideo | 8,73 | [M+H]⁺ | 256,262 | 256,2635 | 6,0 | 0,94 | POS (+) | Amida graxa |
| Ác. estearidônico | 8,82 | [M+H-H2O]⁺ | 259,205 | 259,2058 | 3,0 | 0,74 | POS (+) | Ác. graxo insaturado |
| Ác. linoleico conjugado (10E,12Z) | 9,41 | [M+H-H2O]⁺ | 263,237 | 263,2372 | 0,0 | 0,87 | POS (+) | Ác. graxo |

APÊNDICE F – Tabela das substâncias anotadas na fração em hexano das folhas de *F. maxima*

| Nome | TR (min) | Aduto | *m/z* bibliot. | *m/z* Exp. | Erro ppm | Cosseno | Modo Ionização | Classe Química |
|---|---|---|---|---|---|---|---|---|
| Ác. 16-hidroxipalmítico | 8,53 | [M-H]⁻ | 271,227 | 271,2278 | 3,0 | 0,78 | NEG (-) | Ác. graxo |
| 9-OxoOtre | 3,98 | [M+H-H2O]⁺ | 275,201 | 275,2008 | 0,0 | 0,88 | POS (+) | Octadecanoide |
| Ác. 13-ceto-9,11-octadecadienoico | 9,10 | [M+H-H2O]⁺ | 277,215 | 277,2162 | 4,0 | 0,88 | POS (+) | Ác. graxo |
| Ác. 9-hidroxi-10,12,15-octadecatrienoico | 6,14 | [M+H-H2O]⁺ | 277,216 | 277,2164 | 1,0 | 0,89 | POS (+) | Ác. graxo |
| 9(10)-Epome | 8,64 | [M+H-H2O]⁺ | 279,231 | 279,2320 | 3,0 | 0,91 | POS (+) | ND |
| 9-octadecenamida | 9,09 | [M+H]⁺ | 282,279 | 282,2791 | 1,0 | 0,80 | POS (+) | Amida primária |
| Luteolina | 0,54 | [M-H]⁻ | 285,040 | 285,0405 | 1,0 | 0,88 | NEG (-) | Flavonoide |
| Prunetina | 3,41 | [M+H]⁺ | 285,076 | 285,0759 | 0,0 | 0,94 | POS (+) | Isoflavona |
| 6-Gingerol | 3,79 | [M-H]⁻ | 293,176 | 293,1758 | 0,0 | 0,91 | NEG (-) | Fenólico |
| 13-HOTrE | 6,14 | [M-H]⁻ | 293,212 | 293,2120 | 0,0 | 0,95 | NEG (-) | Octadecanóide |
| Ác. 9-oxo-10,12-octadecadienoico | 7,14 | [M+H]⁺ | 295,226 | 295,2268 | 0,0 | 0,82 | POS (+) | Ácido graxo |
| FA 18:2+1O | 6,68 | [M-H]⁻ | 295,227 | 295,2277 | 2,0 | 0,92 | NEG (-) | Octadecanóide |
| Haemateina | 1,52 | [M+H]⁺ | 301,070 | 301,0710 | 2,0 | 0,77 | POS (+) | Pterocarpano |
| 5-Hete | 8,43 | [M+H-H2O]⁺ | 303,231 | 303,2320 | 4,0 | 0,81 | POS (+) | ND |
| FA 18:5+2O | 3,84 | [M-H]⁻ | 305,175 | 305,1758 | | 0,76 | NEG (-) | Octadecanóide |
| Éster etílico do ác. estearidônico | 8,36 | [M+H]⁺ | 305,247 | 305,2475 | 1,0 | 0,84 | POS (+) | Monoéster |
| Éster etílico do ác. linolênico | 8,80 | [M+H]⁺ | 307,262 | 307,2632 | 3,0 | 0,88 | POS (+) | Monoéster |
| Éster etílico do ác. 9,11,13-octadecatrienoico | 8,80 | [M+H]⁺ | 307,263 | 307,2632 | 1,0 | 0,86 | POS (+) | Monoéster |

APÊNDICE F – Tabela das substâncias anotadas na fração em hexano das folhas de *F. maxima*

| Nome | TR (min) | Aduto | *m/z* bibliot. | *m/z* Exp. | Erro ppm | Cosseno | Modo Ionização | Classe Química |
|---|---|---|---|---|---|---|---|---|
| 3,4-Difenil-7-hidroxicumarina | 4,38 | [M+H]$^+$ | 315,102 | 315,0864 | 49 | 0,77 | POS (+) | Neoflavonoide |
| Ác. 5-[(8as)-2,5,5,8a-tetrametil-3-oxo-4a,6,7,8-tetrahidro-4H-naftalen-1-il]-3-metilpentanoico | 7,34 | [M+H]$^+$ | 321,242 | 321,2426 | 1,0 | 0,71 | POS (+) | ND |
| Éster etílico do ác. ω-3-araquidônico | 9,27 | [M+H]$^+$ | 333,279 | 333,2788 | 1,0 | 0,74 | POS (+) | Monoéster |
| Carpacromeno | 6,56 | [M+H]$^+$ | 337,107 | 337,1071 | 0,0 | 0,72 | POS (+) | Flavonoide |
| 13-docosenamida, (Z) | 10,99 | [M+H]$^+$ | 338,341 | 338,3419 | 2,0 | 0,83 | POS (+) | ND |
| Ác. (+/-)-4-hidroxi-5,7,10,13,16,19-docosahexaenoico | 5,73 | [M-H]$^-$ | 343,230 | 343,2490 | 55 | 0,77 | NEG (-) | Docosanoide |
| Monolinolenina (9c,12c,15c) | 6,33 | [M+H]$^+$ | 353,268 | 353,2687 | 0,0 | 0,82 | POS (+) | Monoglicerídeo |
| 1-linoleoilglicerol | 6,06 | [M+H]$^+$ | 355,284 | 355,2845 | 1,0 | 0,83 | POS (+) | Monoglicerídeo |
| 2-(2-hidroxibut-3-en-2-il)-3a,6,6,9a-tetrametil-2,4,5,5a,7,8,9,9b-octahidro-1H-benzo[E][1]benzofurano-4,5-diol | 7,18 | [M+Na]$^+$ | 361,235 | 361,2352 | 1,0 | 0,85 | POS (+) | ND |
| Cortisol | 6,63 | [M+H]$^+$ | 363,217 | 363,2145 | 3,0 | 0,76 | POS (+) | Esteróide |
| Lupeol | 13,63 | [M+H-H2O]$^+$ | 409,382 | 409,3832 | 2,0 | 0,74 | POS (+) | Triterpeno |
| Pomiferina | 6,59 | [M-H]$^-$ | 419,150 | 419,1503 | 1,0 | 0,72 | NEG (-) | Flavonoide |
| Glochidona | 16,44 | [M+H]$^+$ | 423,362 | 423,3626 | 0,0 | 0,82 | POS (+) | Triterpeno |
| Isovitexina | 0,53 | | 431,100 | 431,0981 | 0,0 | 0,93 | NEG (-) | Flavonoide |

APÊNDICE G – Tabela das substâncias anotadas na fração em DCM das folhas de *F. maxima*

| Nome | TR (min) | Aduto | *m/z* bibliot. | *m/z* Exp. | Erro ppm | Cosseno | Modo Ionização | Classe Química |
|---|---|---|---|---|---|---|---|---|
| Niacinamida | 20,51 | [M-H]⁻ | 121,041 | 121,0407 | 95 | 0,82 | NEG (-) | Aminoácido |
| 3-hidroxibenzoato | 0,71 | [M-H]⁻ | 137,024 | 137,0243 | 2,0 | 0,84 | NEG (-) | Fenilpropanoide |
| Aldeído protocatecuico | 0,71 | [M-H]⁻ | 137,024 | 137,0243 | 0,0 | 0,72 | NEG (-) | Fenilpropanoide |
| Esculetina | 0,39 | [M-H]⁻ | 177,019 | 177,0192 | 1,0 | 0,93 | NEG (-) | Cumarina |
| Ác. Azelaico | 0,48 | [M-H]⁻ | 187,098 | 187,0974 | 2,0 | 0,91 | NEG (-) | Ácido graxo |
| Ác. Traumático | 1,11 | [M-H]⁻ | 227,129 | 227,1288 | 1,0 | 0,96 | NEG (-) | Ácido graxo |
| Ác. abscísico | 0,65 | [M-H]⁻ | 263,129 | 263,1288 | 1,0 | 0,96 | NEG (-) | Apocarotenoides |
| Genisteína | 0,89 | [M-H]⁻ | 269,046 | 269,0459 | 0,0 | 0,86 | NEG (-) | Flavonoide |
| Ác. 16-hidroxipalmítico | 8,53 | [M-H]⁻ | 271,227 | 271,2278 | 3,0 | 0,78 | NEG (-) | Ác. graxo |
| Luteolina | 0,54 | [M-H]⁻ | 285,04 | 285,0405 | 1,0 | 0,88 | NEG (-) | Flavonoide |
| 6-Gingerol | 3,79 | [M-H]⁻ | 293,176 | 293,1758 | 0,0 | 0,91 | NEG (-) | Fenólico |
| Cinchonina | | [M-H]⁻ | 293,166 | 293,1770 | 36 | 0,79 | NEG (-) | Alcaloide de cinchona |
| 13-HOTrE | 6,14 | [M-H]⁻ | 293,212 | 293,2120 | 0,0 | 0,95 | NEG (-) | Octadecanóide |
| FA 18:5+2O | 3,84 | [M-H]⁻ | 305,175 | 305,1758 | 2,0 | 0,76 | NEG (-) | Octadecanóide |

APÊNDICE G – Tabela das substâncias anotadas na fração em DCM das folhas de *F. maxima*

| Nome | TR (min) | Aduto | *m/z* bibliot. | *m/z* Exp. | Erro ppm | Cosseno | Modo Ionização | Classe Química |
|---|---|---|---|---|---|---|---|---|
| FA 18:4+2O | 2,61 | [M-H]⁻ | 307,19 | 307,1916 | 4,0 | 0,81 | NEG (-) | Ácido graxo |
| Ác. 9,12,13-trihidroxioctadeca-10,15-dienoico | 0,79 | [M-H]⁻ | 327,218 | 327,2176 | 0,0 | 0,88 | NEG (-) | Octadecanóide |
| FA 18:1+3O | 0,79 | [M-H]⁻ | 329,231 | 329,2327 | 7,0 | 0,91 | NEG (-) | Octadecanóide |
| Sucrose | 0,30 | [M-H]⁻ | 341,109 | 341,1085 | 2,0 | 0,96 | NEG (-) | Sacarídeo |
| Ác. (+/-)-4-hidroxi-5,7,10,13,16,19-docosahexaenoico | 5,73 | [M-H]⁻ | 343,23 | 343,2490 | 55 | 0,77 | NEG (-) | Docosanoide |
| α,α-Trehalose | 0,30 | Unknown | 387,115 | 387,1143 | 1,0 | 0,92 | NEG (-) | Sacarídeo |
| Pomiferina | 6,59 | [M-H]⁻ | 419,15 | 419,1503 | 1,0 | 0,72 | NEG (-) | Flavonoide |
| 3-(3,4-dihidroxifenil)-5,7-dihidroxi-6,8-bis(3-metilbut-2-enil)cromen-4-ona | 5,98 | [M-H]⁻ | 421,166 | 421,1657 | 0,0 | 0,82 | NEG (-) | ND |
| Enoxolona | 6,70 | [M+H]⁺ | 471,347 | 471,3476 | 1,0 | 0,92 | NEG (-) | Triterpenos |
| Éster 3-(hexopiranosiloxi)-2-hidroxipropil do ác. 9,12,15-octadecatrienoico | 6,34 | M+FA-H | 559,312 | 559,3126 | 0,0 | 0,89 | NEG (-) | Glicerolipídeo |
| 2-[4-[-3-(4-hidroxi-3,5-dimetoxifenil)-1,3,3a,4,6,6a-hexahidrofuro[3,4-c]furan-6-il]-2,6-dimetoxifenoxi]-6-(hidroximetil)oxano-3,4,5-triol | 0,39 | [M-H]⁻ | 579,209 | 579,2086 | 0,0 | 0,79 | NEG (-) | Lignana |
| Ác. 2-O-E-p-coumaroil alftólico | 8,00 | [M-H]⁻ | 617,385 | 617,3847 | 1,0 | 0,87 | NEG (-) | Triterpenos |

APÊNDICE H – Tabela das substâncias anotadas na fração em AcOEt das folhas de *F. maxima*

| Nome | TR (min) | Aduto | *m/z* bibliot. | *m/z* Exp. | Erro ppm | Cosseno | Modo Ionização | Classe Química |
|---|---|---|---|---|---|---|---|---|
| Nicotinamida | 0,36 | [M+H]⁺ | 123,055 | 123,0556 | 0,0 | 0,97 | POS (+) | Alcaloide de piridina |
| 3-hidroxibenzoato | 0,71 | [M-H]⁻ | 137,024 | 137,0243 | 2,0 | 0,84 | NEG (-) | Fenilpropanoide |
| Aldeído protocatecuico | 0,71 | [M-H]⁻ | 137,024 | 137,0243 | 0,0 | 0,72 | NEG (-) | Fenilpropanoide |
| 2,3-dihidroxibenzoato | 0,41 | [M-H]⁻ | 153,019 | 153,0192 | 1,0 | 0,95 | NEG (-) | Fenilpropanoide |
| Esculetina | 0,39 | [M-H]⁻ | 177,019 | 177,0192 | 1,0 | 0,93 | NEG (-) | Cumarina |
| Ác. Azelaico | 0,48 | [M-H]⁻ | 187,098 | 187,0974 | 2,0 | 0,91 | NEG (-) | Ácido graxo |
| D-(-)-ácido quínico | 0,41 | [M-H]⁻ | 191,056 | 191,0559 | 0,0 | 0,95 | NEG (-) | ND |
| Ác. traumático | 1,06 | [M+H]⁺; [M-H]⁻ | 229,143 | 229,1437 | 1,0 | 0,83 | POS (+); NEG (-) | Ác. dicarboxílico |
| Ác. abscísico | 0,65 | [M-H]⁻ | 263,129 | 263,1288 | 1,0 | 0,96 | NEG (-) | Apocarotenoides |
| Luteolina | 0,54 | [M-H]⁻ | 285,04 | 285,0405 | 1,0 | 0,88 | NEG (-) | Flavonoide |
| Ác. 9,12,13-trihidroxioctadeca-10,15-dienoico | 0,79 | [M-H]⁻ | 327,218 | 327,2176 | 0,0 | 0,88 | NEG (-) | Octadecanóide |
| Metoxihaemoventosina | 0,34 | [M+H]⁺ | 335,075 | 335,0946 | 52 | 0,87 | POS (+) | Naftaleno |
| Sucrose | 0,30 | [M-H]⁻ | 341,109 | 341,1085 | 2,0 | 0,96 | NEG (-) | Sacarídeo |
| 4-(2,6,6-Trimetil-4-oxo-2-ciclohexen-1-il)-2-butanil β-D-glicopiranosídeo | 0,37 | [M+H]⁺ | 373,222 | 373,2225 | 0,0 | 0,77 | POS (+) | Apocarotenóide |
| α,α-Trehalose | 0,30 | ND | 387,115 | 387,1143 | 0,0 | 0,92 | POS (+) | Sacarídeo |
| Vitexina | 0,34 | [M-H]⁻ | 431,098 | 431,0980 | 0,0 | 0,91 | NEG (-) | Flavonoide |
| Isovitexina | 0,60 | [M+H]⁺ | 433,113 | 433,1134 | 0,0 | 0,92 | POS (+) | Flavona |
| Isoorientina | 0,40 | [M+H]⁺ | 449,108 | 449,1084 | 0,0 | 0,85 | POS (+) | Flavona |

APÊNDICE H – Tabela das substâncias anotadas na fração em AcOEt das folhas de *F. maxima*

| Nome | TR (min) | Aduto | *m/z* bibliot. | *m/z* Exp. | Erro ppm | Cosseno | Modo Ionização | Classe Química |
|---|---|---|---|---|---|---|---|---|
| 6-[4,5-dihidroxi-6-(hidroximetil)-3-[-3,4,5-trihidroxi-6-metiloxan-2-il]oxioxan-2-il]-5,7-dihidroxi-2-(4-hidroxifenil)cromen-4-ona | 0,34 | [M-H]⁻ | 577,156 | 577,1559 | 0,0 | 0,77 | NEG (-) | ND |
| 6-[4,5-dihidroxi-6-(hidroximetil)-3-[-3,4,5-trihidroxi-6-metiloxan-2-il]oxioxan-2-il]-5,7-dihidroxi-2-(4-hidroxifenil)cromen-4-ona | 0,33 | [M+H]⁺ | 579,171 | 579,1716 | 0,0 | 0,8 | POS (+) | Flavona |
| Glc-Glc-octadecatrienoil-sn-glicerol | 21,13 | [M-H]⁻ | 721,361 | 721,3618 | 5,0 | 0,72 | NEG (-) | ND |

APÊNDICE I – Tabela das substâncias anotadas na subfração R3+4 da fração em DCM das folhas de *F. maxima*

| Nome | TR (min) | Aduto | *m/z* bibliot. | *m/z* exp. | Erro ppm | Cosseno | Modo Ionização | Classe Química |
|---|---|---|---|---|---|---|---|---|
| Pipecolato | 0,84 | [M+H]⁺ | 130,086 | 130,0864 | 3,0 | 0,88 | POS (+) | Aminoácido |
| Suberato | 8,30 | [M-H]⁻ | 173,082 | 173,0818 | 1,0 | 0,74 | NEG (-) | Ác. carboxílico |
| Ác. acetilsalicílico | 16,24 | [M+H]⁺ | 181,049 | 181,0496 | 0,0 | 0,84 | POS (+) | Ác. fenólico |
| Ác. azelaico | 9,88 | [M-H]⁻ | 187,098 | 187,0976 | 2,0 | 0,92 | NEG (-) | Ác. carboxílico |
| Loliolideo | 7,91 | [M+H]⁺ | 197,117 | 197,1170 | 0,0 | 0,96 | POS (+) | Apocarotenoide |
| N-metildodecilamina | 13,20 | [M+H]⁺ | 200,237 | 200,2374 | 1,0 | 0,74 | POS (+) | |
| Hidroxi-1-(2-hidroxietil)-2,2,6,6-tetrametilpiperidinea | 0,83 | [M+H]⁺ | 202,180 | 202,1802 | 1,0 | 0,91 | POS (+) | Alcaloide de lisina |
| Vitamina K1 | 12,49 | M+2H] | 226,182 | 226,1801 | 8,0 | 0,79 | POS (+) | Meroterpeno |
| Palmitamideo | 15,70 | [M+H]⁺ | 256,262 | 256,2634 | 5,0 | 0,97 | POS (+) | Ác. graxo |
| Ác. linoleico conjugado | 15,57 | [M+H-H2O]⁺ | 263,237 | 263,2369 | 0,0 | 0,89 | POS (+) | Ác. graxo |
| Ác. oleico | 15,90 | [M+H-H2O]⁺ | 265,253 | 265,2525 | 1,0 | 0,96 | POS (+) | Ác. graxo |
| 9(10)-EpOME | 14,80 | [M+H-H2O]⁺ | 279,231 | 279,2320 | 2,0 | 0,91 | POS (+) | ND |
| 9-octadecenamida | 15,81 | [M+H]⁺ | 282,279 | 282,2789 | 0,0 | 0,84 | POS (+) | Amida graxa |
| Octadecanamida | 16,07 | [M+H]⁺ | 284,295 | 284,2949 | 2,0 | 0,90 | POS (+) | ND |
| Lauril dietanolamida | 14,39 | [M+H]⁺ | 288,253 | 288,2533 | | 0,93 | POS (+) | Amida graxa |
| Cinchonina | 12,94 | [M-H]⁻ | 293,166 | 293,1789 | 42 | 0,73 | NEG (-) | Alcaloide de cinchona |
| trans-EKODE-(E)-Ib | 14,55 | [M+H-H2O]⁺ | 293,211 | 293,2091 | 6,0 | 0,71 | POS (+) | Octadecanoide |
| Aurapteno | 16,81 | [M-H]⁻ | 297,150 | 297,1528 | 9,0 | 0,85 | NEG (-) | Éter cumarínico |

APÊNDICE I – Tabela das substâncias anotadas na subfração R3+4 da fração em DCM das folhas de *F. maxima*

| Nome | TR (min) | Aduto | *m/z* bibliot. | *m/z* exp. | Erro ppm | Cosseno | Modo Ionização | Classe Química |
|---|---|---|---|---|---|---|---|---|
| Ác. 8-hidroxi-8-(3-octiloxiran-2-il)octanoico | 14,91 | [M-H2O+H]$^+$ | 297,243 | 297,2427 | 1,0 | 0,79 | POS (+) | Octadecanoide |
| Hidroquinidina | 13,65 | [M-H]$^-$ | 325,192 | 325,1842 | 23 | 0,90 | NEG (-) | ND |
| 13-docosenamida | 16,63 | [M+H]$^+$ | 338,341 | 338,3412 | 2,0 | 0,84 | POS (+) | ND |
| 3,6,9,12-Tetraoxatetracosan-1-ol | 15,48 | [M+H]$^+$ | 363,310 | 363,3104 | 0,0 | 0,80 | POS (+) | Glicerolipídeo |
| Desmosterol | 17,61 | [M+H-H2O]$^+$ | 367,336 | 367,3363 | 1,0 | 0,91 | POS (+) | Esteroide |
| Dioctil sulfosuccinato | 18,54 | [M-H]$^-$ | 421,227 | 421,2266 | 0,0 | 0,86 | NEG (-) | ND |
| Feoforbídeo A | 16,20 | [M+H]$^+$ | 593,274 | 593,2758 | 3,0 | 0,89 | POS (+) | Alcaloide de triptofano |

APÊNDICE J – Tabela das substâncias anotadas na subfração R5+6 da fração em DCM das folhas de *F. maxima*

| Nome | TR (min) | Aduto | *m/z* bibliot. | *m/z* exp. | Erro ppm | Cosseno | Modo Ionização | Classe Química |
|---|---|---|---|---|---|---|---|---|
| 4-Hidroxibenzaldeído | 5,59 | [M+H]⁺ | 123,044 | 123,0443 | 2,0 | 0,83 | POS (+) | Ác. fenólico |
| Pipecolato | 0,84 | [M+H]⁺ | 130,086 | 130,0864 | 3,0 | 0,88 | POS (+) | Aminoácido |
| Suberato | 8,30 | [M-H]⁻ | 173,082 | 173,0818 | | 0,74 | NEG (-) | Ác. carboxílico |
| Coniferaldeído | 10,17 | [M+H]⁺ | 179,070 | 179,0704 | 0,0 | 0,70 | POS (+) | Fenilpropanoide |
| Loliolideo | 7,91 | [M+H]⁺ | 197,117 | 197,1170 | 0,0 | 0,96 | POS (+) | Apocarotenoide |
| Ác. laurico leelamida | 14,48 | [M+H]⁺ | 200,201 | 200,2011 | 0,0 | 0,81 | POS (+) | ND |
| N-metildodecilamina | 13,20 | [M+H]⁺ | 200,237 | 200,2374 | 1,0 | 0,74 | POS (+) | ND |
| Sebacato | 10,96 | [M-H]⁻ | 201,113 | 201,1131 | | 0,71 | NEG (-) | Ác. carboxílico |
| Vitamina K1 | 12,49 | M+2H] | 226,182 | 226,1801 | 8,0 | 0,79 | POS (+) | Meroterpeno |
| Dimetil sebacato | 12,91 | [M+H]⁺ | 231,159 | 231,1592 | 0,0 | 0,85 | POS (+) | Éster graxo |
| Palmitamideo | 15,70 | [M+H]⁺ | 256,262 | 256,2634 | 5,0 | 0,97 | POS (+) | Ác. graxo |
| Ác. linoleico conjugado | 15,57 | [M+H-H2O]⁺ | 263,237 | 263,2369 | 0,0 | 0,89 | POS (+) | Ác. graxo |
| Ác. oleico | 15,90 | [M+H-H2O]⁺ | 265,253 | 265,2525 | 1,0 | 0,96 | POS (+) | Ác. graxo |
| Galaxolidona | 14,46 | [M+H]⁺ | 273,185 | 273,1849 | 1,0 | 0,91 | POS (+) | Sesquiterpeno |
| MoNA:936691 DBP | 16,23 | [M+H]⁺ | 279,159 | 279,1590 | 0,0 | 0,92 | POS (+) | ND |
| 9(10)-EpOME | 14,80 | [M+H-H2O]⁺ | 279,231 | 279,2320 | 2,0 | 0,91 | POS (+) | ND |
| 9-octadecenamida | 15,81 | [M+H]⁺ | 282,279 | 282,2789 | 0,0 | 0,84 | POS (+) | Amida graxa |
| Octadecanamida | 16,07 | [M+H]⁺ | 284,295 | 284,2949 | 2,0 | 0,90 | POS (+) | ND |

APÊNDICE J – Tabela das substâncias anotadas na subfração R5+6 da fração em DCM das folhas de *F. maxima*

| Nome | TR (min) | Aduto | *m/z* bibliot. | *m/z* exp. | Erro ppm | Cosseno | Modo Ionização | Classe Química |
|---|---|---|---|---|---|---|---|---|
| Prunetina | 12,66 | [M+H]⁺ | 285,076 | 285,0758 | 0,0 | 0,93 | POS (+) | Isoflavonoide |
| Octinoxato | 13,89 | [M+H]⁺ | 291,195 | 291,1954 | 0,0 | 0,85 | POS (+) | Fenilpropanoide |
| Cinchonina | 12,94 | [M-H]⁻ | 293,166 | 293,1789 | 42 | 0,73 | NEG (-) | Alcaloide de cinchona |
| trans-EKODE-(E)-Ib | 14,55 | [M+H-H2O]⁺ | 293,211 | 293,2091 | 6,0 | 0,71 | POS (+) | Octadecanoide |
| Ác. 13-ceto-9,11-octadecadienoico | 14,69 | [M+H]⁺ | 295,226 | 295,2265 | 0,0 | 0,85 | POS (+) | Ác. graxo |
| Ác. 8-hidroxi-8-(3-octiloxiran-2-il)octanoico | 14,91 | [M-H2O+H]⁺ | 297,243 | 297,2427 | 1,0 | 0,79 | POS (+) | Octadecanoide |
| FA 18:1+3O | 13,48 | [M-H]⁻ | 329,231 | 329,2332 | 7,0 | 0,89 | NEG (-) | Octadecanoide |
| Metoxihaemoventosina | 13,92 | [M+H]⁺ | 335,075 | 335,0925 | 52 | 0,86 | NEG (-) | Naftoquinona |
| Carpacromeno | 14,66 | [M+H]⁺ | 337,107 | 337,1070 | 0,0 | 0,78 | POS (+) | ND |
| metil 3-oxo-2-[(4-prop-2-eniloxifenil)metileno]benzo[b]furan-5-carboxilato | 14,66 | [M+H]⁺ | 337,107 | 337,1070 | 0,0 | 0,73 | POS (+) | Flavonoide |
| 13-docosenamida | 16,63 | [M+H]⁺ | 338,341 | 338,3412 | 2,0 | 0,84 | POS (+) | ND |
| Desmosterol | 17,61 | [M+H-H2O]⁺ | 367,336 | 367,3363 | 1,0 | 0,91 | POS (+) | Esteroide |
| Éster bis(2-etilhexil) do ác. hexadióico | 16,26 | [M+H]⁺ | 371,313 | 371,3160 | 6,0 | 0,87 | POS (+) | ND |
| Lup-20(29)-en-3-ol, (3alpha)- | 17,79 | [M-H2O+H]⁺ | 409,382 | 409,3828 | 2,0 | 0,74 | POS (+) | Triterpeno |
| Dioctil sulfosuccinato | 18,54 | [M-H]⁻ | 421,227 | 421,2266 | 0,0 | 0,86 | NEG (-) | ND |
| Glochidona | 16,44 | [M+H]⁺ | 423,362 | 423,3626 | 1,0 | 0,82 | POS (+) | Triterpeno |

APÊNDICE J – Tabela das substâncias anotadas na subfração R5+6 da fração em DCM das folhas de *F. maxima*

| Nome | TR (min) | Aduto | *m/z* bibliot. | *m/z* exp. | Erro ppm | Cosseno | Modo Ionização | Classe Química |
|---|---|---|---|---|---|---|---|---|
| Betulina | 16,34 | [M+H-H2O]⁺ | 425,378 | 425,3777 | 0,0 | 0,91 | POS (+) | Triterpeno |
| Bis[2-(2-butoxietoxi)etil] hexandioato | 14,24 | [M+H]⁺ | 435,295 | 435,2953 | 1,0 | 0,90 | POS (+) | Glicerolipídeo |
| 4-N-metil-N-[2,4,6-octatrienoil]valilalanilprolinamida | 18,64 | [M+Na] | 441,247 | 441,2525 | 11 | 0,79 | NEG (-) | Tripeptídeo |
| Ác. ursólico | 15,69 | [M-H]⁻ | 455,353 | 455,3532 | 0,0 | 0,97 | NEG (-) | Triterpeno |
| Feoforbídeo A | 16,20 | [M+H]⁺ | 593,274 | 593,2758 | 3,0 | 0,89 | POS (+) | Alcaloide de triptofano |
| 3-[(6-deoxi-beta-D-talopiranosil)oxi]-1,5,11,14,19-pentahidroxicard-20(22)-enolídeo | 15,64 | [M+Na] | 607,273 | 607,2550 | 29 | 0,71 | POS (+) | Esteroide |

APÊNDICE K – Tabela das substâncias anotadas na subfração R7 da fração em DCM das folhas de *F. maxima*

| Nome | TR (min) | Aduto | *m/z* bibliot. | *m/z* exp. | Erro ppm | Cosseno | Modo Ionização | Classe Química |
|---|---|---|---|---|---|---|---|---|
| Pipecolato | 0,84 | [M+H]⁺ | 130,086 | 130,0864 | 3,0 | 0,88 | POS (+) | Aminoácido |
| Ác. (4S,5Z,6S)-4-(2-metoxi-2-oxoetil)-5-[2-[(E)-3-fenilprop-2-enoil]oxietilidene]-6-[(2S,3R,4S,5S,6R)-3,4,5-trihidroxi-6-(hidroximetil)oxan-2-yl]oxi-4H-pirano-3-carboxílico | 7,40 | [M-H]⁻ | 161,0240 | 161,0242 | 1,0 | 0,72 | NEG (-) | ND |
| Suberato | 8,30 | [M-H]⁻ | 173,0820 | 173,0818 | | 0,74 | NEG (-) | Ác. carboxílico |
| Esculetina | 8,15 | [M+H]⁺ | 179,034 | 179,0337 | 1,0 | 0,75 | POS (+) | Cumarina |
| N-metildodecilamina | 13,20 | [M+H]⁺ | 200,237 | 200,2374 | 1,0 | 0,74 | POS (+) | ND |
| Vitamina K1 | 12,49 | M+2H] | 226,182 | 226,1801 | 8,0 | 0,79 | POS (+) | Meroterpeno |
| Dimetil sebacato | 12,91 | [M+H]⁺ | 231,159 | 231,1592 | 0,0 | 0,85 | POS (+) | Éster graxo |
| Palmitamideo | 15,70 | [M+H]⁺ | 256,262 | 256,2634 | 5,0 | 0,97 | POS (+) | ND |
| Ác. linoleico conjugado | 15,57 | [M+H-H2O]⁺ | 263,237 | 263,2369 | 0,0 | 0,89 | POS (+) | Ác. graxo |
| MoNA:936691 DBP | 16,23 | [M+H]⁺ | 279,159 | 279,1590 | 0,0 | 0,92 | POS (+) | ND |
| 9(10)-EpOME | 14,80 | [M+H-H2O]⁺ | 279,231 | 279,232 | 2,0 | 0,91 | POS (+) | ND |
| 9-octadecenamida | 15,81 | [M+H]⁺ | 282,279 | 282,2789 | 0,0 | 0,84 | POS (+) | Amida graxa |
| Octadecanamida | 16,07 | [M+H]⁺ | 284,295 | 284,2949 | 2,0 | 0,90 | POS (+) | ND |

APÊNDICE K – Tabela das substâncias anotadas na subfração R7 da fração em DCM das folhas de *F. maxima*

| Nome | TR (min) | Aduto | *m/z* bibliot. | *m/z* exp. | Erro ppm | Cosseno | Modo Ionização | Classe Química |
|---|---|---|---|---|---|---|---|---|
| Prunetina | 12,66 | [M+H]⁺ | 285,076 | 285,0758 | 0,0 | 0,93 | POS (+) | Isoflavonoide |
| Octinoxato | 13,89 | [M+H]⁺ | 291,195 | 291,1954 | 0,0 | 0,85 | POS (+) | Fenilpropanoide |
| trans-EKODE-(E)-Ib | 14,55 | [M+H-H2O]⁺ | 293,211 | 293,2091 | 6,0 | 0,71 | POS (+) | Octadecanoide |
| Ác. 13-ceto-9,11-octadecadienoico | 14,69 | [M+H]⁺ | 295,226 | 295,2265 | 0,0 | 0,85 | POS (+) | Ác. graxo |
| 13-docosenamida | 16,63 | [M+H]⁺ | 338,341 | 338,34115 | 2,0 | 0,84 | POS (+) | ND |
| 3,6,9,12-Tetraoxatetracosan-1-ol | 15,48 | [M+H]⁺ | 363,31 | 363,3104 | 0,0 | 0,80 | POS (+) | Glicerolipídeo |
| Desmosterol | 17,61 | [M+H-H2O]⁺ | 367,336 | 367,3363 | 1,0 | 0,91 | POS (+) | Esteroide |
| Éster bis(2-etilhexil) do ác. hexadióico | 16,26 | [M+H]⁺ | 371,313 | 371,316 | 6,0 | 0,87 | POS (+) | ND |
| 7-cetocolesterol | 16,46 | [M+H]⁺ | 401,341 | 401,3416 | 0,0 | 0,73 | POS (+) | Esteroide |
| 17(21)-Hopen-6-ona | 16,60 | [M+H]⁺ | 425,378 | 425,378 | 0,0 | 0,90 | POS (+) | Triterpeno |
| Bis[2-(2-butoxietoxi)etil] hexandioato | 14,24 | [M+H]⁺ | 435,295 | 435,2953 | 1,0 | 0,90 | POS (+) | Glicerolipídeo |
| Ác. betulônico | 15,33 | [M-H2O+H]⁺ | 437,341 | 437,3412 | 1,0 | 0,86 | POS (+) | Triterpeno |
| Ác. sumaresinólico | 15,15 | [M+H-H2O]⁺ | 455,352 | 455,3521 | 0,0 | 0,86 | POS (+) | Triterpeno |
| Enoxolona | 16,98 | [M-H]⁻ | 469,332 | 469,3297 | 5,0 | 0,95 | NEG (-) | Triterpeno |
| AEG(o-16:2/16:0) | 16,77 | [M+H]⁺ | 551,503 | 551,5038 | 1,0 | 0,86 | POS (+) | ND |

APÊNDICE K – Tabela das substâncias anotadas na subfração R7 da fração em DCM das folhas de *F. maxima*

| Nome | TR (min) | Aduto | *m/z* bibliot. | *m/z* exp. | Erro ppm | Cosseno | Modo Ionização | Classe Química |
|------|----------|-------|----------------|------------|----------|---------|----------------|----------------|
| 1-Palmitoil-2-oleoil-sn-glicerol | 16,99 | [M+H-H2O]$^+$ | 577,516 | 577,5193 | 6,0 | 0,81 | POS (+) | ND |
| AEG(o-18:3/18:1) | 17,32 | [M+H]$^+$ | 603,534 | 603,5354 | 1,0 | 0,84 | POS (+) | ND |